\documentclass[a4paper,11pt]{article}
\usepackage{fullpage}
\usepackage{color}
\usepackage{latexsym}
\usepackage{mathrsfs}
\usepackage{enumitem}

\usepackage{amsmath,amssymb,amsthm}
\usepackage{bbm} 
\usepackage{tikz-cd} 
\usepackage{xspace,bbold}
\usepackage[margin=1in]{geometry} 
\usepackage[utf8]{inputenc}
\usepackage{stmaryrd}
\usepackage{comment}
\usepackage{multirow}

\numberwithin{equation}{section}       % equation numbers in each section
\allowdisplaybreaks                    % allow pagebreaks in multiline eqs

\usepackage{jheppub}                   % jheppub.sty

\makeatletter
\gdef\@fpheader{\ }                    % hack the jhep header
\makeatother

\usepackage{diagbox}
\newcommand{\be}{\begin{equation}}
\newcommand{\ee}{\end{equation}}

\newcommand{\SU}[1]{\mathrm{SU}( #1 )}
\newcommand{\SUs}[1]{\mathrm{SU}^*\!(#1)}
\newcommand{\SOs}[1]{\mathrm{SO}^*\!(#1)}
\newcommand{\SL}[1]{\mathrm{SL}( #1 )}
\newcommand{\SO}[1]{\mathrm{SO}( #1 )}
\newcommand{\SOo}[1]{\mathrm{SO_0}( #1 )}
\newcommand{\ISO}[1]{\mathrm{ISO}( #1 )}
\newcommand{\Spin}[1]{\mathrm{Spin}(#1)}
\newcommand{\Spins}[1]{\mathrm{Spin}^*\!(#1)}

\newcommand{\USp}[1]{\mathrm{USp}( #1 )}
\newcommand{\U}[1]{\mathrm{U}(#1)}
\DeclareMathOperator{\GL}{\mathit{GL}}
\newcommand{\Ex}[1]{\mathrm{E}_{#1(#1)}}
\newcommand{\Fff}{\mathrm{F}_{4(4)}}
\newcommand{\Gtt}{\mathrm{G}_{2(2)}}

\DeclareMathOperator{\im}{Im}
\DeclareMathOperator{\so}{\mathfrak{so}}
\DeclareMathOperator{\su}{\mathfrak{su}}

\newcommand{\ex}[1]{\mathfrak{e}_{#1(#1)}}
\newcommand{\GM}[2]{\big<#1,#2\big>}
\newcommand{\obf}[1]{\overline{\mathbf{#1}}}
\newcommand{\mbf}[1]{\mathbf{#1}}

\newcommand{\Gvt}{{G_{\rm VT}}}
\newcommand{\Gh}{{G_{\rm H}}}
\newcommand{\Hvt}{H_{\rm VT}}
\newcommand{\Hh}{H_{\rm H}}
\newcommand{\gh}{\mathfrak{g}_{\rm H}}

\newcommand{\tr}{\mathrm{tr}}

\numberwithin{equation}{section}

\newcommand{\mtw}[2]{\left(\mbf{#1},\mbf{#2}\right)}

\newcommand\Tstrut{\rule{0pt}{2.7ex}}         % = `top' strut
\newcommand\Bstrut{\rule[-1.4ex]{0pt}{0pt}}   % = `bottom' strut

\newcommand{\Esing}{E_{\rm singlet}}

\newcommand{\Leib}[2]{\llbracket #1,#2 \rrbracket}

\newcommand{\nn}{\nonumber}

\newcommand{\bpm}{\begin{pmatrix}}
\newcommand{\epm}{\end{pmatrix}}

\newcommand{\diff}{\mathrm{d}}

\newcommand{\rme}{\mathrm{e}} %for exponentials

\newcommand{\dd}{\mathrm{d}}

\newcommand{\id}{\mathbb{1}}

\DeclareMathOperator{\adj}{ad}

\DeclareMathOperator{\Comm}{C}
\newcommand{\Com}[2]{\Comm_{#2}(#1)}

\newcommand{\Gst}{G_S}
\newcommand{\Giso}{G_{\text{iso}}}
\newcommand{\giso}{\mathfrak{g}_{\text{iso}}}
\newcommand{\Tint}{{T_{\text{int}}}}

\newcommand{\bbZ}{\mathbb{Z}}
\newcommand{\bbR}{\mathbb{R}}
\newcommand{\bbC}{\mathbb{C}}

\newcommand{\rep}[1]{\mathbf{#1}}

\newcommand{\Lgen}{{L}}
\DeclareMathOperator{\vol}{vol}
\newcommand{\Ggauge}{G_{\text{ext}}}
\newcommand{\ggauge}{\mathfrak{g}_{\text{ext}}}

\newcommand{\gleib}{\mathfrak{a}}

\newcommand{\Gred}{G_{\text{gauge}}}
\newcommand{\gred}{\mathfrak{g}_{\text{gauge}}}
\newcommand{\Heis}{\mathrm{Heis}}

\DeclareMathOperator{\Lie}{Lie}
\DeclareMathOperator{\End}{End}

\def\kk{{\boldsymbol\kappa}}

\newcommand{\EM}[1]{\textcolor{cyan} {{\bf #1}}}
\newcommand{\MP}[1]{\textcolor{red} {{\bf #1}}}
\newcommand{\GJ}[1]{\textcolor{magenta} {{\bf #1}}}

\usepackage{booktabs}

\title{The higher-dimensional origin of five-dimensional $\boldsymbol{{\cal N}\!=\!2}$ gauged supergravities}

\author[a]{Gr\'egoire Josse,}
\emailAdd{josse@lpthe.jussieu.fr}
\author[b]{Emanuel Malek,}
\emailAdd{emanuel.malek@physik.hu-berlin.de}
\author[a]{Michela Petrini,}
\emailAdd{petrini@lpthe.jussieu.fr}
\author[c]{and Daniel Waldram}
\emailAdd{d.waldram@imperial.ac.uk}

\affiliation[a]{Sorbonne Universit\'e, UPMC Paris 05, UMR 7589, LPTHE, 75005 Paris, France}
\affiliation[b]{Institut f\"{u}r Physik, Humboldt-Universit\"{a}t zu Berlin, IRIS Geb\"{a}ude, Zum Gro{\ss}en Windkanal 2, 12489 Berlin, Germany}
\affiliation[c]{Department of Physics, Imperial College London,
London, SW7 2AZ, UK}

\subheader{\hfill\textrm{Imperial/TP/21/DW/4}\\
	\hspace*{\fill}\textrm{HU-EP/21/52}}

\abstract{Using exceptional generalised geometry, we classify which five-dimensional ${\cal N}=2$ gauged supergravities can arise as a consistent truncation of 10-/11-dimensional supergravity. Exceptional generalised geometry turns the classification into an algebraic problem of finding subgroups $\Gst \subset \USp{8} \subset \Ex{6}$ that preserve exactly two spinors. Moreover, the intrinsic torsion of the $\Gst$ structure must contain only constant singlets under $\Gst$, and these, in turn, determine the gauging of the five-dimensional theory. The resulting five-dimensional theories are strongly constrained: their scalar manifolds are necessarily symmetric spaces and only a small number of matter multiplets can be kept, which we completely enumerate. We also determine the largest reductive and compact gaugings that can arise from consistent truncations.}

\begin{document}

\maketitle
%\tableofcontents
%%%%%%%%%%%%%%%%%%%%%%%%%%%%%%%%%%%%%%%%%%%%%%%%%%%%%%%%%%%%%%%%%%%%%%%%%%

\section{Introduction}

When studying compactifications of 10- and 11-dimensional supergravities, the low-energy limits of string theory, it is useful to have a lower-dimensional theory which captures key aspects of the physics. If the compactification leads to a separation of scales, we can obtain a lower-dimensional low-energy effective supergravity theory by integrating out modes above the cut-off scale. This is the case for compactifications on special holonomy manifolds to Minkowski space-time, where the effective theory is obtained by keeping only the massless modes, namely the zero-modes of appropriate differential operators on the internal space.

However, when there is no separation of scales, or if we want to keep both some light and massive modes within the truncation, we must instead resort to a \textit{consistent truncation} of 10-/11-dimensional supergravity %to a lower-dimensional  supergravity 
\cite{Duff:1984hn}. A consistent truncation ensures that all solutions of the lower-dimensional theory also satisfy the equations of motion of 10-/11-dimensional supergravity. Consistent truncations are therefore particularly relevant for anti-de Sitter (AdS) compactifications, where no explicit scale-separated example is known. This is even conjectured to be true for all AdS compactifications of string theory \cite{Lust:2019zwm}.

Constructing consistent truncations is 
a notoriously difficult problem, due to the highly non-linear equations of motion of 10/11-dimensional supergravity (see e.g. \cite{Duff:1984hn,Cvetic:2000dm}). 
Thus  it might be tempting to use lower-dimensional gauged supergravity models without a clear higher-dimensional origin. However, this is fraught with dangers. For example, a vacuum that appears stable within a lower-dimensional supergravity might suffer from instabilities triggered by modes not kept in the truncation \cite{Malek:2020mlk}, or vacua which appear different within the lower-dimensional model may actually be identified in the full 10/11-dimensional theory \cite{Giambrone:2021zvp}. These examples highlight how important it is to know which lower-dimensional theories can arise as consistent truncations of 10-/11-dimensional supergravity.

Until recently, the only systematic approach to consistent truncations relied on  considering manifolds with  reduced structure group and keeping all the modes that are singlets under the reduced group. For example, group manifolds (and freely-acting discrete quotients thereof) have a trivial structure group and give the classic Scherk--Schwarz reductions \cite{Scherk:1979zr}. Alternatively one can consider, for example,   Sasaki--Einstein and weak-$G_2$ holonomy manifolds of \cite{Gauntlett:2009zw,Cassani:2010uw,Gauntlett:2010vu,Liu:2010sa}, or  tri-Sasakian manifolds \cite{Cassani:2011fu}. However, there are also famous consistent truncations, such as those of 11-dimensional supergravity on $S^7$ \cite{deWit:1986oxb} and $S^4$ \cite{Nastase:1999kf,Nastase:1999cb}, that cannot be explained by this traditional group action argument.

Recently, it has become clear that the appropriate framework for understanding general consistent truncations of 10-/11-dimensional supergravity  is given by generalised $G_S$-structures in exceptional generalised geometry and exceptional field theory. 

Exceptional generalised geometry and exceptional field theory are reformulations of  10/11-dimensional supergravity in a way which unifies fluxes and metric degrees of freedom into exceptional symmetry groups. In exceptional generalised geometry, for instance, the exceptional groups appear as structure groups of the generalised tangent bundle of the compactification manifold, which is an extension of the tangent bundle by appropriate exterior powers of the cotangent bundle. A reduction of the exceptional structure group to a subgroup $G_S$ defines a reduced ``generalised structure group''. Given such generalised a $G_S$ structure one can define its ``intrinsic torsion'' \cite{Coimbra:2014uxa}. In analogy to the case of conventional $G$ structures, this is a differential object that measures the obstruction to finding a torsion-free connection, compatible with the structure. For a given structure, it can decomposed into generalised tensors transforming in particular $G_S$ representations. 

 It is now understood \cite{Cassani:2019vcl}
 that generalised $G_S$ structures provide a systematic and general derivation of consistent truncations:  any generalised $G_S$ structure with constant singlet intrinsic torsion defines a consistent truncation of 10-/11-dimensional supergravity. For instance, all maximally supersymmetric truncations are associated to generalised identity structure and so can be seen as 
 generalised Scherk-Schwarz reductions \cite{Lee:2014mla,Hohm:2014qga,Baguet:2015sma}. This provides a unified description of 
 the  consistent truncations of 11-dimensional supergravity on $S^7$ and $S^4$, as well as of IIB supergravity on $S^5$ and massive IIA on spheres
 \cite{Ciceri:2016dmd,Cassani:2016ncu},
 and give a framework for analysing generic maximally supersymmetric truncations \cite{Inverso:2017lrz,Bugden:2021wxg,Bugden:2021nwl}.
Moreover,  considering larger generalised structure groups, we obtain consistent truncations preserving less supersymmetry \cite{Malek:2017njj,Malek:2019ucd,Cassani:2019vcl,Cassani:2020cod}.

What is particularly interesting in this approach is that a good deal of information about the reduced lower-dimensional theory is derived from purely algebraic considerations. The embedding of the generalised structure group $G_S$ in $\Ex{n}$ completely fixes the field content and the allowed components of the embedding tensor of the reduced theory, as well as the truncation ansatz. 

One is then left with the problem of solving the  differential consistency condition that the $G_S$ structure has constant, singlet intrinsic torsion.
This will determine whether there exists an internal manifold that realises any of the reduced theories allowed by the algebraic analysis. 

In this paper, we apply these ideas to classify consistent truncations of M-theory and type IIB supergravities to five-dimensional ${\cal N}=2$ gauged supergravities. In this case the relevant exceptional group is $\Ex{6}$. 
 We focus on the algebraic part of the problem, that is identifying the possible $G_S\subset\Ex6$ structures, and work under the hypothesis that the differential one is solved. 

We first classify all the continuous subgroups of $\Ex{6}$ that give rise to only two spinor supercharges in five-dimensions, as required by $\mathcal{N}=2$ supersymmetry, and we derive, in each case, the field content of the reduced theory. This allows us to show that the structure of the five-dimensional ${\cal N}=2$ gauged supergravities that can arise from consistent truncations of type II/11-dimensional supergravity is very constrained. 
For example, the scalar manifolds of such gauged supergravities must necessarily be symmetric, and there is a maximum number of vector and hypermultiplets that can be coupled. 
Indeed, we find that only a handful of matter contents can arise from consistent truncations.

We can then further constrain the allowed truncated theories as follows. Under the assumption  that the compactification manifolds satisfy the differential constraint of constant singlet intrinsic torsion, we determine the embedding tensors of the reduced theory and analyse the possible gaugings. Again purely group-theoretical arguments allow us to fully determine the gauging of the reduced theory.  As expected these include as special cases the known truncations \cite{Faedo:2019cvr,Cassani:2020cod} that arise from the $\mathcal{N}=2$ Maldacena--Nu\~nez \cite{Maldacena:2000mw} and ``BBBW'' \cite{Bah:2012dg} backgrounds. We find in general that the embedding tensor is constrained, so that generically not all gaugings that are allowed from a five-dimensional point of view  are realised as consistent truncations. The result is that the gauged supergravities that can be obtained as consistent truncations are a very small subset of those that can be constructed from a purely five-dimensional point of view. 

It is worth stressing that while our analysis gives the list of the reduced theories that can a priori be obtained as consistent truncations, this does not mean that all of them can actually be realised. First, one must find compactification manifolds that admit the appropriate $\Gst$ generalised structure groups. Secondly, we must show that they  %To this extend the knowledge of 
satisfy the condition of constant singlet intrinsic torsion, and then analyse the non-zero components of intrinsic torsion/embedding tensor to see which gauge algebras in fact appear. The analogous condition is known to limit the possible gaugings in the maximally supersymmetric case \cite{Lee:2014mla,Malek:2015hma,Cassani:2016ncu,Inverso:2017lrz,Bugden:2021wxg,Bugden:2021nwl}. 
So it is to be expected that the number of actual truncations is even more restricted than what we present here.

This result is of particular interest for theories with AdS vacua. It is conjectured that no AdS vacua of string theory admit scale separation \cite{Lust:2019zwm}. Hence it is not possible to write an effective  $\mathcal{N}=2$ theory in this case. Thus we are led to conjecture that those gauged supergravities that cannot come from consistent truncations and which have AdS vacua must belong to the ``swampland''. Put differently, these gauged supergravities are lower-dimensional artefacts that are not related to string theory.

This paper is structured as follows. In Section \ref{sugra_review}  we recall the main features of 5-dimensional $\mathcal{N}=2$ gauged supergravity. In particular, we describe  the gauging procedure in terms of Leibniz algebras, as this is the natural language to make the connection to exceptional generalised geometry. The exceptional generalised geometry formalism  for $\mathcal{N}=2$ truncations to five dimensions is reviewed in Section \ref{sec:genN2}.  We first introduce  $\Ex{6}$ generalised geometry, which is the relevant one for compactifications to five dimensions. Then we discuss the $G_S$ structures that are associated to $\mathcal{N}=2$ truncations and establish  the dictionary between the $G_S$ structure data and those of the truncated theory. Section \ref{sec:classres} contains the main results of the paper, namely the classification of the gauged supergravity that can come from consistent truncations of M-theory or type IIB supergravities.
We organise the list according to the field content, first theories with only vector and tensor multiplets, then only hypermultiplets and finally those with both vector/tensor and hypermultiplets. 
Appendix \ref{PreliminariesE66_Mth} contains more details about $\Ex{6}$ exceptional geometry, while in  Appendix \ref{app:intors}, for concreteness,  we provide  the explicit computation of the intrinsic torsion  for the truncation with $n_{\rm VT}$ vector multiplets. 
Finally in Appendix \ref{app:fielddec} we discuss the truncation ansatz.

\section{$5d \,$  $\mathcal{N}=2$ gauged supergravity: moduli spaces and gaugings}
\label{sugra_review}

In this section, we summarise the features of five-dimensional  $\mathcal{N}=2$ gauged supergravity coupled to matter  \cite{Gunaydin:1999zx,Ceresole:2000jd,Bergshoeff:2004kh} that 
we want to reproduce from  consistent truncations of M-theory or type IIB theory.  We follow the conventions of  \cite{Bergshoeff:2004kh}.

We are interested in  5$d$  $\mathcal{N}=2$  supergravity  coupled to  $n_{\rm V}$ vector multiplets, $n_{\rm T}$  tensor multiplets and $n_{\rm H}$  hyper-multiplets. 
The  gravity multiplet consists of the graviton, two  gravitini transforming as a doublet of the R-symmetry group $\SU{2}_R$  and  the graviphoton,
\be
\{g_{\mu\nu},\, \psi_\mu^{\tilde{x}}, \, A_\mu\}. 
\ee
The index $\tilde{x}=1,2$ denotes the $\SU{2}_R$  R-symmetry.   Each vector multiplet
% transform in the adjoint of the gauge group, $\Ggauge_{gauge}$ and each
contains a vector, two spin-1/2 fermions in the fundamental  of $\SU{2}_R$   and a complex scalar  $ \phi$. Since in five dimensions a vector is dual to a two-form, a tensor multiplet has the same number of degrees of freedom. % and consists of an antisymmetric tensor,  two spin-1/2 fermions in the fundamental of $\SU{2}_R$   and a complex scalar  $ \phi$. 
Thus we have vector and tensors multiplets
\be
\{ A_\mu ,\, \lambda^{\tilde{x}} ,\,  \phi \},  \qquad
\{ B_{\mu \nu},\, \lambda^{\tilde{x}} ,\, \phi \},  \qquad \tilde{x}=1,2 \,.
\ee
If we have $n_{\rm V}$ vector multiplets and $n_{\rm T}$ tensor multiplets we will use the notation %$A_\mu^{\tilde{I}}$ with $\tilde{I}=0,\ldots, n_{\rm V}$
$A_\mu^{I}$ with $I = 0, \ldots, n_{\rm V}$ to denote the graviphoton and the vectors fields  and $B^M_{\mu \nu}$ with $M= n_{\rm{ V}} +1, \ldots n_{\rm V} + n_{\rm T}$ for the two-form fields. The  scalars of the vector and tensor multiplets are grouped together into   $\phi^i$,  with $i = 1, \ldots, n_{\rm V} + n_{\rm T}$.
These scalars parametrise a very special  real manifold, $\mathcal{M}_{\rm VT}$. 

There are also hypermultiplets, each of which consists of four real scalars and an R-symmetry doublet of spin-1/2 fermions 
\be
\{ \zeta^{\tilde x},\, q^u \}\,, \qquad  u=1, \ldots, 4\,, \quad   \tilde{x}=1,2 \, . 
\ee
If we have $n_{\rm H}$ hypermultiplets, the scalars are grouped into  $q^X$, with $X=1,\dots,4 n_{\rm H}$, and  parameterise a quaternionic-K\"ahler manifold, $\mathcal{M}_{\rm H}$. It is also convenient  to collect  the spinors $\zeta$ into  $\zeta^{\tilde{\alpha}}$, with $\tilde{\alpha} = 1, \ldots, 2 n_{\rm H}$,  transforming in the fundamental representation of $\USp{2 n_{\rm H}}$.

\medskip

The very special real manifold   can be described  as  an $n_{\rm VT} $-dimensional cubic hypersurface  in  an $(n_{\rm VT} + 1)$-dimensional  ambient space, where $n_{\rm VT}=n_{\rm V}+n_{\rm T}$. Viewing  $h^{\tilde I} = h^{\tilde I}(\phi^i)$, with $\tilde{I} = 0, \ldots, n_{\rm VT}$, as embedding coordinates, $\mathcal{M}_{\rm VT}$ is given by 
\be
\label{scalarm}
C(h) = C_{\tilde{I} \tilde{J} \tilde{K}} h^{\tilde I}  h^{\tilde J} h^{\tilde K} = 1 \, ,
\ee
where $C_{\tilde{I} \tilde{J} \tilde{K}}$ is a completely symmetric constant tensor.

The metric on $\mathcal{M}_{\rm VT}$ is given by
\be\label{from_gxy_to_aIJ}
g_{ij} = h_i^{\tilde I} h_j^{\tilde J} \,a_{\tilde{I} \tilde{J} }\,,
\ee
where  $a_{\tilde{I} \tilde{J}}$  is the metric on the ambient space 
\be
\label{aIJ_general_formula}
 a_{\tilde I \tilde J} =  3 h_{\tilde I} h_{\tilde J}  -2 C_{\tilde I  \tilde J \tilde K} h^{\tilde K} \,,
\ee
and 
\be
\begin{aligned}
& h_{i}^{\tilde I} = - \sqrt{\tfrac{3}{2}}\, \partial_{i} h^{\tilde I}\, , \\
& h_{\tilde I} =  C_{\tilde I  \tilde J \tilde K} h^{\tilde K} h^{\tilde L} = a_{\tilde I \tilde K} h^{\tilde K}\, .
\end{aligned}
\ee

The   homogeneous ``very special real''   manifolds  have been classified in  \cite{deWit:1991nm}.  For the symmetric ones, which are the only ones we will need, a classification is possible based on whether the polynomial %\eqref{constraint_Chhh} 
\eqref{scalarm}
or, equivalently,  the tensor  $C_{{\tilde I} {\tilde J} {\tilde K}}$  is  factorisable or not  \cite{Gunaydin:1983bi, Gunaydin:1984ak, Gunaydin:1986fg}.  We will discuss this classification in Section \ref{sec:classres}.

The $4 n_H$ scalars of the hypermultiplets parameterise a
quaternionic K\"ahler manifold $\mathcal{M}_H$, of real dimension $4 n_H$ with metric
\be
\label{Mhmet}
g_{XY} = C_{\tilde{\alpha} \tilde{ \beta}}  \epsilon_{\tilde{x}\tilde{y}}   f_X{}^{\tilde{\alpha} \tilde{x}}  f_Y{}^{\tilde{\beta} \tilde{y}} \,,
\ee
where $f_X{}^{\tilde{\alpha} \tilde{x}}$ are the quaternionic vielbeine and  $C_{\tilde{\alpha} \tilde{ \beta}}$ is the flat metric on $\USp{ 2n_{\rm H}}$.
On $\mathcal{M}_{\rm H}$  there exist a (local) triplet of complex structures  $\vec{J}_X{}^Y$  satisfying 
\be
[J^\alpha, J^\beta]  = 2 \epsilon^{\alpha \beta \gamma} J_q \,, \qquad (J^\alpha)^2 = - {\rm Id} \,, \qquad \alpha, \beta =1,2,3 \, ,
\ee
with respect to which the metric $g_{XY}$ is hermitian. 

As for the vector multiplets, only symmetric spaces will be relevant for consistent truncations. The Riemannian symmetric quaternionic--K\"ahler spaces were first considered by Wolf in~\cite{Wolf65} and then classified by Alekseevsky in~\cite{Alek68}. This was then extended to the pseudo-Riemannian class by Alekseevsky and Cort\'es in~\cite{AC05}. We will discuss the relevant ones in Section \ref{sec:classres}.

Together, the  scalar  manifold of the theory is the direct product 
\be
\mathcal{M} = \mathcal{M}_{\rm VT}\times  \mathcal{M}_{\rm H} \,,
\ee
with isometries   $G_{\rm iso} = G_{\rm VT} \times G_{\rm H}$, 
where $G_{\rm VT}$ and $G_{\rm H}$ are the isometry groups of $\mathcal{M}_{\rm VT}$ and $\mathcal{M}_{\rm H} $ respectively and define the global symmetries of the ungauged theory.\footnote{\label{fn:SU2} In the case of no hypermultiplets, we define $G_{\rm H}=\SU2$ so that $\Giso$ still matches the global symmetries.}

\medskip
The most general gauged theory is described in \cite{Bergshoeff:2004kh}. It is useful for what follows to translate it into the language of Leibniz algebras (or more precisely ``G-algebras'' \cite{Bugden:2021wxg}). In doing so we also see how the gauging picks out the space of vector and tensor multiplets. Let $\mathcal{V}$ be  the vector space  of dimension $n_{\rm VT} + 1$  formed by the graviphoton, the $n_{\rm V} $  vectors and $n_{\rm T}$ tensors. The gauging defines a Leibniz algebra $\gleib$, on $\mathcal{V}$, that is a bilinear bracket $\Leib{v}{w}$ that satisfies a Leibniz-relation 
\be
\label{leib-cond}
\Leib{u}{\Leib{v}{w}} = \Leib{\Leib{u}{v}}{w} + \Leib{v}{\Leib{u}{w}}\,, \qquad
\forall\, u,v,w \in \mathcal{V}. 
\ee
Choosing a basis, the algebra defines a set of structure constants $ t_{{\tilde J} {\tilde K}}{}^{\tilde I}$ via 
\be
\label{leib5d}
\Leib{v}{w}^{\tilde I}  = t_{{\tilde J} \tilde{K}}{}^{\tilde I} v^{\tilde J} w^{\tilde K}\,,  \qquad \forall \, v, w \in \mathcal{V}. 
\ee
Note that in general $\Leib{v}{w}\neq-\Leib{w}{v}$ (that is $t_{{\tilde J} {\tilde K}}{}^{\tilde I}\neq -t_{{\tilde K} {\tilde J}}{}^{\tilde I}$) so the bracket does not define a Lie algebra. 

We then define the subspace $\mathcal{T}\subset\mathcal{V}$ as the image of the symmetrised bracket
\be \label{eq:symbracket}
\mathcal{T} = \left\{ \Leib{v}{w} + \Leib{w}{v} : v, w \in \mathcal{V} \right\} \, , 
\ee
and identify elements of $\mathcal{T}$ with tensor multiplets, so that $\dim\mathcal{T}=n_{\rm T}$. Note that the Leibniz condition \eqref{leib-cond} implies that
\be
\label{tensorcond}
\Leib{b}{v} = 0 \,, \qquad \forall\, b\in\mathcal{T},\, v\in\mathcal{V} \,.
\ee
Thus $\Leib{v}{b}=\Leib{v}{b}+\Leib{b}{v}\in\mathcal{T}$ and hence $\mathcal{T}$ forms a two-sided ideal. As a consequence, if we identify the space of vector multiplets as the quotient $\mathcal{R}=\mathcal{V}/\mathcal{T}$, then the bracket descends to an ordinary Lie bracket on $\mathcal{R}$ defining what we will call the ``extended Lie algebra'' $\ggauge$. Note that by construction $\mathcal{V}$ is a reducible representation of $\ggauge$ where $\mathcal{T}$ forms an invariant subspace. 

If one chooses a particular splitting so $\mathcal{V}=\mathcal{R}\oplus\mathcal{T}$ and fixes a basis, where $I=0,1,\dots,n_{\rm V}$ labels components in $\mathcal{R}$ and $M=n_{\rm V}+1,\dots,n_{\rm VT}$ labels components in $\mathcal{T}$, this structure means that one has 
\be
%\label{tconst2}
  t_{I J}{}^{K} = f_{IJ}{}^K \,,  \qquad  t_{M \tilde{I}}{}^{\tilde J} =0 \, , \qquad t_{(I {\tilde J})}{}^I = 0 \, , 
\ee
where $f_{IJ}{}^K=-f_{JI}{}^K$ are the structure constants of the Lie algebra $\ggauge$. In summary, we see that the splitting into vector and tensor multiplets is defined by the choice of Leibniz algebra. 

The choice of Leibniz algebra is not completely general if it is to lead to a consistent gauging. Note first that we can define the adjoint action given some $v \in {\cal V}$ as
\be
\label{ggaugemap}
\begin{aligned}
t_v \,  : \,  \,  \mathcal{V} &\to \mathcal{V} \,, \\
  w &\mapsto t_v w := \Leib{v}{w}  \,,
\end{aligned} 
\ee
so that, in components $t_v$ is the matrix $(t_v)^{\tilde{J}}{}_{\tilde{I}}=v^{\tilde K}   t_{{\tilde K} \tilde{I}}{}^{\tilde J}$. From the Leibniz condition, the commutator is given by 
\be
\label{commutator}
[ t_v, t_w] = t_{\Leib{v}{w}} \,,
\ee
and furthermore $t_b=0$ for all $b\in\mathcal{T}$. Hence the adjoint action defines a Lie algebra.
In terms of the split basis, we have the generators \cite{Bergshoeff:2002qk} (see also \cite{Louis:2016qca}),
\be
\label{matgen}
(t_I)_{\tilde J}{}^{\tilde K} = \begin{pmatrix}  (t_I)_{J}{}^K &  (t_I)_J{}^N \\ 0 &  (t_I)_M{}^N  \end{pmatrix}  \,,
 \qquad \qquad  
 \begin{array}{l}
  I , J, K = 0, \ldots, n_{\rm{V}}\,, \\
  M,N  =  n_{\rm{V}} +1, \ldots, n_{\rm{T}} \,,
\end{array}
\ee
such that 
\be
[t_I,t_J]=-f_{IJ}{}^K t_K \,,
\ee
where $f_{IJ}{}^K$ are the structure constants of $\ggauge$. The components $(t_I)_{\tilde J}{}^N$  give the representation of the gauge group on the tensors. The off-diagonal components $(t_I)_J{}^N$ can be non-zero only in the case of non-compact groups since these allow for non-completely reducible representations \cite{Bergshoeff:2002qk,Louis:2016qca}.

Consistency requires that the symmetric tensor $C$ in \eqref{scalarm} is invariant under the action of $t_v$
% and that $C$ vanishes on the space of tensors 
\be
\label{tconst}
\begin{aligned}
C(t_v(w),  w, w) &= 0 \,,  &\qquad \forall \,   v, w \in \mathcal{V}   \,,
% C(b, b^{\prime},  b^{\prime   \prime}) &= 0   &\qquad \forall \,   b, b^{\prime} , b^{\prime \prime} \in T \\
\end{aligned}
\ee
and that the expression
\be
\label{tconst2}
C(b,v,w) = \Omega(\tfrac{1}{2} t_v(w) + \tfrac{1}{2} t_w(v) , b)\,,    \qquad  \forall \,  b \in T,  \, v,w  \in \mathcal{V}   \, , 
\ee
defines a symplectic form $\Omega$ on $\mathcal{T}$. This implies, in particular, that the bracket defines a symplectic representation of $\ggauge$ on $\mathcal{T}$. Invariance of $C$ in turn means that the action of $t_v$ is an isometry of the metric on $\mathcal{M}_{\rm VT}$. 
In components, these conditions read
\be
\label{tconst3}
t_{(I {\tilde J})}{}^M= \Omega^{MN} C_{N I {\tilde J}}\,,  \qquad
t_{I(\tilde{J}}{}^{\tilde{H}} C_{\tilde{K}\tilde{L})\tilde{H}}=0       \, . 
\ee
Note that the first condition is equivalent to requiring that the map $\mathcal{V}\otimes\mathcal{V}\to\mathcal{T}$ defined by $(v,w)\mapsto \Leib{v}{w}+\Leib{w}{v}$ factors through $\mathcal{V}^*$ via
\be
\label{eq:factor-map}
\begin{tikzcd}
    \mathcal{V}\otimes\mathcal{V} \arrow{r}{C} & \mathcal{V}^* \arrow{r}{\Omega^{-1}} & \mathcal{T} \,. 
\end{tikzcd}
\ee
That is, it can be viewed as a map to $\mathcal{V}^*$ given by $(v,w)\mapsto C(v,w,\cdot)$ followed by the action of $\Omega^{-1}$.

The gauging of the five-dimensional theory can be expressed in terms of the embedding tensor \cite{Samtleben:2008pe,Trigiante:2016mnt}. This is a map
\be
\label{eq:Theta}
\Theta \, : \, \mathcal{V} \to \mathfrak{g}_{\rm iso} \,,
\ee
where $\mathfrak{g}_{\rm iso}$ is the Lie algebra of isometries of the scalar manifold of the underlying \emph{rigid} supersymmetric theory. In this case, this means the product of the real cone over $\mathcal{M}_{\rm VT}$ (that is the ambient space $\mathcal{V}$) with the hyper-K\"ahler cone over $\mathcal{M}_{\rm H}$. Given $v\in\mathcal{V}$ the embedding tensor $\Theta(v)$ specifies how the action of $t_v$ %\in\gred$
gauges the isometries. That is, it defines the embedding of the 
gauge algebra, $\gred$, inside the isometry algebra of the scalar manifold, where we define
\begin{equation}\label{eq:redalgebra}
 \gred = \gleib / \textrm{Ker}\, \Theta \,.
\end{equation}
The matter fields of the ${\cal N}=2$ gauged supergravity are charged under $\gred$ rather than the larger $\ggauge$, which generically is a central extension of $\gred$.

For $\mathcal{N}=2$ supersymmetry the isometry algebra splits  $\mathfrak{g}_{\rm iso}=\mathfrak{g}_{\rm VT}\oplus\mathfrak{g}_{\rm H}$ where $\mathfrak{g}_{\rm VT}$ and $\mathfrak{g}_{\rm H}$ are the Lie algebras of isometries on the vector and hypermultiplet rigid moduli spaces respectively. For very special real and quaternionc K\"ahler homogeneous spaces these are just the Lie algebras of the numerator groups $G_{\rm VT}$ and $G_{\rm H}$, except when there are no hypermultiplets in which case $G_{\rm H}=\SU2$, in line with footnote~\ref{fn:SU2}. The embedding tensor thus splits into two parts \cite{deWit:2011gk,Louis:2012ux}.
For the vector multiplets the isometries on the cone are generated by a basis composed of Killing vectors $k_a^{i}$ on $\mathcal{M}_{\rm VT}$, where $a =1 , \ldots, {\rm dim} \, \mathfrak{g}_{\rm VT}$. For the hypermultiplets the isometries on the hyper-K\"ahler cone are generated by a basis composed of Killing vectors $\tilde{k}_m^X$, with $m =1 , \ldots, {\rm dim} \, \mathfrak{g}_{\rm H}$, together with $\su(2)_R$ elements $\vec{P}_m$ that are the Killing prepotentials\footnote{In the case where there are no hypermultiplets, one can still have constant prepotentials $\vec{P}_m$ with $m=1,2,3$ that can lead to Fayet--Iliopoulos terms.} for each $\tilde{k}_m^X$. The generators that are gauged are then given by 
\be
\label{eq:gauged-Killing}
 k^{i}_{\tilde{I}}(\phi)=\Theta_{\tilde{I}}{}^a k_a^{i}(\phi) \, ,
 \qquad 
 \tilde{k}^X_{\tilde{I}}(q)=\Theta_{\tilde{I}}{}^m \tilde{k}_m^X(q) \, , 
 \qquad 
 \vec{P}_{\tilde{I}} = \Theta_{\tilde{I}}{}^m \vec{P}_m \, , 
 \ee
 where ${\tilde I}, {\tilde J}, {\tilde K} = 0, \ldots, n_{\rm VT}$ and $X,Y = 1, \ldots, 4 n_{\rm H}$.
The two pieces of the embedding tensor $ \Theta_I{}^a $ and $ \Theta_I{}^m$ are thus constant
 $( {\tilde n}_{\rm V}+1)  \times  {\rm \dim} \, \mathfrak{g}_{\rm VT}$ and  $( {\tilde n}_{\rm V}+1)  \times  {\rm \dim} \, \mathfrak{g}_{\rm H}$  matrices, whose rank determines the dimension of the gauge group. The $k_{\tilde{I}}$ vectors are required to act linearly on the embedding coordinates $h^{\tilde{I}}$ such that
\be
k^{i}_{\tilde{I}} \partial_{i} h^{\tilde{J}}
    = t_{\tilde{I}\tilde{K}}{}^{\tilde{J}} h^{\tilde{K}} \,,
\ee
thus relating $\Theta_{\tilde I}{}^a$ to the structure constants $t_{\tilde{I}\tilde{J}}{}^{\tilde{K}}$. Given a splitting $\mathcal{V}=\mathcal{R}\oplus\mathcal{T}$, one then has $k_M=\tilde{k}_M=0$ and
\begin{equation}
\label{eq:Killing-alg}
\begin{aligned}[]
 [ k_I ,  k_J ]^{i}  &=  f_{IJ}{}^K\, k_K^{i} \,, \\
 [ \tilde{k}_I ,  \tilde{k}_J ]^{X}  &=  f_{IJ}{}^K\, \tilde{k}_K^{X} \,,
\end{aligned}
\end{equation}
realising the gauge algebra $\gred$. 
 
\medskip
 
On the scalars the gauging defines covariant derivatives 
\be
\begin{aligned}
\mathcal{D}_\mu \phi^{i} & = \partial_\mu  \phi^{i} + g\, k^{i}_I \mathcal{A}_\mu^I  \,,\\[1mm]
\mathcal{D}_\mu q^X & = \partial_\mu  q^X + g \, \tilde{k}^X_I \mathcal{A}_\mu^I \,.\\
\end{aligned}
\ee

The bosonic Lagrangian of the gauged theory is then given by 
\be
\label{eq:5dlagrangian}
\begin{aligned}
e^{-1} \mathcal{L} &=\frac{1}{2} R  -  \mathcal{V}(\phi, q) - \frac{1}{4} a_{\tilde{I} \tilde{J} } H^{\tilde I}_{\mu \nu}  H^{ \tilde{J} \mu \nu} - \frac{3}{4} a_{\tilde I \tilde J}  \mathcal{D}_\mu h^{\tilde I} \mathcal{D}^\mu  h^{\tilde J}  - \frac{1}{2} g_{XY} \mathcal{D}_\mu q^X  \mathcal{D}_\mu q^Y   \\
& \quad +  \frac{e^{-1}}{16 g}  \epsilon^{\mu \nu \rho \sigma \tau} \Omega_{MN} B^M_{\mu \nu} (\partial_\rho B_{\sigma \tau}^N + 2 g  t^N_{IJ} A^I_\rho F^J_{\sigma \tau} + g  t^N_{IP} A^I_\rho B^P_{\sigma \tau} ) \\
& \quad + \frac{1}{12} \sqrt{\frac{2}{3}} e^{-1}  \epsilon^{\mu \nu \rho \sigma \tau} C_{IJK} A^{I}_\mu \left[ F^J_{\nu \rho} F^K_{\sigma \tau} + f_{FG}^J A^F_\nu A^G_\rho \Bigl( - \frac{1}{2} F^K_{\sigma \tau} + \frac{g^2}{10} f^K_{HL} A^H_\sigma A^L_\tau \Bigr) \right]  \\
& \quad - \frac{1}{8} e^{-1}   \epsilon^{\mu \nu \rho \sigma \tau} \Omega_{MN} t^M_{IK} t^N_{FG} A^I_\mu A^F_\nu A^G_\rho \left(-\frac{g}{2} F^K_{\sigma \tau} + \frac{g^2}{10} 
f^K_{HL} A^H_\sigma  A^L_\tau\right)  \, .
\end{aligned}
\ee
The kinetic terms for the vector/tensor\footnote{The vector multiplet scalar kinetic term can also be written in terms of the scalar fields $\phi^i$ and the metric $g_{ij}$ on $\mathcal{M}_{\rm VT}$ 
using 
\be\label{eq:covderhI}
\mathcal{D}_\mu h^{\tilde{I}} = \partial_\mu h^{\tilde{I}} + g\, f_{\tilde{J}\tilde{K}}^{\tilde{I}} \mathcal{A}_\mu^{\tilde{J}} \, h^{\tilde{K}}  
=  \partial_ih^{\tilde{I}}  \mathcal{D}_\mu\phi^i \,,
\ee
and the identity
\be
\tfrac{3}{2}\,a_{\tilde{I}\tilde{J}}  \mathcal{D}_\mu h^{\tilde{I}}  \mathcal{D}^\mu h^{\tilde{J}} = g_{ij}  \mathcal{D}_\mu \phi^{i}  \mathcal{D}^\mu \phi^{j} \, .
\ee}
 and hypermultiplets are controlled by the metrics $a_{\tilde{I} \tilde{J} }$ and $g_{XY}$,  defined in \eqref{aIJ_general_formula} and \eqref{Mhmet}.
The gauge field strengths
\be
\mathcal{F}^I = \diff A^I - \tfrac{1}{2} g f^I_{JK} A^J  \wedge A^K  \, , 
\ee
and the anti-symmetric tensors $B_{\mu \nu}$  are grouped into the tensors $H^{\tilde I}_{\mu \nu} = ( \mathcal{F}^I_{\mu \nu}, B^M_{\mu \nu})$.  We see that the gauge field strengths are indeed elements of the extended algebra $\ggauge$, while the matter fields have non-trivial charge only under the action of the gauge algebra $\gred$.

In general, the scalar potential $\mathcal{V}$ is a function of the Killing vectors on the scalar manifolds ${\cal M}_{\rm VT}$ and ${\cal M}_{\rm H}$, and on the Killing prepotentials,
$\vec{P}_I$, on $\mathcal{M}_{\rm H}$\footnote{The  Killing prepotentials $\vec{P}_I$   are defined by
\be\label{KillingPrepGeneral}
4n_{\rm H} \vec{P}_I \,=\,  \vec{J}_X{}^Y \nabla_Y \tilde{k}_I^X\,,
\ee
where $\vec{J}_X{}^Y$ is the triplet of (local) complex structures on $\mathcal{M}_{\rm H}$.}
\be \label{eq:scalpot}
\mathcal{V} =  2 g^2 \left(   g^{ij} \vec{P}_{i} \cdot  \vec{P}_{j} -  2 \vec{P} \cdot \vec{P}  +  g_{ij} \mathcal{K}^{i} \mathcal{K}^{j}  +   \mathcal{N}_{\tilde{\alpha}  \tilde{x}} \mathcal{N}^{\tilde{\alpha}  \tilde{x}} 
\right) \, ,
\ee
where the arrow  denotes a triplet of $\su(2)_R$ elements and
\be
\label{scalarf}
\begin{array}{l}
\vec{P}  = h^I \vec{P}_I\,,\\
\vec{P}_{i}   = \partial_{i}  \vec{P} =   h^I_{i} \vec{P}_I \,, 
\end{array}
\qquad \qquad 
\begin{array}{l}
\mathcal{K}^{i} =  \tfrac{\sqrt{6}}{4} h^I k^{i}_I  \, , \\
\mathcal{N}^{\tilde{\alpha} \tilde{x}} = \tfrac{\sqrt{6}}{4} h^I \tilde{k}_I^X f_X{}^{\tilde{\alpha}\tilde{x}}\, . 
\end{array}
\ee
Notice that due to the identity $h^{\tilde I} k_{\tilde I}^i = 0$, the Killing vectors on $\mathcal{M}_{\rm VT}$  do not contribute to the potential when there are no tensor multiplets.

The functions  in \eqref{scalarf} also control the bosonic part of the supersymmetry variations:
\be
\begin{aligned}
& \delta \psi^{\tilde{x}}_\mu = D_\mu \epsilon^{\tilde{x}} + \frac{i g}{\sqrt{6}} P^{\tilde{x}\tilde{y}} \gamma_\mu \epsilon_{\tilde{y}} + \ldots \,, \\
& \delta \lambda^{i \tilde{x}}  = g \mathcal{K}^{i} \epsilon^{\tilde{x}}  +  g  P^{i \tilde{x}\tilde{y}}  \epsilon_{\tilde{y}} + \ldots \,, \\
& \delta \zeta^{\tilde\alpha}=  g \mathcal{N}^{\tilde{\alpha}}{}_{\tilde{x}}  \epsilon^{\tilde{x}}  + \ldots  \, . 
\end{aligned} 
\ee
where we have written out the explicit adjoint action of the $\su(2)_R$ elements $\vec{P}$ and $\vec{P}_i$. 

%%%%%%%%%%%%%%%%%%%%%%%%%%%%%%%%%%%%%%%%%%%%%%%%%%%%%%%%%%%%%%%%%%%%%%%%%%

%%%%%%%%%%%%%%%%%%%%%%%%%%%%%%%%%%%%%%%%%%%%%%%%%%%%%%%%%%%%%%%%%%%%%%%%%%

\section{$\mathcal{N}=2$ supergravities from generalised geometry }
\label{sec:genN2}

In the language of exceptional generalised geometry, consistent truncations are associated to generalised $G_S$-structures. 
If a  $d$-dimensional manifold $M$ admits a generalised $G_S$-structure, namely a set of globally defined generalised invariant tensors, with constant intrinsic torsion,  a  consistent truncation of  type II or eleven-dimensional supergravity on $M$ is obtained by expanding all  supergravity fields on such tensors and keeping only the $G_S$-singlet modes.
Knowing the generalised structure is enough to determine all the data of the truncated theory. 
This approach has been successfully applied to the study of consistent truncations with several amount of supersymmetry \cite{Lee:2014mla,Cassani:2016ncu,Cassani:2019vcl} (see also for the exceptional field theory version of this approach \cite{Hohm:2014qga,Ciceri:2016dmd,Malek:2017njj,Malek:2019ucd}).
In particular,   \cite{Cassani:2020cod} provides the generic framework to study  type IIB or M-theory  consistent truncations  to five dimensions with $\mathcal{N}=2$ supersymmetry.
The purpose of this paper is to use this formalism to classify the possible consistent truncations of  type IIB or M-theory  to five-dimensional  $\mathcal{N}=2$ supergravity.

In this section,  we give a brief summary of  the exceptional generalised geometry relevant for type IIB or M-theory reductions  to five dimensions  and
then in the next section we review the formalism  of \cite{Cassani:2020cod}. 

\subsection{Generalised $G_S$ structures and $\mathcal{N}=2$ supersymmetry} 
\label{GSgen}

Type IIB or M-theory supergravity on a  $d$-dimensional manifold $M$, with $d=5$ for type IIB and $d=6$ for M-theory, are conveniently reformulated in terms of   $\Ex{6} \times \mathbb{R}^+$ generalised geometry\footnote{ See  Appendix~\ref{PreliminariesE66_Mth} for a more detailed review  of  $\Ex{6} \times \mathbb{R}^+$  generalised geometry}. For definiteness, we will focus on the M-theory case, though  the formalism is equally applicable in type IIB.

To the  manifold $M$ we associate a generalised tangent bundle  $E$, whose sections  transform  in the real $\mbf{27}^*$ representation\footnote{Given a representation $\mbf{n}$ we will use $\mbf{n}^*$ and $\obf{n}$ for the dual and conjugate representations, respectively. For non-compact groups these may not be equivalent.} of $\Ex{6}$, the generalised structure group, with weight one under $\bbR^+$.  
The ordinary structure group $GL(d)$ embeds in $\Ex{6}\times\bbR^+$ and can be used to decompose the generalised tangent bundle as 
 \begin{equation}
 \label{gentan} 
   E \,\simeq \, TM \oplus \Lambda^2T^*M \oplus \Lambda^5T^*M  \, .
\end{equation}
The sections of $E$  are called  generalised vectors and, using \eqref{gentan}, can be seen  as (local) sums of a vector, a two-form and a five-form on $M$,
\be
V = v + \omega + \sigma \, . 
\ee

The frame bundle $F$ for $E$ defines an $\Ex6\times\bbR^+$ principal bundle. By considering bundles whose fibres transform in different representations of $\Ex6\times\bbR^+$, we can then define other generalised tensors. To describe the bosonic sector of the supergravity theories we will need, besides the generalised vectors,  weighted dual vectors, adjoint tensors and the generalised metric. Adjoint tensors $R$ are sections of the adjoint bundle $\adj F$ of the form
\be
\begin{aligned}
  {\adj} F &\simeq \bbR \oplus (TM \otimes T^* M) \oplus \Lambda^3 T^* M \oplus \Lambda^6 T^* M  \oplus \Lambda^3 T M \oplus \Lambda^6 T  M  \, , \\
  R &= l + r + a + \tilde a + \alpha + \tilde \alpha \, ,
\end{aligned}
\ee
and hence transform in the ${\bf 1} + {\bf 78}$ of $\Ex6$ with weight zero under the $\bbR^+$ action. Locally $l$ is a function, $r$ a section of $\End(TM)$, $a$ is a three-form and so on. One notes that in the exceptional geometric  reformulation,  the internal components of the gauge potentials of  type II or M-theory, are embedded in the adjoint bundle. 

It will be useful to also define weighted dual vectors $Z$ as sections of the bundle   $N\simeq \det T^*M \otimes E^*$ which has $\bbR^+$ weight two\footnote{Note that $\det T^*M$ is just a different notation for the top-form bundle $\Lambda^6T^*M$ that stresses that it is a real line bundle. In the following we will assume that the manifold is orientable and hence $\det T^*M$ is trivial. Thus, we can define arbitrary powers $(\det T^*M)^p$ for any real $p$.}. Concretely one finds 
\be
\begin{aligned}
  N &\simeq T^*M \oplus \Lambda^4T^*M \oplus  (T^*M \otimes  \Lambda^6 T^*M )  \,, \\
  Z &= \lambda  +   \rho  + \tau  \, .
\end{aligned}
\ee

Finally the generalised metric $G$  is a positive-definite, symmetric rank-2 tensor
\be
\label{genmeta}
G \in \Gamma(\det T^*M \otimes S^2E^*) \, , 
\ee
so that, given two generalised vectors $V,W\in \Gamma(E)$, the inner product $G(V,W)$ is a top form. Just as an ordinary metric $g$, at each point on $M$,  parameterises the coset $\mathrm{GL}(6)/\mathrm{O}(6)$%GL(6)/O(6)$
, a generalised metric at a point $p \in M$ corresponds to an element of the coset
\be
 \left.G\right|_p \in \frac{ \Ex{6} \times \mathbb{R}^+}{\USp{8}/\mathbb{Z}_2} \,  . 
\ee
The generalised metric encodes the internal components of all  bosonic fields of type II or M-theory on $M$.

\medskip

The fermionic fields of type IIB or M-theory are arranged into representations of  $\USp{8}$, 
the double cover of the maximal compact subgroup $\USp{8}/\bbZ_2$ of $\Ex{6}$.  
For instance, supersymmetry parameters are section of the generalsied spinor bundle $\mathcal{S}$, transforming in the ${\bf 8}$ of 
$\USp{8}$.  The R-symmetry of the reduced  five-dimensional theory is in general then some subgroup $G_R \subseteq \USp{8}$.

\bigskip

A generalised $G_S$ structure is the reduction of the generalised structure group  $\Ex{6}\times\bbR^+$ to a subgroup $G_S$. For all the structure groups that we discuss here, this is equivalent to the existence on $M$ of globally defined generalised tensors that are invariant under $G_S$.\footnote{For non-simple (and discrete) groups, you can in principle have $G_S$ groups that are not defined as stabilizer groups of tensors.} For example, the generalised metric $G$ in~\eqref{genmeta} defines an $\USp{8}/\bbZ_2$ structure. In what follows, since we always assume the existence of a generalised metric, we will consider $G_S$ structures that are subgroups of  $\USp{8}/\bbZ_2$. Moreover, we are interested in generalised structures preserving some amount of supersymmetry and hence we need the structure group to lift to a subgroup $\tilde{G}_S$ of $\USp8$ acting on the spinor bundle $\mathcal{S}$ and to 
keep track of how many spinors are singlets of $\tilde{G}_S$. In all the cases considered here we have $\tilde{G}_S\simeq G_S$. Hence for simplicity we will from now on write $G_S$ for both. For $\mathcal{N}=2$ supersymmetry we need two invariant supercharges in the spinor bundle $\mathcal{S}$ implying that we need subgroups $G_S \subset\USp{8}$ that give only two singlets when decomposing the ${\bf 8}$ of $\USp{8}$.

\medskip

The largest structure group giving $\mathcal{N}=2$ supersymmetry is  $G_S = \USp{6}$:  under the breaking 
\be
\label{Usp6br}  
\USp{8} \supset \USp{6} \times \SU{2}_R \, , 
\ee 
the spinorial representation decomposes as
\be
\label{eq:8-USp6}
{\bf 8} = ({\bf 6}, {\bf 1}) \oplus ({\bf 1}, {\bf 2}) \, . 
\ee
The  $\SU{2}_R$ factor in  \eqref{Usp6br} is the R-symmetry of the reduced theory  under which the two spinors singlets form
 a doublet,  as expected for $\mathcal{N}=2$ supersymmetry parameters. One also has the decompositions for the $\Ex6$ representations
\be
\begin{aligned}
   \label{eq:27-78-USp6}
   \mbf{27}^* &= (\rep{1} , \rep{1})  \oplus ( \rep{14},\rep{1}) 
      \oplus ( \rep{6}, \rep{2} ) \, , \\
    \rep{78} &= (\rep{1} , \rep{3}) \oplus ( \rep{6},\rep{2})
       \oplus ( \rep{21}, \rep{1} )   
       \oplus ( \rep{14},\rep{1}) \oplus ( \rep{14}',\rep{2}) \, .
\end{aligned}
\ee
Note that the embedding of the structure $\USp6\subset\Ex6$, in contrast to \eqref{Usp6br}, defines the subgroup
\begin{equation}
    \Ex6 \supset \USp6 \cdot \SU2_R \, , 
\end{equation}
where we are using the ``central product'' between $\USp6$ and $\SU2_R$. By definition, for any group $G$ and subgroup $H$, the commutant\footnote{Throughout this paper we will use the notation $\Com{H}{G}$, with $H \subset G$, for the commutant (or centralizer) of $H$ within $G$.} $\Com{H}{G}$ of $H$ in $G$ includes the centre $Z(H)$ of $H$. The central product is defined to be $H\cdot \Com{H}{G}=(H\times \Com{H}{G})/Z(H)$ where one modes out by the diagonal $Z(H)$ subgroup. In this case $Z(\USp6)=\bbZ_2$ and the central product reflects the fact that the maximal compact subgroup of $\Ex6$  is $\USp8/\bbZ_2$ and not $\USp8$. 
 
The  $G_S = \USp{6}$ structure  is often called an {\it HV structure}~\cite{Grana:2009im,Ashmore:2015joa,Ashmore:2016qvs} and can also be defined in terms of  non-vanishing invariant adjoint  tensors and a generalised vector, corresponding to the singlets under $\Gst = \USp{6}$ in \eqref{eq:27-78-USp6}.
As they will be useful in the rest of the paper, let us first introduce the vector and hypermultiplet structures that these tensors separately define. 

A vector-multiplet structure, or  {\it V structure},  is given by a globally defined  generalised vector $K\in \Gamma(E)$ of positive norm with respect to the $\Ex{6}$ cubic invariant,
\be\label{eq:cKKK}
c(K,K,K) \,:=\, 6\,\kk^2 > 0  \, , 
\ee
where $\kk$ is a section of $(\det T^*M)^{1/2}$.
The vector $K$ is the $(\rep{1},\rep{1})$ singlet in the decomposition of the $\rep{27}^*$ in \eqref{eq:27-78-USp6} and is stabilised by $\Fff \subset \Ex{6}$. 
A hypermultiplet structure, or  {\it H structure}, is determined by a pair $(J_\alpha,\kk^2)$ where $J_\alpha \in \Gamma({\adj} F)$ ($\alpha =1,2,3$)   is  a triplet that define a basis for a highest root $\frak{su}_2$ subalgebra of $\frak{e}_{6(6)}$ and hence satisfy
\be 
\label{normJ}
[J_\alpha , J_\beta] \,=\,\ 2 \epsilon_{\alpha \beta \gamma} J_\gamma\,, \qquad \tr (J_\alpha J_\beta) \,=\,  - \delta_{\alpha \beta}  \, , 
\ee
while $\kk$ is a section of $(\det T^*M)^{1/2}$ as above. The $J_\alpha$ correspond to the $(\rep{1},\rep{3})$ triplet in the decomposition of the $\rep{78}$ in \eqref{eq:27-78-USp6} and are stabilised by $\SUs{6}\subset \Ex6$. 

The HV structure corresponds to a V and an H structure, such that the two $\kk$ densities are the same and in addition  compatibility constraint 
\be
\begin{aligned}
\label{compJK}
 & J_\alpha  \cdot K \,=\, 0\,, 
\end{aligned}
\ee
is satisfied, where $\cdot$ denotes the adjoint action (see Appendix \ref{PreliminariesE66_Mth} for all relevant  definitions).  The  common stabiliser of compatible $K$ and $J_\alpha$ is 
\be
\SUs{6} \cap \Fff \simeq \USp{6} \, .
\ee

As shown in \cite{Cassani:2020cod}, given an $\USp{6}$ structure, one can construct a generalised metric as 
\be
\begin{aligned}
\label{USp6_gm}
 G(V,V) \,=\, 3\left(3 \, \frac{c(K,K,V)^2}{c(K,K,K)^2}  - 2  \frac{c(K,V,V)}{c(K,K,K)} +  4 \, {\frac{c(K,J_3\cdot V,J_3\cdot V)}{c(K,K,K)}}\right)\,,
\end{aligned}
\ee
where $c$ is the $\Ex{6}$ cubic invariant and $V$ is a generalised vector.

\medskip

As we will discuss later,  in  terms of the multiplets of the truncated theory, an HV structure, that is one where $G_S=\USp6$, implies that there are neither vector multiplets 
nor hypermultiplets present; the reduced theory is minimal ${\cal N}=2$ supergravity. 
To allow for vectors or hypermultiplets, one has to look for reduced structure groups $G_S\subset\USp6$ such that in the decomposition
\be
\label{eq:GSsubgroup}
\USp{8} \supset \USp{6} \times \SU{2}_R \supset G_S \times \SU{2}_R\,,
\ee
additional $G_S$ singlets beyond those defined by the $\USp{6}$ structure appear in $\mbf{27}^*$ and the  ${\bf 78}$, but none in the ${\bf 8}$. This means the $\rep{6}$ in the decomposition \eqref{eq:8-USp6} cannot admit any singlets, and hence that all the singlets in the $\rep{27}^*$ must transform trivially under $\SU{2}_R$. 

Each $G_S \subset \USp{6}$ singlet will give a $G_S$-invariant generalised tensor in the corresponding bundle. In particular, the singlets in $\rep{27}^*$ will span a sub-bundle $\Esing$ 
\be
\label{eq:Esing}
   E \supset \Esing \simeq M \times \mathcal{V} \, .
\ee
The bundle is by definition trivial and hence can be written as a product where $\mathcal{V}$ is the fibre. The vector space $\mathcal{V}$ transforms as a representation of the commutant $\Com{G_S}{\Ex{6}}$ of $G_S$ in $\Ex{6}$. In particular, from the discussion above, there must be an R-symmetry subgroup $\SU{2}_R\subset \Com{G_S}{\Ex{6}}$ that acts trivially on $\mathcal{V}$ (and hence $\Esing$). Furthermore, the corresponding Lie algebra $\su(2)$ must correspond to a highest root in $\ex6$. Let us define $\Gh$ as the simple subgroup of $\Com{G_S}{\Ex{6}}$ that contains such a highest root $\SU2$. We can then also identify the corresponding trivial sub-bundle of the adjoint bundle\footnote{Note that there are  singlets in the adjoint bundle that are not in $\adj F_{\Gh}$. In addition to elements generating the other possible factors in $\Com{G_S}{\Ex{6}}$ there are also elements of the form $V \otimes_{\adj} W$, where $V$ is a section of $\Esing$, $W$ is a section of the dual bundle $E^*_{\rm singlet}$ and $\otimes_{\adj}$ is the projection onto the adjoint bundle. However these will not play a relevant role in our construction.\label{foot:KtimesK*}}
\be
\label{eq:FH}
   \adj F \supset \adj F_{\Gh} \simeq M \times \gh \, , 
\ee
where $\gh$ is the Lie algebra of $\Gh$. Note that by definition $R\cdot v=0$ for all $v\in\Gamma(\Esing)$ and $R\in \Gamma(\adj F_{\Gh})$. 

Given any trivial $\Gst$-invariant vector bundle $P\simeq M\times \bbR^n$ and $G_S$-compatible generalised connection $\tilde{D}$, one can define a constant  section $s\in\Gamma(P)$ by $\tilde{D}s=0$. Furthermore, the definition is independent of the choice of $\tilde{D}$ since the bundle transforms trivially under $G_S$. For the sub-bundles $\Esing$ and $\adj F_{\Gh}$ we can identify $\mathcal{V}$ and $\mathcal{U}\simeq\gh$ with the spaces of constant sections 
\begin{equation}
\label{eq:UV-def}
\begin{aligned}
   \mathcal{V} &= \left\{ v\in\Gamma(\Esing) : \tilde{D}v=0 \right\} \, , \\
   \gh \simeq \mathcal{U} &=
        \left\{ R\in\Gamma(\adj F_{\Gh}) : \tilde{D}R=0 \right\} \, , \\
\end{aligned}
\end{equation}
giving a natural realisation of the isomorphisms \eqref{eq:Esing} and \eqref{eq:FH}. Note that the elements of $\mathcal{U}$ generate a global $\Gh$ symmetry. The $G_S$-structure also defines a constant invariant section $\kk^2\in \Gamma(\det T^*M)$. Hence for each $v\in\mathcal{V}$ the expression
\begin{equation}
\label{eq:Cdef}
    C(v,v,v) = \kk^{-2} c(v,v,v) \, , 
\end{equation}
where $c$ is the $\Ex6$ cubic invariant, defines a map into $\bbR$ (or more precisely to constant functions on $M$). We can always choose a basis of normalised nowhere-vanishing linearly independent vectors and adjoint elements for $\mathcal{V}$ and $\mathcal{U}$
\be
%\label{eq:JdotK=0}
\{ K_{\tilde I} \,,  J_A \} \,,
\qquad  \tilde{I} = 0, \dots,  \dim\mathcal{V}-1  , \quad  A =1, \dots , \dim \Gh \, ,
\ee
where by definition we have 
\be
 \label{eq:JdotK=0}
 J_A \cdot K_{\tilde I}  =0 \,, \qquad  \qquad  \forall \, \tilde{I},\, A   \, .
 \ee
In this basis, the components $C_{\tilde{I} \tilde{J} \tilde{K}}$ of the map \eqref{eq:Cdef} are given by 
\be\label{eq:CKKK_cond}
c(K_{\tilde I}, K_{\tilde J}, K_{\tilde K} ) = 6\,\kk^2 C_{\tilde{I} \tilde{J} \tilde{K}} \,,
\ee
and define a symmetric, constant tensor, while the adjoint tensor basis $J_A$ satisfy
\be\label{commutorJAJB}
[ J_A, J_B ] = f_{AB}{}^C J_C \, ,
\ee
where $f_{AB}{}^C$ are the structure constants of $\gh$. 
Finally, we can normalise 
\be
\label{nomrJ}
\tr(J_A  J_B) = \,\eta_{AB} \,,
\ee
where $\eta_{AB}$ is a diagonal matrix with $-1$ and $+1$ entries in correspondence with compact and non-compact generators of $\Gh$, respectively. Note that in the ``minimal'' case of $G_S=\USp6$ with the HV structure $(K,\,J_\alpha)$ the spaces $\mathcal{V}$ and $\mathcal{U}$ are one- and three-dimensional, with basis vectors $K$ and $J_\alpha$, respectively.

\subsubsection{Moduli space of HV structures}
\label{mod-space}

A strict $\USp{6}$ structure is rigid, up to an overall scaling of $\kk^2$. However, a reduced $G_S \subset \USp{6}$ structure group naturally leads to a moduli space of $G_S$-invariant HV structures. Note that the moduli do not necessarily consist of massless scalar fields from the point of view of the reduced $\mathcal{N}=2$ five-dimensional theory, but rather will lead to a consistent truncation.

Out of the invariant tensors $K_{\tilde I}$ and $J_A$ defining the $\Gst$ structure, we can  define an HV structure by constructing a vector $K\in\mathcal{V}$ and a triplet $J_\alpha\in\mathcal{U}$ that form  a basis for a highest root $\su(2)$ algebra in $\gh$.
Any  such HV structure is related to another by the local action of $g\in\Ex6\times \bbR^+$. The $\bbR^+$ factor rescales $\kk^2$ and can be absorbed by rescaling of the metric in the reduced theory. It therefore does not define a modulus and we can consider only $g\in\Ex6$. In order for the deformed  HV structure to remain in $\mathcal{V}$ and $\mathcal{U}$, the action $g$ needs to lie in the commutant group $\Com{\Gst}{\Ex{6}}$ and to be constant in the sense that $Dg=0$ for any $G_S$ compatible connection $\tilde{D}$. In other words, different points in the moduli space of $\Gst$-invariant HV structures are related by global $\Com{\Gst}{\Ex{6}}$ transformations.

However, the actual physical moduli come from the generalised metric. Given a reference  $\USp{6}$  structure, we can build a reference generalised metric using the  definition  \eqref{USp6_gm}. The physical moduli are then generated  by acting on the structure with elements of $\Ex{6}$ that commute with $G_S$,  modulo elements of $\USp{8}/\bbZ_2$, that leave the generalised metric invariant.  The  moduli obtained this way hence parameterise the coset
\begin{equation}
    \label{eq:GS-coset}
   \mathcal{M} \,=\, \frac{\Com{\Gst}{\Ex{6}}}{\Com{\Gst}{\USp{8}/\bbZ_2}} \, . 
\end{equation}

By definition we are only considering $\Gst$ that only  admits $\mathcal{N}=2$ supersymmetry, in other words we are not interested in theories that are subsectors of more supersymmetric ones. This means there are no elements of $\Com{\Gst}{\Ex{6}}$ that lead to two different $\USp6$ structures with the same generalised metric. Hence $\Com{\Gst}{\Ex{6}}$ must factorise into groups that act separately on $\mathcal{V}$ and $\mathcal{U}$, that is
\begin{equation}
  \label{eq:factorise}
  \Com{\Gst}{\Ex{6}} = \Com{G_S}{G_{\mathcal{U}}} \times
  \Com{G_S}{G_{\mathcal{V}}} \, , 
\end{equation}
where $G_{\mathcal{U}}$ and $G_{\mathcal{V}}$ are the subgroups of $\Ex6$ that leave fixed all elements of $\mathcal{U}$ and $\mathcal{V}$, respectively.
Consequently, the moduli space $\mathcal{M}$ factorises into V structure and H structure moduli spaces, as expected from $\mathcal{N}=2$ supergravity,
\be
\label{eq:MVTxMH}
\mathcal{M} \,=\, \mathcal{M}_{\rm VT} \times \mathcal{M}_{\rm H} = \frac{\Com{G_S}{G_{\mathcal{U}}}}{\Com{G_S}{H_{\mathcal{U}}}} \times  \frac{\Com{G_S}{G_{\mathcal{V}}}}
{\Com{G_S}{H_{\mathcal{V}}}} =  \frac{\Gvt}{\Hvt} \times \frac{\Gh}{\Hh}  \, ,
\ee
where, similarly, $H_{\mathcal{U}}$ and $H_{\mathcal{V}}$ are the subgroups of $\USp8/\bbZ_2$ that leave $\mathcal{U}$ and $\mathcal{V}$ fixed, respectively. In general there are common factors that cancel between the numerators and denominators in the commutator group expression for the cosets; for example the centre $C(G_S)$ is always a subgroup common to both. Thus it is useful to introduce the notation  $\Gvt$, $\Gh$, $\Hvt$ and $\Hh$ for the numerators and denominators that remain in the quotients in \eqref{eq:MVTxMH} once all the common factors have been cancelled (except when there are no hypermultiplets in which case we take $\Gh=\Hh=\SU2$). For $\mathcal{M}_{\rm H}$, one finds $\Gh$ is the simple subgroup of $\Com{G_S}{\Ex6}$ that contains a highest root $\SU2$, consistent with our definition of $\Gh$ above .

The V structure moduli space corresponds to deformations of $K$ that leave $J_\alpha$ invariant, while the  H structure moduli space  describes deformations of $J_\alpha$ that leave $K$ invariant. When given a dependence on the external spacetime coordinates, these deformations provide the scalar fields in the truncated theory, with $\mathcal{M}_{\rm VT}$ and $\mathcal{M}_{\rm H}$ being identified with the vector multiplet and the hypermultiplet scalar manifolds, respectively.

\medskip
We can identify the moduli explicitly as follows. Consider first  $\mathcal{M}_{\rm VT}$. Using the basis $K^{\tilde{I}}$, a general vector $K\in\mathcal{V}$ can be written as a linear combination
\be\label{eq:Kdressing}
K = h^{\tilde{I}} K_{\tilde{I}} \,,
\ee 
where $h^{\tilde{I}}$, $\tilde{I}= 0,\ldots, n_{\rm VT}$%$h^I$, $I= 0,\ldots, n_{\rm VT}$
, are real parameters. Fixing $\kk^2$ in  \eqref{eq:cKKK}, and using \eqref{eq:CKKK_cond}, gives
\begin{equation}
\label{constraint_Chhh}
C_{\tilde{I}\tilde{J}\tilde{K}}h^{\tilde{I}}h^{\tilde{J}}h^{\tilde{K}}=  1\,,
\end{equation}
showing that the $n_{\rm VT}+1$ parameters $h^{\tilde{I}}$ are constrained by one real relation and thus define an $n_{\rm VT}$-dimensional hypersurface, just as in \eqref{scalarm}, 
\be
\mathcal{M}_{\rm VT} = \{\,  h^{\tilde{I}} :\   C_{\tilde{I}\tilde{J}\tilde{K}}h^{\tilde{I}}h^{\tilde{J}}h^{\tilde{K}}=1 \,\}\,.
\ee
 The space  $\mathcal{M}_{\rm VT}$ is the moduli space of the V structure and, in the truncation, will determine  
  the vector multiplet scalar manifold of the  five-dimensional theory. The metric on $\mathcal{M}_{\rm VT}$ is obtained by evaluating the generalised metric on the invariant generalised vectors,
 \be
 a_{\tilde{I}\tilde{J}} = \tfrac{1}{3} \,G(K_{\tilde{I}},K_{\tilde{J}})\,.
 \ee
 It is straightforward to verify that, using  \eqref{USp6_gm}, the expression above reproduces the  five-dimensional expression  \eqref{aIJ_general_formula}.

Consider now $\mathcal{M}_{\rm H}$.
The family of H structures is obtained by parameterising the possible choices of $\su_2$ algebra. Recall that by definition $\mathcal{U}\simeq\gh$, so we are interested in the space of highest root $\su(2)\subset\gh$ subalgebras. Fixing $\kk^2$ and modding out by the $\SU2$ symmetry that relates equivalent triples $J_\alpha$ we have the moduli space
\be\label{Hstr_space_general}
\mathcal{M}_{\rm H} \,=\, \frac{\Gh}{\SU{2}_R\cdot \Com{\SU{2}_R}{\Gh}}\, , 
\ee 
that is, comparing with~\eqref{eq:MVTxMH}, we have $\Hh=\SU{2}_R\cdot \Com{\SU{2}_R}{\Gh}$. Points in $\mathcal{M}_{\rm H}$ can be parameterised by starting from a reference subalgebra $\mathfrak{j} \simeq\su_2 \subset \gh$ and then acting on a basis $\{ j_1,j_2,j_3 \}$ of $\mathfrak{j}$ by the adjoint action of group elements $h \in \Gh$, defined as 

\be\label{adjoint_action_on_j}
 J_\alpha = 
 {\adj}_\Gh \, j_\alpha = h \, j_\alpha\, h^{-1} \,.
\ee 
One has to mod out by the elements of  $\Gh$ that have a trivial action, namely   $h\in \SU{2}_R\simeq \exp(\mathfrak{j})$ and $h\in \Com{\SU{2}_R}{\Gh}$. 
The resulting symmetric spaces \eqref{Hstr_space_general} and are all quaternionic--K\"ahler, in agreement with the identification of $\mathcal{M}_{\rm H}$ with the hyperscalar manifold in five-dimensional supergravity.

\subsubsection{Singlet generalised intrinsic torsion}
\label{sec:gentors}

Any generalised $G_S$ structure has an associated intrinsic torsion \cite{Coimbra:2014uxa}. Given a $G_S$-compatible generalised connection, $\tilde{D}$, its torsion $T$ is defined as
\begin{equation}
   \label{eq:gen-torsion}
   \big(\Lgen^{\tilde{D}}_V - \Lgen_V\big)\, \alpha = T(V) \cdot \alpha \, , 
\end{equation}
where $\alpha$ is a generic generalised tensor,   $\Lgen$ is the generalised Lie derivative (see  Appendix \ref{PreliminariesE66_Mth}), $\Lgen^{\tilde{D}}$ is the generalised Lie derivative calculated using $\tilde{D}$ and  $\cdot$ is the  adjoint action on $\alpha$.
 
As a generalised tensor, the torsion $T$ belongs to the  sub-bundle  
\be
W   \in  E^* \oplus K  \subset E^*\otimes \adj F\,, \\
\ee
with $ E^*$ transforming in the $\rep{27}$ representation and  $K$ in the $\rep{351}$ representation. 

Let $\Sigma =  \tilde{D}- \tilde{D}^\prime$ be the difference between two  $G_S$-compatible generalised connections. It is a generalised tensor, specifically a section of $K_{G_S} =  E^*\otimes \adj F_{G_S}$, where $\adj F_{G_S}\subset \adj F$ is the $G_S$-adjoint subbundle defined by the structure.
Using~\eqref{eq:gen-torsion}  one can define a map from $K_{G_S}$ to $W$, the  space of generalised torsions,
\be 
\begin{split}
\tau    :  \, K_{G_S} &\to W \,, \\
 \Sigma &\mapsto  \tau(\Sigma) = T - T^\prime\,,
\end{split}
\ee
as  the difference of the torsions of the connections $\tilde{D}$ and  $\tilde{D}^\prime$.  The image of the map $\tau$ is not necessarily surjective, that is $\im \tau = W_{G_S} \subset W$. The part of $W$ that is not spanned by $W_{G_S}$ is the intrinsic torsion of the generalised structure $G_S$, i.e.
 \be
 W_{\text{int}}^{G_S}=W/W_{G_S} \, . 
\ee
The intrinsic torsion $T_{\rm int}$ is  the component of $T$ that is independent of the choice of compatible connection $\tilde{D}$ and is fixed only by the choice of generalised structure. When $G_S \in \USp{8}/\mathbb{Z}_2$ and therefore defines a generalised metric, the norm defined by the generalised metric $G$ allows one to decompose the space of generalised torsions as\footnote{See Appendix \ref{app:intors} for an explict example.}
 \be
 \label{metdector}
W= W_{G_S} \oplus W_{\text{int}}^{G_S} \, .
\ee

We can always decompose the intrinsic torsion into representations of $\Gst$. For a consistent truncation we will be interested in generalised structures  whose only non-zero components are in  singlet representations of $G_S$. 

\medskip
 
As for ordinary $G$-structures, the intrinsic torsion of a generalised structure $G_S$ can be encoded in first-order differential expressions in the $G_S$ invariant generalised tensors. Recall that $K_{\tilde{I}}$ and $J_A$ form a basis for the invariant tensors and by definition, $\tilde{D}K_{\tilde{I}}=\tilde{D}J_A=0$ for any $G_S$-compatible connection. It was shown in \cite{Cassani:2020cod} that the intrinsic torsion is encoded in the expressions 

\begin{equation}
   \label{eq:GS-int-tor-1}
   \Lgen_{K_{\tilde I}} K_{\tilde J}   \, , \qquad  \Lgen_{K_{\tilde I}} J_A  \,,
\end{equation}

and \begin{equation}
   \label{eq:GS-int-tor-2}
   \int_M\kk^2 \tr(J_A(L_WJ_B))\, , 
\end{equation}
where the generalised vector $W$ is orthogonal to the generalised vectors in $\mathcal{V}$ in the sense that
\be
c(K_{\tilde{I}},K_{\tilde{J}},W)=0 \, . 
\ee
Note that the expressions~\eqref{eq:GS-int-tor-1} and~\eqref{eq:GS-int-tor-2} are in general not independent, but are sufficient to determine the intrinsic torsion. 

\medskip

For a consistent truncation we need to require that the intrinsic torsion lies only in the singlet representation of $G_S$ and is constant. This is equivalent to requiring 

\begin{equation}
   \label{eq:GS-sing-tor}
   \begin{aligned}
      \Lgen_{K_{\tilde I}} K_{\tilde J}  &= - \Tint(K_{\tilde I}) \cdot  K_{\tilde J}
      =  t_{{\tilde I} {\tilde J}}{}^{\tilde K} K_{\tilde K}   \, , \\
      \Lgen_{K_{\tilde I}} J_A &= - \Tint(K_{\tilde I}) \cdot J_A
       = p_{\tilde{I}A}{}^B J_B \, , 
   \end{aligned}
\end{equation}
where the $t_{{\tilde I} {\tilde J}}{}^{\tilde K}$ and $p_{\tilde{I}A}{}^B$ are constants and that \eqref{eq:GS-int-tor-2} vanishes for all $W$. The latter  follows from the fact that the condition on $W$ implies that it transforms non-trivially under $G_S$ and hence, since $J_A$ are singlets, the corresponding intrinsic torsion cannot be a singlet and so must vanish. Recall that $\Tint(V)$ is a section of the adjoint bundle $\adj F$. For singlet torsion, $\Tint(K_{\tilde{I}})$ must act in sub-bundle defined by the commutant\footnote{Note that strictly speaking the singlet torsion also allows $\Tint(K_{\tilde{I}})$ to act in the $\bbR^+$ factor of $\adj F$. This would correspond to a gauging of the ``trombone symmetry'' in the 5d theory \cite{LeDiffon:2008sh}. Such theories do not have an action and for simplicity we do not consider them here.} $\Com{\Gst}{\Ex{6}}$. From the factorisation \eqref{eq:factorise} we see that we can view the matrices $(t_{{\tilde I}})_{{\tilde J}}{}^{\tilde K}$ and $(p_{\tilde{I}})_A{}^B$ as elements of Lie algebras of $\Gvt$ and $\Gh$ respectively.

\subsubsection{The data of the truncation}
\label{sec:truncdata}

Any generalised $G_S$ structure on a manifold $M$ with only constant,
singlet intrinsic torsion gives rise to a consistent truncation of eleven-dimensional or type II supergravity with spacetime $X\times M$ to a gravitational theory on $X$ \cite{Cassani:2019vcl,Cassani:2020cod}.
In this section we focus on truncations to five-dimensional $\mathcal{N}=2$ supergravity and recall how 
the the generalised $G_S\subseteq \USp{6}$ structure encodes the data of the truncated theory, as summarised in Section~\ref{sugra_review}. 
 
\medskip

The field content of the truncated theory is completely determined by the $G_S$-invariant spaces $\mathcal{U}$ and $\mathcal{V}$ and the moduli space of HV structures,\footnote{For completeness we give in 
 Appendix \ref{app:fielddec}  the explicit form of the  truncation ansatz.} while the gauging is determined by the singlet torsion.
 
The scalars of the truncated theory are given by the moduli space \eqref{eq:GS-coset} of generalised metrics on $M$ that factors \eqref{eq:MVTxMH} into 
\begin{equation}
  \begin{aligned}
     \text{VM scalars:} && \phi(x)^i &\leftrightarrow  \mathcal{M}_{\rm VT} = \frac{\Gvt}{\Hvt}
    \, , \\
    \text{HM scalars:} && q(x)^X & \leftrightarrow  \mathcal{M}_{\rm H} = \frac{\Gh}{\Hh}
    = \frac{\Gh}{\SU{2}_R\times \Com{\SU{2}_R}{\Gh}}
    \, ,
  \end{aligned}
\end{equation}
where  $x^\mu$ are the coordinates on $X$.

By construction, both spaces are homogeneous and so correspond to one of the cases listed in Section \ref{sugra_review}. As discussed in Section \ref{mod-space}, the metrics can be explicitly constructed in terms of the basis vectors $K_{\tilde{I}}$ and $J_A$. In particular, the cubic invariant on $\mathcal{V}$, which  fixes the metric on $\mathcal{M}_{\rm VT}$, is given by \eqref{eq:Cdef}. 

The other bosonic fields are the vectors and two-forms. As we will see in a moment, the singlet intrinsic torsion allows  one to decompose the space of constant vectors as $\mathcal{V}=\mathcal{R}\oplus\mathcal{T}$ so that the basis vectors split
\begin{equation}
  \{ K_{\tilde{I}} \} = \{ K_I \} \cup  \{K_M \} \, , 
\end{equation}
where $\{K_I\}$ with $I=0,\dots,n_{\rm V}$ are a basis for $\mathcal{R}$ and $\{K_M\}$ with $M=n_{\rm V}+1, \dots,n_{\rm VT}$ are a basis for $\mathcal{T}$. The vector fields and two-forms are in one-to-one correspondence with a basis in  $\mathcal{R}$ and $\mathcal{T}$ respectively\footnote{In the general formalism given in \cite{Cassani:2019vcl,Cassani:2020cod} the two-forms were valued in  constant sections of the singlet sub-bundle of $N\simeq\det T^*M\otimes E^*$, written using dual basis vectors $K^{\flat\tilde{I}}$, and isomorphic to elements of $\mathcal{V}^*$. The relation to the fields here is that the $\tilde{I}$ index is raised using the symplectic form $\Omega^{-1}$ defined by the singlet torsion. Note also that one can consider $A^{\tilde{I}}_\mu$ and $B^{\tilde{I}}_{\mu\nu}$ defined for all values of $\tilde{I}$. However, once the non-propagating fields are eliminated only $A^I_{\mu}$ and $B^M_{\mu \nu}$ are dynamical and the Lagrangian takes the form \eqref{eq:5dlagrangian}.}
\begin{equation}
  \begin{aligned} \label{eq:VecTwoFormAnsatz}
    \text{vectors:} &&  A^I_{\mu}(x) 
       & \leftrightarrow  K_I \, , \\
    \text{two-forms:} &&  B^M_{\mu \nu}(x)
       & \leftrightarrow K_M \, .
  \end{aligned}
\end{equation}

\medskip

The gauge interactions of the truncated theory are determined by the intrinsic torsion of the $\Gst$-structure, which in turn is captured by the constants appearing in \eqref{eq:GS-sing-tor}. The first relation defines a bracket $\Leib{\cdot}{\cdot}:\mathcal{V}\otimes\mathcal{V}\to\mathcal{V}$ on $\mathcal{V}$ given by
\begin{equation} \label{eq:gen-gauging}
  \Leib{v}{w}^{\tilde{I}} := (\Lgen_vw)^{\tilde{I}} = t_{{\tilde J} \tilde{K}}{}^{\tilde I}
      v^{\tilde J} w^{\tilde K} \, , \qquad \forall\, v,w \in\mathcal{V} \, , 
\end{equation}
just as in \eqref{leib5d}. Since the generalised Lie derivative satisfies $\Lgen_u(\Lgen_vw)=\Lgen_{\Lgen_uv}w+\Lgen_v(\Lgen_uw)$ the bracket defines a Leibniz algebra. As in Section \ref{sugra_review}, one can then choose a splitting $\mathcal{V}=\mathcal{R}\oplus\mathcal{T}$, where $\mathcal{T}$ is the image of the symmetrised bracket, such that $\mathcal{R}$ is the space of vector multiplets and $\mathcal{T}$ the space of tensors. 

For a consistent gauging we need to check the conditions \eqref{tconst} and \eqref{tconst2}. They each follow from the properties of the generalised Lie derivative as we now show. Recall first, from the discussion below \eqref{eq:GS-sing-tor}, that  $(t_v)^{\tilde{J}}{}_{\tilde{I}}=v^{\tilde{K}}t_{\tilde{K}\tilde{I}}{}^{\tilde{J}}$ is an element of the Lie algebra of $\Gvt$ and so
\begin{equation}
   \gred \subset \Lie \Gvt \subset \ex6 \, . 
 \end{equation}
Since $c$ and $\kk^2$ are $\Ex6$ invariants, the action of $\gred$ must preserve the cubic tensor $C$ given by \eqref{eq:Cdef} and hence we satisfy \eqref{tconst}. Furthermore, by definition
\begin{equation}
\label{eq:symm-dorf}
  \Lgen_v w + \Lgen_w v = \dd (v \otimes_N w) \, ,
\end{equation}
where $\dd$ is the exterior derivative and $\otimes_N$ is the projection \eqref{N'prod_Mth} onto $N\simeq\det T^*M \otimes E^*$ given by $v\otimes_N w=c(v,w,\cdot)$. If $v,w\in\mathcal{V}$ then the left-hand side of \eqref{eq:symm-dorf} is by definition an element of $\mathcal{T}$. Using \eqref{eq:Cdef}, the right-hand side is just the sequence of maps in \eqref{eq:factor-map}, where the symplectic form on $\mathcal{T}$ is defined by the composition $\Omega^{-1}=\dd \circ \kk^2$. Hence \eqref{eq:symm-dorf} implies we satisfy the second condition \eqref{tconst2} required for a consistent gauging. 

To complete the description of the gauging we identify the embedding tensor and the Killing vector fields on $\mathcal{M}_{\rm VT}$ and $\mathcal{M}_{\rm H}$. Since both manifolds are coset spaces, from \eqref{eq:factorise}, the group of isometries is $\Giso=\Gvt \times  \Gh$ and the embedding tensor is a map 
\begin{equation}
\label{eq:EmbedT}
  \Theta : \mathcal{V} \to \giso = \Lie \Gvt
       \oplus \Lie \Gh \, . 
\end{equation}
The corresponding gauged Killing vectors $k_{\tilde{I}}^{i}(\phi)$ and $\tilde{k}_{\tilde{I}}^X(q)$ on $\mathcal{M}_{\rm VT}$ and $\mathcal{M}_{\rm H}$ are given by \eqref{eq:gauged-Killing}. If we view $K=h^{\tilde{I}}(\phi)K_{\tilde{I}}$ as giving the embedding of $\mathcal{M}_{\rm VT}$ in $\mathcal{V}$ and $J_\alpha=m^A_\alpha(q) J_A$ as giving the embedding of $\mathcal{M}_{\rm H}$ in $\mathcal{U}$ then, from \eqref{eq:GS-sing-tor}, we can identify the Killing vectors explicitly from the relations
\begin{equation}
  \begin{aligned}
    k^{i}_{\tilde{I}} \partial_{i} h^{\tilde{J}}
    &= \Theta_{\tilde{I}}{}^a k_a^{i} \partial_{i} h^{\tilde{J}}
    = t_{\tilde{I}\tilde{K}}{}^{\tilde{J}} h^{\tilde{K}} \, , \\
    \tilde{k}^X_{\tilde{I}} \partial_X m^A_\alpha
    &= \Theta_{\tilde{I}}{}^m \tilde{k}_m^X \partial_X m^A_\alpha
    = p_{\tilde{I}B}{}^A m^B_\alpha \,.
  \end{aligned}
\end{equation}
Thus we can identify the embedding tensor as an element of $\Lie\Gvt\oplus\Lie\Gh$
\begin{equation}
    \Theta_{\tilde I} = \begin{pmatrix} (t_{{\tilde I}})_{{\tilde J}}{}^{\tilde K} & 0 \\ 0 & (p_{\tilde{I}})_A{}^B \end{pmatrix} \,.
\end{equation}
Using the Leibniz property that $\Lgen_{K_{\tilde{I}}}(\Lgen_{K_{\tilde{J}}}\alpha)=\Lgen_{(\Lgen_{K_{\tilde{I}}}K_{\tilde{J}})}\alpha+ \Lgen_{K_{\tilde{J}}}(\Lgen_{K_{\tilde{I}}}\alpha)$ for any generalised tensor $\alpha$, it follows that each set of vectors forms a representation of $\gred$ as in \eqref{eq:Killing-alg}. In other words, we have 
\begin{equation}
\label{eq:tp-alg}
    [ t_{\tilde{I}}, t_{\tilde{J}}] = t_{\tilde I \tilde J}{}^{\tilde K} t_{\tilde K} \,, \qquad 
    [ p_{\tilde{I}}, p_{\tilde{J}}] = t_{\tilde I \tilde J}{}^{\tilde K} p_{\tilde K} \,. 
\end{equation}
Finally, it is worth noting that the Killing prepotentials descend directly from the moment maps for generalised diffeomorphisms that appear in integrability conditions for an HV structure \cite{Ashmore:2015joa} and are given by
\begin{equation}\label{KillingPrep_from_mommaps}
  \begin{aligned}
 g\, P_{\tilde{I}}^\alpha 
      &= \tfrac18 \, \epsilon^{\alpha\beta\gamma}
       \tr \left( J_\beta (L_{K_{\tilde{I}}} J_\gamma ) \right)\,,
     \end{aligned}
\end{equation}
where as above $J_\alpha=m^A_\alpha(q) J_A$ is the dressed triplet. 

It is important to note that generic $\mathcal{N}=2$ supergravity allows gaugings defined by an embedding tensor $\Theta$ that is a general element of $\mathcal{V}^*\otimes \giso$. However, the fact that our theory comes from a consistent truncations will typically restrict the form of $\Theta$ to only lie in certain $\Gvt\times\Gh$ representations in the decomposition of $\mathcal{V}^*\otimes \giso$. For this reason, in the following we will use $T$ to denote the embedding tensor that appears in the consistent truncations to distinguish it from the more general $\Theta$. As a consequence, we will see that not all the allowed $\mathcal{N}=2$ gaugings can arise from consistent truncations.

\section{Classification of $\mathcal{N}=2$ truncations to five-dimensions }\label{sec:classres}

In this section, we discuss the main result of the paper, namely the classification of the consistent truncations to $\mathcal{N} =2$ gauged supergravity in five-dimensions that can a priori be obtained from  M-theory and type IIB. 

As the data of a consistent truncation  are encoded in the generalised structure $G_S$  that is defined on the compactification manifold $M$, the problem reduces to classifying the possible $G_S \subset  \Ex{6}$ structures with constant intrinsic torsion that preserve $\mathcal{N}=2$ supersymmetry. Therefore, the classification consists of an algebraic problem -- the existence of an appropriate $G_S \subset \Ex{6}$ structure -- and a differential one -- the existence of constant singlet intrinsic torsion. In the following, we will study the algebraic problem in general, but will simply assume that the differential condition of having constant singlet intrinsic torsion can be solved. From the example of maximally supersymmetric gauged supergravity we know that the differential condition puts important restrictions on the allowed gauged supergravities \cite{Inverso:2017lrz,Bugden:2021wxg,Bugden:2021nwl}. Strikingly, even when ignoring this additional constraint, we find that for $\mathcal{N}=2$ theories the algebraic conditions alone significantly constrain the possible gaugings that can arise.

Let us recall from Section \ref{GSgen} what the main idea is. 
Demanding that the truncated theory is supersymmetric implies that the internal manifold must be spin and that the structure group must be a subgroup of $\USp{8}$, the maximal compact subgroup of $\Ex{6}$.  The largest  structure giving $\mathcal{N}=2$ supersymmetry is  $G_S = \USp{6}$. Under the breaking
\be
\label{Usp6brbis}  
\USp{8} \supset \USp{6} \times \SU{2}_R \,,
\ee 
the spinorial representation of  $\USp{8}$ decomposes as  
\be
{\bf 8} = ({\bf 6}, {\bf 1}) \oplus ({\bf 1}, {\bf 2}) \, , 
\ee
where the $({\bf 1}, {\bf 2})$ is associated to the two supersymmetry parameters of the truncated theory.  Since under 
\be
\label{E6brb}  
\Ex{6} \supset \USp{6} \cdot \SU{2}_R \,,
\ee  
the only singlets in the  $\rep{27}^*$  and  $\rep{78}$ are the $K$ and $J_\alpha$ of the {\rm HV} structure, the theory obtained form a  $G_S = \USp{6}$  only  contains the gravity multiplet.  To have extra vector- or hyper-multiplets we need the structure group $\Gst$ to be a subgroup of  $\USp{6}$.

The algebraic problem then consists of the following steps. We first scan for  all possible inequivalent ways of breaking $\USp{8}$ to $G_S \subset \USp{6}$ that admit only two singlets in the fundamental representation of $\USp{8}$. Given a $G_S$ with these features, it will embed in $\Ex{6}$ as
\be
\label{E6brc}  
\Ex{6} \supset G_S\cdot \Com{G_S}{\Ex{6}}  \,,
\ee  
where $\Com{G_S}{\Ex{6}}$ is the commutant group. We then check whether under this breaking the  $\rep{27}^*$  and  $\rep{78}$ of $\Ex{6}$ contain $G_S$ singlets, which will determine the vector and hyper-multiplets of the truncated theory. In each case the singlets will transform under $\Com{G_S}{\Ex{6}}$ which also determines the form of the scalar manifold $\mathcal{M}$ of the truncated theory
\be
\label{scalarmanbis}
\mathcal{M} = \frac{ \Com{G_S}{\Ex{6}} }{ \Com{G_S}{\USp{8}/\bbZ_2} } \, . 
\ee
We stress again that by construction the scalar manifolds are always necessarily symmetric spaces and furthermore are always a product $\mathcal{M}=\mathcal{M}_{\rm VT}\times\mathcal{M}_{\rm H}$ of vector-tensor multiplet and hypermultiplet scalar manifolds as in \eqref{eq:MVTxMH}. 

We have performed a complete scan for all Lie subgroups\footnote{In the following section we will also discuss a few  examples of $G_S = \mathbb{Z}_2$ structures that are easily identified, but we do not provide an exhaustive analysis of discrete subgroups of $\USp{6}$.} $G_S \subset \USp{6}$. 
We find that there are only a small number of inequivalent $G_S$ structures with  the properties above.  We list them here according to the type of breaking of $\USp{6}$ that they correspond to. 
All other cases either give rise to extra singlets in the $\mathbf{6}$ of $\USp{6}$ or can be obtained as subgroups of  the $\Gst$-structures   listed below without giving rise to any new fields in the consistent truncation.

\begin{enumerate}
[label=$\mathbf{Br}$\textbf{.\arabic*},ref=${\rm Br}$.\arabic*]
\item  \label{GstGen} $\Gst = \SU{2} \times \Spin{p}$, $ 2 \leq p \leq 5$.

These are obtained  from the embedding
\be
\USp{6} \supset \USp{4} \times \SU{2}  \simeq \Spin{5} \times \SU{2} \, , 
\ee	
which gives
\be
\rep{6} = (\rep{4}, \rep{1}) \oplus  (\rep{1}, \rep{2}) \, , 
\ee
and  by further breaking the  $\USp{4}$ factor
\be
\begin{aligned}
& \USp{4}  \supset \SU{2} \times \SU{2} \simeq \Spin{4} \,, \\ 
& \USp{4}  \supset \SU{2} \times \SU{2} \supset \SU{2}_D  \simeq \Spin{3} \,, \\ 
& \USp{4}  \subset \SU{2} \times \SU{2} \supset \SU{2}_D \subset \U{1}_D   \simeq \Spin{2}  \, . 
\end{aligned}
\ee

The corresponding branching of the $\rep{6}$ of $\USp{6}$ are 
\be
\begin{aligned}
& \quad \rep{6} = (\rep{2}, \rep{1} , \rep{1}) \oplus (\rep{1}, \rep{2} ,  \rep{1})  \oplus  (\rep{1}, \rep{1} ,  \rep{2}) \,,  \\ 
& \quad  \rep{6} = 2 \cdot   (\rep{2}, \rep{1} )   \oplus  (\rep{1},  \rep{2}) \,, \\ 
& \quad  \rep{6} = 2 \cdot  \rep{1}_1 \oplus     2 \cdot  \rep{1}_{-1}   \oplus  \rep{2}_0  \,,
\end{aligned}
\ee
for the breaking to $\Spin{4} \times \SU{2}$, $\Spin{3} \times \SU{2}$ and $\Spin{2} \times \SU{2}$, respectively.

\item \label{GstSU2}  $\Gst = \SO3$ and $\Gst=\SU2$. 

The relevant breaking is 
\be
\USp{6} \supset \SO3\times \SU{2} \,,
\ee
with the $\mathbf{6}$ of $\USp{6}$ branching as
\begin{equation}
\label{Su2str}
\mathbf{6} = \mbf{(3,2)} \, . 
\end{equation}
Taking $\Gst = \SO3$ or  $\Gst = \SU{2}$ leads to two different consistent truncations.

\item  \label{GstSU3U1}  $G_S = \SU{3}$.

This  comes from the breaking
\be
\label{Usp6toSU3U1} 
\USp{6} \supset \SU{3} \times \U{1}
\ee
which gives
\begin{equation} \label{eq:SU3U1}
\mathbf{6} = \mathbf{3}_1 \oplus \obf{3}_{-1} \,.
\end{equation}

\item \label{GstSU2U1} $\Gst = \SU{2} \times \U{1}$

This truncation is obtained by further breaking the $\SU{3}$ group of the previous case. Under $ \SU{3} \supset \SU{2} \times \U{1}$, we get
\be
\rep{6} =  \mathbf{2}_{1,1} \oplus  \mathbf{1}_{-2,1}  \oplus  \mathbf{2}_{-1, -1} \oplus  \mathbf{1}_{2, -1} \, . 
\ee

\item \label{GstU1a} $\Gst = \U{1}$.

This comes from  the same breaking $ \SU{3} \supset \SU{2} \times \U{1}$ as \ref{GstSU2U1} but taking only the $\U{1}$ factor as the structure group.

\item \label{GstU1b} $\Gst = \U{1}$.

This  comes from the same breaking as  {\ref{GstSU3U1}} and taking the $\U{1}$ factor as structure group.

\end{enumerate}

\medskip

Once the possible $G_S$ structure have been identified, we need to study their singlet intrinsic torsion as this determines the embedding tensor and thus the gaugings of the truncated theory. The details of this calculation for $G_S=\SU2\times\Spin{p}$ are discussed in Appendix \ref{app:intors}. The condition of having only components of the intrinsic torsion that are singlets of $G_S$ imposes differential constraints on the compactification manifold that can be complicated to solve in general. In our analysis, we assume that these differential conditions are satisfied and instead solely study the intrinsic torsion's algebraic properties. We will see that this is still enough to significantly 
restrict the possible gaugings obtainable by a consistent truncation.

We will first decompose the singlet intrinsic torsion into representations of the global isometry group $\Giso$. This will allow us to identify the various components of the embedding tensor of the  truncated theory. 
We then impose the Leibniz condition on these singlets\footnote{If the differential conditions on the intrinsic torsion are satisfied, the Leibniz condition is also automatically satisfied as discussed in Section \ref{sec:truncdata}. However, since here we are not analysing whether the differential conditions can be solved, we must impose the Leibniz condition as a restriction. Put differently, only for those singlet components of the intrinsic torsion which obey the Leibniz condition, can the differential conditions on the compactification manifold be satisfied.}. The resulting embedding tensor components determine the Leibniz algebra $\gleib$ and hence the extended Lie algebra $\ggauge$. As discussed in Section~\ref{sec:genN2}, the matter in the theory is charged under the gauge algebra $\gred$ that is generically a quotient of $\ggauge$ by a central subalgebra. The embedding tensor then also describes the embedding of $\gred$ into the Lie algebra $\giso$ of the isometry group. In the following we will  refer to a group as ``gauged'' if it is part of the corresponding group $\Gred$.

In general, solving the Leibniz conditions for all singlets of $\Gst$ can be very cumbersome. It is hence sometimes useful to streamline the search for possible gauge groups $\Ggauge$ as follows. First, consider the decomposition of $\mathcal{V}$ under a putative gauge group $\Gred\subset\Gvt\subset\Giso$. We must impose that there is a subset of the $n_{\rm VT}$ vectors transforming in the adjoint of $\Gred$. Once this condition is satisfied, we keep only those components of the intrinsic torsion that are $\Gred$-singlets. Finally, we impose the Leibniz condition \eqref{leib-cond} on the singlet intrinsic torsion. From the resulting Leibniz bracket, we can read off the gauge groups and tensor multiplets. In particular, if $n_{\rm VT} > \dim\Gred$ and the $n_{\rm VT} - \dim\Gred$ extra vectors are uncharged under $\Ggauge$, then they are central elements filling out the full gauge algebra $\ggauge$, while if they are charged they either enlarge the $\gred$ algebra or correspond to charged tensor multiplets. The two charged cases are distinguished by whether or not the extra vectors are in the image of the symmetrised Leibniz bracket, as discussed in \eqref{eq:symbracket}. 

It is worth stressing that we do not mean to give an exhaustive list of all possible gaugings. Where we cannot solve the Leibniz condition in general, we will instead limit ourselves to the largest reductive groups and largest compact groups that can be gauged.  We will find that only a handful of gaugings are possible.

Finally, the computation we perform bears some resemblance to the purely five-di\-men\-sion\-al analysis that would have to be performed to find possible gaugings. However, crucially, in order to have a consistent truncation, we are analysing the intrinsic torsion that descends from the $\Ex{6}$ generalised Lie derivative, and thus lives in the $W_{\rm int}\subset\rep{351}$ of $\Ex{6}$. By contrast, the five-dimensional computation would search for gaugings living in the $\mathcal{V}^* \otimes \rep{\giso}\supseteq W_{\rm int}$, where $\mathcal{V}$ denotes the space of vector fields of the five-dimensional supergravity. Therefore, it is not a priori clear whether all gaugings that are allowed from a five-dimensional perspective can also arise from consistent truncations. In fact, as we will see, some five-dimensional gaugings cannot arise from consistent truncations. For example, in theories with scalar manifolds ${\cal M}_{\rm VT} = \mathbb{R}^+ \times \frac{\SO{n_{\rm VT}-1,1}}{\SO{n_{\rm VT}-1}}$ and no hyper multiplets, consistent truncations only lead to gaugings where the tensor multiplets are charged under the graviphoton and not any of the other $n_{\rm VT}-1$ vector fields.

\medskip

In the following sections, we will derive the consistent truncations associated to the  $\Gst$  structures listed here and derive their field content and invariant tensors. For sake of exposition, 
we will  first discuss the consistent truncations including only vector and tensor multiplets in Section  \ref{sec:vectoronly}, then only hypermultiplets in  Section \ref{sec:hyperonly}, before giving the mixed cases with vector/tensor and hypermultiplets in  Section \ref{sec:mixed}.

We summarise the matter content of the consistent truncations that arise from our scan  in Table \ref{t:summary}: we list the $\Gst$ structure group, the number of vector/tensor multiplets $n_{\rm  VT}$ and hypermultiplets $n_{\rm H}$, and the associated scalar manifolds. 
We see that the possible consistent truncations are limited. In particular,  

we find the largest possible truncation consists of only 14 vector/tensor multiplets.

\begin{table}[h!]

\centering
\makebox[\textwidth][c]{
\begin{tabular}{|c|c|c|c|}
\hline%[width=3]
\diagbox[innerwidth=2.5em]{$n_{\rm VT}$}{$n_{\rm H}$} & 0 & 1 & 2  \\
\hline
0&\begin{minipage}[c]{0.35\linewidth} \vspace{0.2cm} \centering $G_S= \USp{6}$ \\  $\mathcal{M} =1$ \vspace{0.2cm} \end{minipage}
&\begin{minipage}[c]{0.35\linewidth} \vspace{0.2cm} \centering$G_S=\SU{3}$\\ $\mathcal{M}=\frac{\SU{2,1}}{\mathrm{S}(\U{2}\times\U{1})}$  \vspace{0.2cm} \end{minipage}&\begin{minipage}[c]{0.15\linewidth}
 \vspace{0.2cm}\centering$G_S=\SO{3}$ \\ $\mathcal{M}=\frac{\text{G}_{2(2)}}{\SO{4}}$  \vspace{0.2cm} \end{minipage}\\
\hline
1&\begin{minipage}[c]{0.25\linewidth}  \vspace{0.2cm} \centering$G_S=\SU{2}\times\Spin{5}$ \\$\mathcal{M}= \mathbb{R}^+$  \vspace{0.2cm} \end{minipage}&\begin{minipage}[c]{0.35\linewidth}  \vspace{0.2cm} \centering$G_S=\SU{2}\times\U{1}$\\$ \mathcal{M}=\mathbb{R}^+\times\frac{\SU{2,1}}{\mathrm{S}(\U{2}\times\U{1})}$  \vspace{0.2cm} \end{minipage}&-\\ 
\hline
2&\begin{minipage}[c]{0.25\linewidth}  \vspace{0.2cm} \centering$G_S=\SU{2}\times\Spin{4}$\\$ \mathcal{M}=\mathbb{R}^+\times\SO{1,1}$  \vspace{0.2cm} \end{minipage}& \begin{minipage}[c]{0.38\linewidth}  \vspace{0.2cm} \centering$G_S=\U{1}$\\$\mathcal{M}=\mathbb{R}^+\times \SO{1,1} \times\frac{\SU{2,1}}{\mathrm{S}(\U{2}\times\U{1})}$  \end{minipage} &-\\ 
\hline
3&\begin{minipage}[c]{0.25\linewidth}  \vspace{0.2cm} \centering$G_S=\SU{2}\times\Spin{3}$\\$ \mathcal{M}=\mathbb{R}^+\times\frac{\SO{2,1}}{\SO{2}}$  \vspace{0.2cm} \end{minipage}& \begin{minipage}[c]{0.35\linewidth}   \centering$G_S=\U{1}$\\$\mathcal{M}=\mathbb{R}^+\times\frac{\SO{2,1}}{\SO{2}}\times\frac{\SU{2,1}}{\mathrm{S}(\U{2}\times\U{1})}$  \end{minipage} &-\\
\hline
4& \begin{minipage}[c]{0.25\linewidth}  \vspace{0.2cm} % \vspace{0.2cm}
\centering$G_S=\SU{2}\times\Spin{2}$\\$\mathcal{M}=\mathbb{R}^+\times\frac{\SO{3,1}}{\SO{3}}$  \vspace{0.2cm}
\end{minipage}&\begin{minipage}[c]{0.35\linewidth} % \vspace{0.2cm}
\centering$G_S=\U{1}$\\$\mathcal{M}=\mathbb{R}^+\times\frac{\SO{3,1}}{\SO{3}}\times\frac{\SU{2,1}}{\mathrm{S}(\U{2}\times\U{1})}$  \end{minipage}&-\\
\hline
5& \begin{minipage}[c]{0.25\linewidth}  \vspace{0.2cm} \centering$G_S=\SU{2}$\\$\mathcal{M}=\frac{\SL{3,\bbR}}{\SO{3}}$\\[0.2cm]
\centering$G_S= \SU{2} \times \mathbb{Z}_2$\\$\mathcal{M}= \mathbb{R}^+\times\frac{\SO{4,1}}{\SO{4}}$
  \vspace{0.2cm} \end{minipage}&-&-\\
\hline
6&\begin{minipage}[c]{0.25\linewidth}  \vspace{0.2cm} \centering$G_S=\SU{2}\times\mathbb{Z}_2$\\$ \mathcal{M}=\mathbb{R}^+\times\frac{\SO{5,1}}{\SO{5}}$  \vspace{0.2cm} \end{minipage}&-&-\\
\hline
8&\begin{minipage}[c]{0.25\linewidth}  \vspace{0.2cm} \centering$G_S=\U{1}$\\$ \mathcal{M}=\frac{\SL{3,\mathbb{C}}}{\SU{3}}$  \vspace{0.2cm} \end{minipage}&-&-\\
\hline
14&\begin{minipage}[c]{0.25\linewidth}  \vspace{0.2cm} \centering$G_S=\mathbb{Z}_2$\\$\mathcal{M}=\frac{{\rm SU}^\ast(6)}{\USp{6}}$  \vspace{0.2cm} \end{minipage}&-&-\\
\hline
\end{tabular}}

\caption{List of  all possible consistent truncation with $n_{\rm VT} $ vector/tensor multiplets, $n_H$ hypermultiplets, and the required $G_S \subset \Ex{6}$ structure group, as well as the associated scalar manifold ${\cal M}$.} \label{t:summary}
\end{table}

Let us again reiterate that the consistent truncations that can be actually realised will be a subset of those presented in the group-theoretic analysis here. This is because the requirement that a given $G_S$ structure has singlet intrinsic torsion will introduce non-trivial differential constraints that a given manifold $M$ must satisfy and which we do not analyse here.

However some of the cases  listed in Table \ref{t:summary} do have an explicit geometric realisation. For instance the mixed cases with 
$n_{\rm H}=1$ and $n_{\rm VT}=1$,  $n_{\rm VT}=2$ and  $n_{\rm VT}=4$  correspond to consistent truncations of eleven-dimensional 
supergravity that have recently obtained. These are truncations around backgrounds with $\mathcal{N}=2$ supersymmetry describing the near-horizon limit of M5-branes wrapping a Riemann surface: the Maldacena--Nu\~nez (MN) solution \cite{Maldacena:2000mw}  and its generalisations called the BBBW solutions \cite{Bah:2012dg}. In particular the truncation with $n_{\rm VT} =4$ and $n_{\rm H}=1$ with gauge group $\Gred = \SO{3} \times \U{1}_R \times \mathbb{R}^+$ is the largest possible truncation around the MN solution , while the case with $n_{\rm VT} =2$ and $n_{\rm H}=1$ and gauge group $\Gred = \U{1}_R \times \mathbb{R}^+$ gives the consistent truncation around
the BBBW solutions \cite{Cassani:2020cod}. The subtruncation to  $n_{\rm VT} =1$ and $n_{\rm H}=1$ was obtained in \cite{Faedo:2019cvr}. 

\subsection{Truncations to only vector and tensor multiplets}
\label{sec:vectoronly}

We analyse first the possible consistent truncations that give rise to a theory with only vector/tensor multiplets. 
Since  a  consistent truncation necessarily gives rise to a symmetric scalar manifold (see Section \ref{sec:genN2}),   
the vector/tensor scalar manifolds that one can obtain must be symmetric ``very special real'' manifolds, as classified in \cite{Gunaydin:1983bi,Gunaydin:1984ak,Gunaydin:1986fg}.

This classification consists of a generic case, possible for arbitrary number of vector/tensor multiplets, where the tensor $C_{\tilde{I}\tilde{J}\tilde{K}}$ factorises, with the only non-zero components given by
\begin{equation} \label{eq:GenC}
	C_{0ij} = \eta_{ij} \,, \qquad i, j = 1, \ldots, n_{\rm VT} \,.
\end{equation}
Here $\eta_{ij}$ has signature $(1, n_{\rm VT}-1)$ and the scalar manifold is given by
\begin{equation} \label{eq:GenList}
	{\cal M}_{\rm VT} = \bbR^+ \times \frac{\SO{n_{\rm VT}-1,1}}{\SO{n_{\rm VT}-1}} \,.
\end{equation}

Additionally, there are a number of ``special'' cases that only exist for specific values of $n_{\rm VT}$ and for which $C_{\tilde{I}\tilde{J}\tilde{K}}$ does not factorise. These are given by
\be \label{eq:SpecialList}
\begin{aligned}
	\mathcal{M}_{\rm VT} &=  \frac{\SL{3, \bbR}}{\SO{3}} \,, &  \qquad n_{\rm VT}  &=5   \, ,  \\
	\mathcal{M}_{\rm VT} &=  \frac{\SL{3 ,\bbC}}{\SU{3}}  \,,  & \qquad n_{\rm VT}  &=  8  \, ,  \\
	\mathcal{M}_{\rm VT} &=  \frac{\SUs6}{\USp{6}}  \,,  & \qquad n_{\rm VT}  &=  14   \, ,  \\
	\mathcal{M}_{\rm VT} &=  \frac{\mathrm{E}_{(6 ,-26)}}{\mathrm{F}_4}  \,,   & \qquad n_{\rm VT}  &= 26    \, .
\end{aligned}
\ee
Finally, there is a second ``generic case'', which exists for arbitrary $n_{\rm VT} > 1$, but where the tensor $C_{\tilde{I}\tilde{J}\tilde{K}}$ does not factorise \cite{Gunaydin:1986fg}. The associated scalar manifolds are given by
\begin{equation} \label{eq:specialgeneric}
	{\cal M}_{\rm VT} = \frac{\SO{n_{\rm VT},1}}{\SO{n_{\rm VT}}} \,. 
\end{equation}

We want to determine which of these gauged supergravities can arise from a consistent truncation and how can they be classified in terms of the structure groups $\Gst$ listed in the
previous section. 

In order to have a consistent truncation with only vector/tensor multiplets, the  generalised tensors defining the $G_S$ structure must consist of the triplet of adjoint tensor  $J_\alpha$,  $\alpha =1,2,3$ corresponding to an  H-structure  (see Section \ref{GSgen}) and of $n_{\rm VT} + 1$ generalised vectors  $K_{\tilde I}$, ${\tilde I} =  0, 1, \ldots n_{\rm VT}$ satisfying 
\begin{equation}
\label{onlyvecstr}
 \begin{split}
 J_\alpha \cdot K_{\tilde I} &= 0 \,, \\
 {\kk}^{-2} c(K_ {\tilde I},\, K_{\tilde J} ,\, K_{\tilde K}) &= C_{{\tilde I} {\tilde J} {\tilde K}} \,,
 \end{split} 
\end{equation}
with constant $C_{\tilde{I}\tilde{J}\tilde{K}}$.

Since the  $J_\alpha$ are stabilised by  $\SUs{6} \subset \Ex{6}$, the structure group must be a subgroup of $\SUs{6}$. 
Under  the breaking  $\Ex{6} \supset \SUs{6} \cdot \SU{2}_R$, we have 
\be
\begin{aligned}
\label{E6SU6split}
\mbf{27}^* &  = \mtw{15^*}{1}   \oplus   \mtw{6}{2}  \,, \\ 
\mbf{78} & =  \mtw{35}{1}   \oplus  \mtw{20}{2}  \oplus   \mtw{1}{3}  \,, \\
\end{aligned}
\ee
where the triplet of $J_\alpha$ belong to $\mtw{1}{3}$ and generate the $\SU{2}_R$ symmetry.  
 Then,  the first condition in \eqref{onlyvecstr},  implies that the vectors  $K_{\tilde I}$ must be invariant under $\SU{2}_R$ and therefore must lie in the real vector space 
 \begin{equation}
     \mathcal{V} \subseteq \mtw{15^*}{1} \, . 
 \end{equation}
Thus, we can have at most $n_{\rm VT} = 14$ vector/tensor multiplets and 
we can immediately rule out the case $n_{\rm VT} = 26$ in \eqref{eq:SpecialList}, as well as the case $n_{\rm VT} > 14$ in \eqref{eq:GenList}.

The family  \eqref{eq:specialgeneric}  is also ruled out, because the isometries of the corresponding scalar manifolds are  not  linearly realised.
As we discussed in Section \ref{sec:genN2}, the isometry group of the scalar manifold is the commutant in $\Ex{6}$ of the structure group and by construction it acts linearly on the set of singlet generalised vectors.   As a result, the  gauged supergravities with vector/tensor scalar manifolds \eqref{eq:specialgeneric} do not arise from consistent truncations.

All other cases can in principle arise in consistent truncations and in the next subsection we will discuss from which generalised structure $G_S$  they can be obtained and 
then use $G_S$ to study the intrinsic torsion and hence find the admissible gaugings.

\subsubsection{Generic case}
 \label{sec:GenV}
The generic case with scalar manifold \eqref{eq:GenList} corresponds to the structure groups
\begin{equation}
	\label{gencstrg} 
	G_S = \Spin{6 - n_{\rm VT}} \times  \SU{2} \,,
\end{equation}
of  item  (\ref{GstGen}) of the list in the previous section,  where for notational convenience we let $\Spin{1} =\Spin{0} = \mathbb{Z}_2$. Note that \eqref{gencstrg} implies that we can have at most $n_{\rm VT} = 6$ vector/tensor multiplets in the truncation. Moreover, the case $n_{\rm VT} = 5$ and $n_{\rm VT} = 6$ have identical structure groups. This means that any background admitting a truncation with $n_{\rm VT} = 5$ actually admits a truncation with $n_{\rm VT} = 6$, with the former truncation being a subtruncation of the latter.

To see how these structure groups arise, note that the structure \eqref{eq:GenC} of the tensor $C_{{\tilde I} {\tilde J} {\tilde K}}$ implies that the vectors $K_{\tilde I}$ can be split into a vector  $K_0$  and $n_{\rm VT}$ vectors $K_{i}$ such that  for any $i,j,k= 1, \ldots, n_{\rm VT}$, %$K_I$ such that  for any $ I,J,K= 1, \ldots, n_{\rm VT}$,
  \be
\label{K0cond}
 c(K_0,K_0, \cdot) = 0 \,, \qquad 
 c(K_{i}, K_{j}, K_{k}) = 0 \,, \qquad 
 c(K_0, K_{i} ,K_{j}) = \eta_{ij}   \,, 
\ee
where $\eta_{ij}$ %$\eta_{IJ}$
has signature $(5,1)$.  The vector $K_0$  corresponds to the graviphoton of the truncated theory.

By studying the form of \eqref{K0cond}, we can deduce the stabiliser group of the generalised vector fields $K_{\tilde{I}}$ as follows.
Being in the $\rep{15}^*$ of $\SUs{6}$, the vectors $K_{\tilde I}$ can be seen as six-dimensional two-forms. 
Then the first condition in  \eqref{K0cond} is equivalent to 
\begin{equation}
 K_0 \wedge K_0 = 0 \,, 
\end{equation}
with $\wedge$ the standard wedge product of $p$-forms. Thus, $K_0$ must be decomposable and we can choose a basis of independent six-dimensional one-forms such that 
\begin{equation}
 K_0 = e_5 \wedge e_6 \, . 
\end{equation}
The stabiliser of $K_0$ is $\SUs{4} \times \SU{2}$, embedded in $\SUs{6}$ as
\be
\begin{aligned}
\label{SU6SU4split}
\SUs{6} & \supset \SUs{4} \times \SU{2} \times \U{1} \,, \\
\rep{15}^*  &=  \mtw{4^*}{2}_{1} \oplus \mtw{6}{1}_{-2}  \oplus  \mtw{1}{1}_{4}  \, , 
\end{aligned}
\ee
with $K_0 \in  \mtw{1}{1}_{4}$. This forces the $G_S$ structure to be a subgroup of $\mathrm{SU}^*(4) \times \SU{2}$. The other  conditions in  \eqref{K0cond}   become
\be
 \label{eq:genericn}
 K_0 \wedge K_{i} \wedge K_{j}  = \eta_{ij} \,, \qquad  K_{i} \wedge K_{j}  \wedge K_{k} = 0 \,,
\ee
where the metric $\eta_{ij}$ % $\eta_{IJ}$ 
is invariant under  $\SUs{4} \simeq \Spin{5,1}$.  From \eqref{eq:genericn} it follows that
\be
\label{sixK}
 K_{i} \in \mtw{6}{1}_{-2}  \, . 
 \ee
 
Thus, there can be at most six vector multiplets of this type.

The structure group $G_S$ can now be easily determined.  Since the $n_{\rm VT}$  singlets $K_{i}$ %$K_I$
satisfy the
 inner product \eqref{eq:genericn} of  signature $(1, n_{\rm VT} - 1)$ they break $\SUs{4}$ to
 \be
 \SUs4 \simeq \Spin{5,1} \supset \Spin{6 -n_{\rm VT}} \times \Spin{n_{\rm VT} - 1,1} \, ,
 \ee
 where  the factor $\Spin{6 -n_{\rm VT}}$ is  the stabiliser of the $K_{i}$  
 while the factor $\Spin{n_{\rm VT} - 1, 1}$ rotates the $K_{i}$ into each other. 
Thus, the structure group  is given by
\begin{equation}
	G_S = \Spin{6 -n_{\rm VT}} \times  \SU{2} \, . 
\end{equation}
Although the structure groups and the isometry groups are  ${\rm Spin}$ subgroups of $\Ex6$, the generalised vectors $K_{i}$ %$K_I$
never appear in spinorial representations of $G_S$ and hence only see the orthogonal groups and not their double covers. This is the reason why the case with $n_{\rm VT}=5$ vectors/tensors can always be enhanced to $n_{\rm VT} = 6$: on  the two-forms $K_{i}$ % $K_I$ 
the $\mathbb{Z}_2$ structure group acts trivially. Moreover, this is why the coset spaces can be reduced to take the form~\eqref{eq:GenList}:

\begin{equation}
   {\cal M} = \mathcal{M}_{\rm VT} = \frac{\Com{\Gst}{\Ex{6}}}{\Com{\Gst}{\USp{8}/\bbZ_2}}  =  \bbR^+\times\frac{\SO{n_{\rm VT} -1,1}}{\SO{n_{\rm VT}  -1}} \,.
\end{equation}
The corresponding isometry group is 
\begin{equation}
\label{gisogenvec}
\Giso = \bbR^+ \times \SO{n_{\rm VT}  -1,1} \times \SU2_R \, ,
\end{equation}
where as discussed above we take $\Gh=\SU2_R$, even though there are no hypermultiplets, in order to include the R-symmetry. Under $\Giso$ the space of vectors transforms as 
\begin{equation}
\begin{aligned}
    \mathcal{V} &= ( \rep{1},\rep{1} )_2 \oplus ( \mbf{1}, \mbf{n})_{-1} 
    \ni (v^0, v^i) \, , 
\end{aligned}
\end{equation}
where  the first entries are the $\SU{2}_R$ representations, $\mbf{n}$ is the vector representation of $\SO{n_{\rm VT}  -1,1}$, the subscripts are the $ \mathbb{R}^+$ charges, and $i=1,\dots,n_{\rm VT}$ denotes $\SO{n_{\rm VT}  -1,1}$ indices. 
\medskip

We can now determine the embedding tensor of the truncated theory and the possible gaugings. These are encoded in the intrinsic torsion of the $G_S$ structure, which must only contain $G_S$ singlets for the truncation to be consistent. We assume that this occurs and decompose the intrinsic torsion in representations of the global isometry group \eqref{gisogenvec} 
\be
\label{dec351genv}
\begin{aligned}
W_{\rm int}  &=  (\mbf{3},\mbf{1} )_{-2}  \oplus   (\mbf{3} ,\mbf{n})_{1}   \oplus  (\mbf{1},\mbf{n})_1  \oplus (\mbf{1},\mbf{ad})_{-2}     \oplus  (\mbf{1},\mbf{X})_{1}    \\
&\ni ( \tau^a_{0  \, b },  \tau^a_{i  \, b},   \tau_i ,  \tau^i_{0 \,  j},   \tau_{[ijk]} ) \,,
\end{aligned}
\ee
where $\mbf{ad}$ and $\rep{X}$ denote the adjoint and the rank-3 anti-symmetric\footnote{In some cases the representation $\rep{X}$ might be reducible.} representations of  $\SO{n_{\rm VT}  -1,1}$, respectively, and $a,b=1,2,3$ are $\SU2_R$ indices. The case $n_{\rm VT}=5$ is different, but can be obtained as a subtruncation of the case $n_{\rm VT}=6$. Therefore, we will not consider $n_{\rm VT}=5$ here.

Now we need the map \eqref{eq:EmbedT}, which gives the generalised geometry embedding tensor, and which we denote by  $T  : \, \mathcal{V} \to \mathfrak{g}_{\rm iso} $ to distinguish it from the generic 5d embedding tensor.  
Given an element $v\in\mathcal{V}$, the intrinsic torsion defines $T$ as having the non-zero components  
\begin{equation}
\label{torsiongencase}
\begin{split}
	T(v)^a{}_b &=    v^0  \tau^a_{0  \, b } +  v^i   \tau^a_{i  \, b}  \in  \,    \su(2)_R \,, \\
	T(v)^i{}_j &=    v^0    \tau^i_{0 \,  j}  + v^k    \tau_{k}{}^{i}{}_{j}\,  \in    \so(n_{\rm VT} -1,1) \,, \\
	T(v)_{(0)}  &=    v^i \tau_i  \in    \mathfrak{u}(1) \,. 
\end{split}
\end{equation}
The adjoint action on the vectors in $\mathcal{V}$
\begin{equation}
\begin{split}
\label{adjointacgnenvec}
(  T(v) \cdot w)^0 & = 2\,T(v)_{(0)}   w^0 \,, \\
(  T(v) \cdot w)^i & = -T(v)_{(0)}   w^i  +  T(v)^i{}_j w^j \,,
\end{split}
\end{equation}
defines  the Leibniz  bracket  $ T(v) \cdot w  = t_v(w) =  \Leib{v}{w}$. 
The Leibniz condition \eqref{leib-cond}  gives a set of constraints  on the torsion components
\begin{equation} \label{eq:GenLeib}
	\tau_{[jk}{}^m \tau_{l]m}{}^i = 0 \,, \qquad \tau_{0\,i}{}^k \tau_{jk}{}^l = 0 \,, \qquad \tau_i = 0 \, , 
\end{equation}
so that the Leibniz bracket simplifies to 
\begin{equation} \label{eq:GenLeibBracket}
\begin{split} 
\Leib{v}{w}^0  &  =   0  \,, \\
\Leib{v}{w}^i  &  = - v^j w^k  \tau_{jk}{}^i   -  v^0 w^k \tau_{0k}{}^i  \, .
\end{split}
\end{equation}
From  \eqref{tensorcond}  we see that the rank of $\tau_{0i}{}^j$  determines the number of tensor multiplets, while $\tau_{ij}{}^k$ form the structure constants of the gauge algebra. Finally, $\tau^a_{0 b}$ and $\tau^a_{i b}$ determine how the $\SU{2}_R$ is gauged.
Moreover, the gauging of the $\SU{2}_R$ R-symmetry must form a representation of $\ggauge$ as in \eqref{eq:tp-alg}. Explicitly, this implies
\begin{equation} \label{eq:Leibniz-Hyp}
 \left( T(v) \cdot T(w) - T(w) \cdot T(v) \right)^a{}_b = - \left( T(\Leib{v}{w}) \right)^a{}_b \,,
\end{equation}
which for \eqref{torsiongencase} imposes
\begin{equation} \label{eq:GenCaseSU2gauging}
\tau_i^a{}_c \tau_0^c{}_b -  \tau_0^a{}_c \tau_i^c{}_b =0 \,, \qquad \tau_{0k}{}^i \tau_i^a{}_b = 0 \,, \qquad \tau_i^a{}_c \tau_j^c{}_b - \tau_j^a{}_c \tau_i^c{}_b = \tau_{ij}{}^k \tau_k^a{}_b \,.
\end{equation}
Then the  first line in \eqref{torsiongencase}  gives the embedding tensor for the $\SU{2}_R$ symmetry 
\begin{equation}
p_{0 a}{}^b = \tau^b_{0 a } \,, \qquad p_{i a}{}^b =  \tau^b_{i a }  \, , 
\end{equation}
while  the  non-zero components of the embedding tensor on the vector isometries are  
\begin{equation}
 t_{0k}{}^i  = -  \tau_{0k}{}^i \,,  \qquad   t_{jk}{}^i =  - \tau_{jk}{}^i   \, . 
\end{equation}

Since $\tau_i = 0$, we note that the $\mathbb{R}^+$ can never be gauged. Also from \eqref{eq:GenLeibBracket}, we see that the graviphoton $v^0$ cannot contribute to non-abelian gaugings. Moreover, from \eqref{dec351genv} and \eqref{eq:GenLeibBracket}, we can already see that not all gaugings of five-dimensional ${\cal N}=2$ supergravity can arise from a consistent truncation. In particular, the  tensor multiplets  can only be charged under the graviphoton $v^0$ and not any of the $n_{\rm VT}-1$ vector fields transforming non-trivially under $\SO{n_{\rm VT}-1,1}$, as for example constructed in \cite{Gunaydin:1999zx}.

From \eqref{eq:GenCaseSU2gauging} we can also determine  in general how the $\SU{2}_R$ global symmetry can be gauged. Whenever an $\SO{3} \subset \SO{n_{\rm VT}-1,1}$ is gauged, those $\SO{3}$ vectors can also be used to gauge the $\SU{2}_R$ via $\tau^a_{i b}$. Alternatively, any combination of abelian vector fields, including the graviphoton can gauge a $\U{1}_R\subset\SU2_R$ subgroup. 

\medskip

Let us now find which gaugings of the $\SO{n_{\rm VT}-1,1}$ global symmetry group of $\mathcal{M}_{\rm VT}$ are possible, beginning with $n_{\rm VT} = 1$ and working up to the maximal case $n_{\rm VT} = 6$.

\paragraph{$\mathbf{n_{\rm VT} = 1}$:}  In this case the isometry group is  $\Giso =\SU{2}_R  \times  \mathbb{R}^+$ and  the structure group is $G_S = \Spin{5} \times  \SU{2}$. 

Any combination of the two vectors can gauge a $\U{1}_R$ subgroup of the R-symmetry.

\paragraph{$\mathbf{n_{\rm VT} = 2}$:}  The structure group is  $G_S = \Spin{4}  \times  \SU{2}$ and the isometry group is   $\Giso =\SU{2}_R  \times  \SO{1,1} \times  \mathbb{R}^+$.
There are three singlet vectors with the following $ \SO{1,1} \times  \mathbb{R}^+\simeq\bbR^+\times\bbR^+$ charges 
\be 
v = (v^0, v^+ , v^-)  \in  \mathcal{V} = \mbf{1_{0, 2}} \oplus  \mbf{1_{2, -1}} \oplus  \mbf{1_{-2, -1}}  \, . 
\ee
The conditions \eqref{eq:GenLeib} are now trivially satisfied since $\tau_{ijk} = 0$.
From the intrinsic torsion
\be
W_{\rm int} \ni ( \tau^a_{0  \, b } \, , \tau^a_{+  \, b } \,  ,  \tau^a_{-  \, b } ,   \tau^+_{0  \, - } )  \, , 
\ee
we see that, when $\tau^+_{0  \, - }=0$, any combination of all three vectors can gauge a $\U{1}_R$ symmetry. Alternatively, when $ \tau^+_{0  \, - } \neq 0$, two vectors are dualised to tensors and the remaining $v^0$ can gauge the $\SO{1,1}$ under which the two tensors are charged, as well as a $\U{1}_R$ symmetry.

\paragraph{$\mathbf{n_{\rm VT} = 3}$:}  The structure group is  $G_S = \Spin3\times  \SU{2}$ and the isometry group is  $G_{\rm iso} =\SU{2}_R \times \SO{2,1} \times  \mathbb{R}^+$. As there are three vector multiplets in the adjoint of $\SO{2,1}$ 
it is a priori  possible to gauge it. 
The conditions \eqref{eq:GenLeib} now imply that either $\tau_{ijk} \neq 0$ or $\tau_{0\,i}{}^j \neq 0$.
\begin{itemize}
\item $\tau_{ijk} := \hat\tau\epsilon_{ijk}\neq 0$, $\tau_{0\,i}{}^j = 0$. 

The Leibniz algebra, given $v=(v^0,v^i)$ and $w=(w^0,w^i)$, takes the form 
\be
\begin{aligned}
\Leib{v}{w}^{i}  &= - \hat\tau v^j w^k   \epsilon_{j k}{}^i \, , \qquad &
\Leib{v}{w}^{0} &= 0 \, . 
\end{aligned} 
\ee
Thus the full  $\SO{2,1}$ can be gauged  and we can use the singlet vector $v^0$ to gauge a $\U{1}_R$. 

\item $\tau_{ijk} = 0$. We now have a purely abelian gauge group. When $\tau_{0\,i}{}^j \neq 0$ two of the vectors are dualised to tensor multiplets.  By choosing the tensors to be both spacelike or one spacelike and one timelike under $\SO{2,1}$ we get different charges for the tensor multiplets under the action of $v^0$, leading to either an $\SO{2}$ or $\SO{1,1}$ gauging. In addition, a linear combination of  $v^0$ and the uncharged vectors  can also gauge the $\U{1}_R$ symmetry. 
\end{itemize}

\paragraph{$\mathbf{n_{\rm VT} = 4}$:}  The structure group is $G_S = \Spin2 \times \SU{2}$ and the  isometry group is $G_{\rm iso} =\SU{2}_R \times \SO{3,1} \times  \mathbb{R}^+$.   

The conditions \eqref{eq:GenLeib} now imply that either $\tau_{ijk} \neq 0$ or $\tau_{0\,i}{}^j \neq 0$. 
We thus have the following possibilities.
\begin{itemize}
\item $\tau_{ijk} \neq 0$, $\tau_{0\,i}{}^j = 0$. We can write $\tau_{ijk} = \epsilon_{ijkl} A^l$. Depending on whether $A^i$ is spacelike, timelike or null with respect to $\SO{3,1}$, we can have the gauge groups $\SO{2,1}$, $\SO{3}$ or $\ISO{2}$, respectively. In all cases, there are no tensor multiplets. This can be seen as follows.

If $A$ is timelike, we can always perform an $\SO{3,1}$ rotation such that it lies along the timelike direction and we have 
\be
\tau_{\alpha \beta \gamma}  = \epsilon_{\alpha \beta \gamma 1}  A^{1} := \hat\tau \epsilon_{\alpha\beta\gamma} \, ,
\ee
where we split the $\SO{3,1}$ indices as  $i= 1$ for the timelike direction and $\alpha, \beta, \gamma =2,3,4$ the spacelike ones. Writing $v = ( v^0, v^\alpha, v^1)\in \mathcal{V}$ we find the brackets
\be
\begin{aligned}
    \Leib{v}{w}^\alpha  &= - \hat\tau v^\beta w^\gamma \epsilon_{\beta\gamma}{}^{\alpha} \, , & \qquad 
    \Leib{v}{w}^{0} &= \Leib{v}{w}^1 = 0 \,,
\end{aligned}
\ee
leading to a gauging of the compact subgroup $\SO{3}\subset\SO{3,1}$.
\begin{comment}
Overall the $\SO{3}$ singlets in the intrinsic torsion are 
\be
%\label{dec351genv4}
W_{\rm int}  \supset    \{ \tau^a_{0  \, b },  \tau^\alpha_{1 \, \beta},      \tau_{\alpha \beta \gamma}  \} \, . 
\ee
The  four singlet vectors decompose into a triplet and a singlet of $\SO{3}$, which are all neutral under  $\mathbb{R}^+$ charges. If 
we parameterise the  elements  in the space $\mathcal{V}$ spanned by the singlet vectors  as  $v = ( v^0, v^\alpha, v^1)$, 
the map $T$ reduces to
\be
\begin{aligned}
	& T(v)^a{}_b =    v^0  \tau^a_{0  \, b } +  v^1   \tau^a_{1  \, b}     \\
	%
	& T(v)^\alpha{}_\beta =   v^\gamma    \tau_{\gamma \, \, \beta}^\alpha   \\
	%
	& T(v)_{(0)}  =   0  \, , 
\end{aligned}
\ee
and gives the brackets
\be
\begin{aligned}
& \Leib{v}{w}^\alpha  = v^\gamma \tau_{\gamma \, \, \beta}^{\, \, \alpha} w^\beta \\ 
& \Leib{v}{w}^{0}   = 0 = \Leib{v}{w}^1
\end{aligned}
\ee
Thus, $\SO{3}$ can be gauged, while  there is no gauging of the $\mathbb{R}^+$ and two singlet vectors can gauge the $\U{1}_R$ symmetry.
\end{comment}
In addition, either a combination of $v^0$ and $v^1$ can be used to gauge a $\U{1}_R$ or the  $v^\alpha$ can  gauge the full $\SU{2}_R$ via $\tau_\alpha^a{}_b$.

For a spacelike $A$ we proceed in the same way. By an $\SO{3,1}$ rotation we bring $\tau_{ijk}$ to the form 
\be
\tau_{\alpha \beta \gamma} = \epsilon_{\alpha \beta \gamma 4}  A^4 := \hat\tau \epsilon_{\alpha\beta\gamma} \, , 
\ee
where now $\alpha, \beta, \gamma = 1,2,3$. The vector decompose as $v=(v^0,v^\alpha,v^4)$  and we get the algebra
\be
\begin{aligned}
    \Leib{v}{w}^\alpha  &= - \hat\tau v^\beta w^\gamma \epsilon_{\beta\gamma}{}^{\alpha} \, , \qquad & 
    \Leib{v}{w}^{0} &= \Leib{v}{w}^4 = 0 \,,
\end{aligned}
\ee
which gauges an $\SO{2,1}$ subgroup. As above, the $v^0$ and $v^4$ can be used to gauge the $\U{1}_R$.
\begin{comment} so that now we have 
\be
\begin{aligned}
	& T(v)^a{}_b =    v^0  \tau^a_{0  \, b } +  v^3   \tau^a_{3  \, b}     \\
	%
	& T(v)^\alpha{}_\beta =   v^\gamma    \tau_{\gamma \, \, \beta}^\alpha   \\
	%
	& T(v)_{(0)}  =   0  \, . 
\end{aligned}
\ee
The  non-zero bracket
\be
\begin{aligned}
& \Leib{v}{w}^\alpha  = v^\gamma \tau_{\gamma \, \, \beta}^{\, \, \alpha} w^\beta \\ 
\end{aligned}
\ee
implies that $\SO{2,1}$ can be gauged and  two singlet vectors can gauge the $\U{1}_R$ symmetry.
\end{comment}

Finally if $A$ is null,  by an $\SO{3,1}$ rotation we can reduce to two non-zero components for $\tau_{ijk}$ 
\be
\tau_{234} = \epsilon_{2341} A^1\,, \qquad  \tau_{123} = \epsilon_{1234} A^4\,, \qquad A^1 = A^4 := \hat\tau \,.
\ee
It is useful to decompose the vectors as $v=(v^0,v^2,v^3,v^-,v^+)$ where $v^\pm=v^1\pm v^4$. The Leibniz algebra then becomes 
\begin{equation}
 \Leib{v}{w}^0 = 0 \quad  
\begin{array}{l}
\Leib{v}{w}^2 = \hat\tau ( v^+ w^3 - w^+ v^3) \, ,  \\ 
    \Leib{v}{w}^3 = - \hat\tau (v^+ w^2 - w^+ v^2) \, , 
\end{array} \quad     
\begin{array}{l}
    \Leib{v}{w}^+ = 0 \, , \\ 
    \Leib{v}{w}^- = 2\, \hat\tau (v^2 w^3 - w^2 v^3) \, , 
\end{array}
\end{equation}

This defines a Lie algebra that is the semi-direct sum of $\mathfrak{so}(2)$ with the 3-dimensional Heisenberg algebra. 
The vector $v^-$ generates the $\so(2)$, under which $v^2$ and $v^3$ are charged. On the other hand, $\{ v^2, \, v^3 ,\, v^+ \}$ form a Heisenberg algebra, with $v^+$ the central element. Since $v^+$ is central, the gauge group \eqref{eq:redalgebra} under which matter is charged is just $\ISO{2}$, generated by $\{v^-,v^2,v^3\}$. Additionally, the graviphoton $v^0$ can gauge the $\U{1}_R$.
	
\item $\tau_{ijk} = 0$. We now have a purely abelian gauging and 0, 2 or 4 tensor multiplets, depending on the rank of $\tau_{0\,i}{}^j$. Depending on whether the tensors are timelike or spacelike we get different charges for the tensor multiplets under the abelian group generated by $v^0$, as discussed for $n_{\rm VT} = 3$. As a result, we either have two tensor multiplets charged under a $\SO{2}$ or $\SO{1,1}$, or four tensor multiplets charged under the $\SO{1,1}$. In addition, $v^0$ and, when present, any combination of the uncharged vectors can also gauge the $\U1_R$. 
\end{itemize}

\paragraph{$\mathbf{n_{\rm VT} = 6}$:} The structure group is  $G_S = \mathbb{Z}_2 \times \SU{2}$ and the isometry group is $\Giso = \SU2_R \times  \SO{5,1} \times \mathbb{R}^+$. 
 
In this case, we will not solve the Leibniz conditions directly but instead we perform a case by case analysis of  the possible gauge groups  with a given number of tensor multiplets. Since there are 6 vectors, if there are no tensors, we can gauge at most the following semi-simple subgroups of the global $\SO{5,1}$ isometries: $\SO{4}$, $\SO{3,1}$ or $\SO{3} \times \SO{2,1}$. These are only possible if the singlet vectors transform in the adjoint of one of these groups and the torsion contains singlets of the gauge groups.

It is straightforward to see that $\SO{4}$ and $\SO{3,1}$ cannot be gauged. Under the breaking $\SO{5,1}\supset\SO{4}\times\SO{1,1}$ (respectively $\SO{5,1}\supset\SO{3,1}\times\SO{2}$) the vector representation $\rep{6}$ decomposes as 
\be
\rep{6} = \rep{4}_0 \oplus \rep{1}_2 \oplus \rep{1}_{-2}  \,,
%\to \rep{5} \oplus \rep{1} \to ( \rep{2} , \rep{2} ) + 2  (  \rep{1} , \rep{1} )  \,,
\ee
where $\rep{4}$ is the vector representation of $\SO{4}$ (respectively $\SO{3,1}$) and the subscripts denote the $\SO{1,1}$ (respectively $\SO2$) charges. Manifestly we see that in each case the decomposition does not include the adjoint represention.

On the other hand, we can gauge $\SO{3} \times \SO{2,1} \subset \SO{5,1}$. In this case, the six vectors decompose as
\begin{equation}
    \rep{6} = (\rep{3}, \rep{1} ) \oplus (\rep{1},\rep{3}) \,,
\end{equation}
containing the adjoint of $\SO{3} \times \SO{2,1}$. We denote these by $v^\alpha$ and $v^{\dot\alpha}$, where $\alpha= 1, 2, 3$ labels the adjoint of $\SO{3}$ and $\dot{\alpha} = 4, 5, 6$ the adjoint of $\SO{2,1}$. This gauging consists of having
\begin{equation}
    \tau_{\alpha\beta\gamma} = A\, \epsilon_{\alpha\beta\gamma} \,, \qquad \tau_{\dot\alpha\dot\beta\dot\gamma} = B\, \epsilon_{\dot\alpha\dot\beta\dot\gamma} \,, \qquad A, B \neq 0 \,.
\end{equation}
Note that therefore \eqref{eq:GenLeib} implies that $\tau_{0\,i}{}^j = 0$ so that we have no tensor multiplets. The Leibniz bracket becomes
\begin{equation}
\begin{split} 
\Leib{v}{w}^\alpha  &  = - v^\beta  w^\gamma  \tau_{\beta \gamma }{}^\alpha \,, \\
\Leib{v}{w}^{\dot \alpha}  &  = - v^{\dot \beta}  w^{\dot \gamma}  \tau_{\dot{\beta}  \dot{\gamma} }{}^{\dot \alpha} \,,
\end{split}
\end{equation}
reproducing the gauge algebra of $\SO{3} \times \SO{2,1}$.
The graviphoton can gauge the $\U{1}_R$ symmetry or the vectors $v^\alpha$ can gauge the diagonal of $\SO{3}$ and $\SU{2}_R$.

Let us now study gaugings that could include tensor multiplets.
These will have $\tau_{0\,i}{}^j \neq 0$. When $\tau_{0\,i}{}^j $ has rank 2, two vectors are dualised into tensors which are charged under $v^0$, and the gaugings can only be given by the other four vectors.  Depending on the signature of the $\SO{5,1}$ metric evaluated in the directions of the tensor multiplets we can have
\begin{equation}
\begin{aligned}
& \SO{3} \times  \SO{1,1}\times  \U{1}_R \,,&& \qquad  \SO{2,1} \times  \U{1} \times  \U{1}_R \,,\\% \qquad
& \SO{3} \times  \U{1} \times  \U{1}_R \,,&& \qquad \ISO{2} \times \U{1} \times \U{1}_R \,, \\
& \SO{1,1} \times \SU{2}_R \,, && \qquad \U{1} \times \SU{2}_R \,,
\end{aligned}
\label{n6tensorg}
\end{equation}
where the factors $\SO{1,1}$ or $\U{1}$ are gauged by the graviphoton and $\U{1}_R$ by any combination of $v^0$ and  the vector that does not gauge the non-abelian factor. Note that in \eqref{n6tensorg} we list the largest group that can be gauged. It is clearly possible to gauge only some factors of the products above.

When $\tau_{0\,i}{}^j$ has rank 4 or 6,  the only possible gauge group is the abelian factor gauged by  $v^0$ and the $\U1_R$. Depending on whether the image of $\tau_{0\,i}{}^j$ includes the negative eigenvalue of the $\SO{5,1}$ signature, or not, we get different charges for the tensor multiplets under the action of $v^0$, which hence   gauges either a $\U{1}$ or $\SO{1,1}$ group.  In addition, $v^0$ and, when present, any combination of the uncharged vectors can also gauge the $\U1_R$.

In Table \ref{summ-gengaugings} we summarise the allowed gaugings for
truncations with only vectors/tensor multiplet of generic type.
Whenever we list a product group, the individual factors can also be gauged separately even though they are not listed as such. Whenever there are abelian factors in $\Gred$, the $\U{1}_R$ can also be gauged diagonally with some combination of these factors.

\begin{table}[h!]
%\begin{equation} \label{eq:sumtable}
\centering
\makebox[\textwidth][c]{\begin{tabular}{|c|c|c|c|}
\hline
$n_{\rm VT}$ & $G_{\rm iso}$  & $\Gred$ & $n_{\rm T}$ \\
\hline
\Tstrut\Bstrut
$1$ & $\SU{2}_R  \times  \mathbb{R}^+$ & $\U{1}_R$ & -- \\
\hline
\multirow{2}{*}{2} & \multirow{2}{*}{$\SU{2}_R  \times \SO{1,1} \times \mathbb{R}^+$ } &   $\U{1}_R$ & -- \\
&  &  $\SO{1,1}$   & $2$  \\
\hline
\multirow{2}{*}{3} & \multirow{2}{*}{$\SU{2}_R  \times \SO{2,1} \times \mathbb{R}^+$ }  & $\SO{2,1} \times \U{1}_R$ & --  \\
& & $\SO{2}$,\; $\SO{1,1}$ & 2 \\
\hline
\multirow{4}{*}{4} & \multirow{4}{*}{$\SU{2}_R  \times \SO{3,1} \times \mathbb{R}^+$ }  &  $\SO{2,1} \times \U{1}_R$,\; 
$\SO{3} \times \U{1}_R$, & \multirow{2}{*}{--} \\
 & & $\ISO{2}\times \U{1}_R$,\; $\SU{2}_R$ %, $\U{1}_R$ 
 & \\ 
 & &  $\SO{2} \times \U{1}_R $,\; $\SO{1,1} \times \U{1}_R$,  %$\U{1}_R$  
 &  2 \\[0.1cm]
  &  &  $\SO{1,1}$  & 4  \\
\hline
\multirow{7}{*}{6} & \multirow{7}{*}{$\SU{2}_R  \times \SO{5,1} \times \mathbb{R}^+$ } &  $\SO{3}\times\SO{2,1} \times \U{1}_R$,\; $\SO{2,1} \times %\SO{3}^\prime
\SU{2}_R$, &  \multirow{2}{*}{--}  \\
&  & $\ISO{2} \times \U{1}_R$   &   \\[0.1cm]
 &  &  $\SO{2,1}\times \U{1} \times \U{1}_R $,\; $\SO{3} \times \SO{2} \times \U1_R$,  & \multirow{3}{*}{2}\\
 & & $\SO{3}\times \SO{1,1}  \times \U{1}_R$,\; $\ISO{2} \times \U{1}  \times \U{1}_R $, & \\
 & &  $\SO{2} \times \SU{2}_R$,\; $\SO{1,1} \times \SU{2}_R$  &  \\[0.1cm]
  &  &  $\U{1} \times \U{1}_R $,\;  $\SO{1,1} \times \U{1}_R $ & 4  \\[0.1cm]
    &  &  $\SO{1,1} $ & 6  \\
\hline
\end{tabular}}
\caption{Allowed gaugings $\Gred$ of the global isometry groups $\Giso$ in the generic cases with $n_{\rm VT} $ vector/tensor multiplets.
The first column gives the total number of vectors and tensor multiplets, the second the global isometry group, the third the allowed gaugings and the last one the number of vectors that are dualised to tensors in each case.}
\label{summ-gengaugings}
\end{table}

\subsubsection{Special cases} 

The special cases \eqref{eq:SpecialList} are also associated to some of  the generalised $G_S$-structures we listed at the beginning of this section. 
We now discuss case by case what the associated structure groups are, we determine the corresponding embedding tensor and hence the possible gaugings of the truncated theory.

Differently from the generic case it is quite cumbersome to analyse in full generality the constraints imposed on the gaugings by the Leibniz condition \eqref{leib-cond} and hence the allowed gaugings. Thus in this section  we will limit ourselves to study what are the  largest reductive groups and largest compact groups that can be gauged. 

\paragraph{$\mathbf{n_{VT} = 5}$:} This truncation is associated to a  $\Gst = \SU{2}$  generalised structure.  The structure group is taken to be the  $\SU{2}$ factor in the breaking 
 (\ref{GstSU2}) of  $\USp{6}$ and it embeds  in $\SUs{6}$ as  $\SUs{6} \supset \SL{3,\bbR} \times \SU{2}$. 
Under this embedding we have 

\begin{equation}
	\mbf{15}^* = \left( \mbf{6}^*,\mbf{1}\right) \oplus \left(\mbf{3},\mbf{3} \right) \,,%,
\end{equation}

so that $\mathcal{V}=\left(\mbf{6}^*,\mbf{1}\right)$ and there are six independent singlet vectors giving rise to $n_{\rm VT}  = 5$ vector multiplets. It is easy to check that we also get the expected scalar manifold
\begin{equation}
	{\cal M} = {\cal M}_{\rm VT} = \frac{\Com{\Gst}{\Ex{6}}}{\Com{\Gst}{\USp{8}/\bbZ_2}} = \frac{\SL{3,\bbR}}{\SO{3}} \,,
\end{equation}

with isometry group  
\be
\Giso = \SU{2}_R \times  \SL{3,\bbR}  \, .
\ee
We can decompose the elements of $\mathcal{V}$ according to 
$\Giso$
\begin{equation}
    \mathcal{V} = (\mbf{1},\mbf{6}^*) \ni v_{ij} \, ,  \qquad i,j=1,2,3 \, . 
\end{equation}

The $G_S$ singlet intrinsic torsion decomposes under $\Giso$ as
\be
\begin{aligned}
\label{dec351n5}
W_{\rm int}   &= (\mbf{3},\mbf{6} )    \oplus (\mbf{1},\mbf{15}^*)  \oplus   \,  \, (\mbf{1},\mbf{3}^*) \\
 & \ni \{ \tau^{(ij)a}{}_{b} ,  \tau^i{}_{(jk)}  ,  \tau_{i}  \} \, , 
\end{aligned}
\ee
where $\tau^i{}_{ik} = 0$. If  $ v_{(ij)}   \in  \mathcal{V}$ the map  $T : \mathcal{V} \to  \mathfrak{g}_{\rm iso} $ is  defined as
\be
\begin{aligned}
	& T(V)^a{}_b =  v_{(ij)}    \tau^{(ij)a}{}_{b} \,, \\ 
	& T(V)^i{}_j =  \epsilon^{imn} v_{(ml)} \tau^l{}_{(nj)}   +  \epsilon^{imn} v_{(mj)} \tau_n \,,
\end{aligned} 
\ee
 and gives the bracket
\be
\label{lbn5}
\begin{aligned}
\Leib{v}{w}_{ij}  &  =  T(v)^k{}_{(i} w_{j)k}  =  t^{(kl) (mn)}{}_{(ij)} v_{kl}  w_{mn} \\ 
& = - \epsilon^{m pk }   [    \tau^l{}_{p(i} \delta^n{}_{j)}   +   \tau_p  \delta^l{}_i \delta^n{}_j  ] v_{kl} w_{mn}  \, . 
\end{aligned}
\ee
Thus the components of the embedding tensor are
\be
 t^{(kl) (mn)}{}_{(ij)}  = \frac12 \epsilon^{pm(k}   [    \tau^{l)}{}_{p(i} \delta^n{}_{j)}   +   \tau_p  \delta^{l)}{}_{(i} \delta^n{}_{j)}] + (m \leftrightarrow n) \,, \qquad  p^{(ij ) a}{}_{b}= \tau^{(ij ) a}{}_{b} \,.
 \ee 

We now want to determine the largest non-abelian gaugings that can arise from the consistent truncation. The compact gaugings are quite limited. It is easy to see that it is not possible to gauge the maximal compact subgroup $\SO{3}$ of  $\SL{3,\bbR}$. Indeed, the 6 vectors  decompose as $\mbf{6}^* = \mbf{5} \oplus \mbf{1}$ and therefore do not contain the adjoint of $\SO{3}$.   
However, the singlet in the decomposition can be used to gauge the $\U{1}_R$ symmetry, as can also be seen from   the intrinsic torsion, which  contains only an $\SO{3}$  singlet in $ \tau^{(ij ) a}{}_{b}$. 
We see that only compact abelian gaugings are a priori possible.

Consider now the non-compact gauging $\SL{2,\mathbb{R}} \simeq \Spin{2,1}$.  This is obtained via the embedding $\SL{3,\bbR} \supset \SL{2,\bbR} \times \mathbb{R}^+$, %$\SU{3} \supset   \SU{2} \times \U{1}$, 
under which  the vectors  decompose as $\rep{6}^* =\rep{3}_{2} \oplus \rep{2}_{-1} \oplus \rep{1}_{-4}$.  Thus we expect to be able to gauge $\SL{2,\bbR}$, with the two vectors that are charged under $\SL{2,\mathbb{R}}$  dualised into tensors.
To see whether this gauging is possible, we must look at the intrinsic torsion and the bracket \eqref{lbn5}. The vectors decompose as
\be
v_{(ij)}  = \{ v_{(\alpha \beta)}, v_\alpha,  v_0 \} \,,
\ee
where $\alpha = 1,2$ are fundamental indices of $\SL{2,\bbR}$. The intrinsic torsion contains the $\SL{2,\bbR}$ singlets 
\be
W_{\rm int} \supset  \mbf{1}_{-4}  \oplus  \mbf{1}_2 \oplus   \in \mbf{1}_{-2}
\ni (\tau^{a}{}_{0 b}  , \hat{\tau}=\tau^0{}_{00}  , \tau =  \tau_0) \, , 
\ee  
and the brackets \eqref{lbn5}  reduce to
\be \label{eq:SL2enh}
\begin{split}
\Leib{v}{w}_{\alpha \beta} &= ( \tau + \hat{\tau} )  \epsilon^{\gamma \delta} v_{\delta (\alpha} w_{\beta) \gamma}  \,, \\ 
\Leib{v}{w}_{\alpha}   & =  \tau  \epsilon^{\gamma \delta}  (v_{\delta \alpha } w_ { \gamma}  +v_{\delta } w_ {\alpha \gamma}  ) + \hat{\tau} \epsilon^{\gamma \delta}  (v_{\delta \alpha } w_ { \gamma} - 3 v_{\delta } w_ {\alpha \gamma}  ) \,, \\
\Leib{v}{w}_{0}  & =   ( \tau - 3 \hat{\tau})  \epsilon^{\gamma \delta} v_{\delta} w_\gamma \,. 
\end{split}
\ee
The Leibniz condition \eqref{leib-cond} now imposes that either $\tau = 3 \hat{\tau}$ or $\hat{\tau} = 0$. These two cases lead to different gaugings.
\begin{itemize}
    \item $\tau = 3 \hat{\tau}$. In this case, the Leibniz bracket \eqref{eq:SL2enh} becomes
    \begin{equation}
    \begin{split}
\Leib{v}{w}_{\alpha \beta} &= 4 \hat{\tau} \epsilon^{\gamma \delta} v_{\delta (\alpha} w_{\beta) \gamma}  \,, \\ 
\Leib{v}{w}_{\alpha}   & = 4 \hat{\tau}  \epsilon^{\gamma \delta} v_{\delta \alpha } w_ { \gamma} \,, \\
\Leib{v}{w}_{0}  & =  0 \,. 
\end{split}
    \end{equation}
    We get an $\SL{2,\bbR}$ gauging, generated by the $v_{\alpha\beta}$ vector fields. The $v_\alpha$ are in the image of the symmetric part of the Leibniz bracket and thus are dualised to tensor fields, charged under the $\SL{2,\bbR}$ gauge group. The graviphoton $v^0$ can gauge $\U{1}_R$.
    
\item $\hat{\tau} = 0$. Now the Leibniz bracket immediately reduces to the Lie bracket
    \begin{equation}
    \begin{split}
\Leib{v}{w}_{\alpha \beta} &= \tau \epsilon^{\gamma \delta} v_{\delta (\alpha} w_{\beta) \gamma}  \,, \\ 
\Leib{v}{w}_{\alpha}   & = \tau  \epsilon^{\gamma \delta}  (v_{\delta \alpha } w_ { \gamma} - w_{\delta\alpha} v_{\gamma }  )  \,, \\
\Leib{v}{w}_{0}  & =   \tau  \epsilon^{\gamma \delta} v_{\delta} w_\gamma \,,
\end{split}
\end{equation}
and the vectors $v_\alpha$ no longer commute with each other. Therefore, the $v_\alpha$'s cannot be dualised to tensor multiplets and instead contribute to a larger non-abelian gauge group. In particular, we find that the algebra enhances to that of $\SL{2,\bbR} \ltimes \Heis$, with $\Heis$ the 3-dimensional Heisenberg group. Here $v_{\alpha\beta}$ generate the semi-simple $\SL{2,\bbR}$ part, $v_\alpha$ transform as doublets of $\SL{2,\bbR}$ and $v_0$ is the central element of $\Heis$. Therefore, $\left\{ v_\alpha,\, v_0 \right\}$ generate the $\Heis$ factor. However, the gauge group under which matter is charged is $\SL{2,\bbR} \ltimes \bbR^2$.
\end{itemize}

We see explicitly that the consistent truncation analysis differs from the purely five-dimensional one. In five dimensions, the embedding tensor belongs to the full bundle 
\begin{equation}
    \mathcal{V}^* \otimes \giso = (\rep{1},\rep{6}) \otimes \left[ (\rep{3},\rep{1}) \oplus (\rep{1},\rep{8}) \right] = (\rep{3},\rep{6}) \oplus (\rep{1},\rep{3}^* \oplus \rep{6} \oplus \rep{15}^* \oplus \rep{24}^*) \,,
\end{equation}
where we are decomposing under  $\SU{2}_R \times \SL{3,\mathbb{R}}$, and therefore contains more representation than those arising in \eqref{dec351n5}. 

As a result, not all five-dimensional gaugings for $n_{\rm VT}=5$ can arise from consistent truncations. For example, in five dimensions, we can have an embedding tensor in $\rep{1}_0 \otimes \rep{3}_0$ of $\SL{2,\mathbb{R}} \times \mathbb{R}^+$, but coming from the $(\rep{1},\rep{15}^*) \oplus (\rep{1},\rep{24}^*)$ of $\SU{2}_R \times \SL{3,\mathbb{R}}$. This embedding tensor would lead to a $\U{1}$ gauging with four tensor multiplets with charges $\pm2$, $\pm4$. However, this gauging cannot arise from a consistent truncation, since the intrinsic torsion \eqref{dec351n5} does not contain the $(\rep{1},\rep{24}^*)$ representation.

\paragraph{$\mathbf{n_{VT} = 8}$:} This truncation arises for the case (\ref{GstU1b})  and corresponds to a $\Gst = \U{1}$ structure group. 
Under the branching $\SUs{6} \supset \SL{3,\mathbb{C}} \times \U{1}$ the vectors decompose as\footnote{Recall that for $\SL{3,\bbC}$ the dual and conjugate representations are not equivalent. Here we denote them by $\mbf{n}^*$ and $\obf{n}$, respectively.}
\begin{equation}
\begin{aligned}
	\mbf{15}^* &= (\mbf{3}\otimes\obf{3})_0 \oplus \mbf{3}^*_2 \oplus \obf{3}^*_{-2} \, , \\
	&\ni ( v^{\alpha  \dot \alpha }, v_\alpha, \bar{v}_{\dot{\alpha}} ) \, , 
\end{aligned}
\end{equation}
where raised $\alpha$ and  $ \dot \alpha$ indices denote  the fundamental representation $\mbf{3}$ and conjugate-fundamental representation $\obf{3}$ of $\SL{3,\bbC}$ respectively. Thus for example, since $\rep{15}^*$ is real, the two components $v^{\alpha}$ and $\bar{v}^{\dot{\alpha}}$ are related by complex conjugation $\left( v^{\alpha} \right)^* = \bar{v}^{\alpha}$ and $(v^{\alpha\dot{\beta}})^* = v^{\beta\dot\alpha}$. We see that the $\U1$-singlet space $\mathcal{V}=(\mbf{3}\otimes\obf{3})_0$ is nine-dimensional giving rise to  $n_{\rm VT} = 8$ vector multiplets. 

It is easy to check that \eqref{scalarmanbis} gives the expected scalar manifold
\begin{equation}
	{\cal M} = {\cal M}_{\rm VT} = \frac{\Com{\Gst}{\Ex{6}}}{\Com{\Gst}{\USp{8}/\bbZ_2}} = \frac{\SL{3,\mathbb{C}}}{\SU{3}} \,,
\end{equation}
with   isometry group 
\be
G = \SL{3, \mathbb{C}} \times  \SU{2}_R   \, .
\ee

The  singlet intrinsic torsion can be written as 
\be
\begin{aligned}
\label{dec351n8b}
 W_{\rm int} & = (\mbf{3}^*\otimes\obf{3}^*,\mbf{3}) \oplus (\mbf{3}^*\otimes\obf{3}^*,\mbf{1})
     \oplus (\mbf{3}^*\otimes\obf{6},\mbf{1}) \oplus (\mbf{6}\otimes\obf{3}^*,\mbf{1}) \\
%  (\mbf{3},  \rep{3} ,  \rep{3}) \oplus  (\rep{3} ,  \rep{3} , \mbf{1}) \oplus    (\rep{3} ,  \rep{\bar 6}, \mbf{1}) \,  \oplus   ( \rep{\bar 6} ,  \rep{3}, \mbf{1})   \\ 
	& \ni \{   \tau_{\alpha  \dot{ \alpha}}{}^a{}_{ b } ,  \tau_{\alpha  \dot{\alpha}}  ,  \tau_\alpha{}^{\dot{\beta} \dot{\gamma}} \, ,  \bar{\tau}^{\alpha \beta}{}_{\dot \gamma}   \} \,, 
\end{aligned}
\ee
where $a,b =1,2,3$ are $\SU{2}_R$ indices. 
Given  $v^{\alpha  \dot {\alpha }}\in\mathcal{V}$ the non-zero components of the map $T$ are
\be
\begin{split}
	T(v)^a{}_b &= v^{\alpha \dot\alpha}   \tau_{\alpha \dot{\alpha}}{}^{a}{}_{b } \,, \\ 
	T(v)^\alpha{}_\beta &= v^{\alpha\dot\beta}  \tau_{\beta\dot\beta}  - \tfrac{1}{3} \delta^\alpha{}_\beta   v^{\delta\dot\delta}  \tau_{\delta\dot\delta}  + \epsilon_{\beta\gamma\rho} v^{\gamma\dot\delta}  \tau^{\rho\alpha}{}_{\dot\delta} \,, \\
	T(v)^{\dot\alpha}{}_{\dot\beta}  &= - v^{\beta\dot\alpha}  \tau_{\beta\dot\beta}  +  \tfrac{1}{3} \delta^{\dot\alpha}{}_{\dot\beta}   v^{\delta\dot\delta}  \tau_{\delta\dot\delta}  -  \epsilon _{\dot\beta\dot\gamma\dot\rho} v^{\delta \dot\gamma} \bar{\tau}_{\delta}{}^{\dot\rho\dot\alpha} \,.
\end{split}
\ee

Let us now consider the possible gaugings. If we focus on maximal simple subgroups of $\SL{3,\bbC}$, there are three possibilities: $\SU3$, $\SU{2,1}$ and $\SL{3,\bbR}$.
It is easy to  show that the real form $\SL{3,\mathbb{R}}$ cannot be gauged. For  the subgroup $\SL{3,\bbR}\subset\SL{3,\bbC}$ the real and conjugate representations are isomorphic, and we can write
\begin{equation}
    \bar{v}^{\dot\alpha} = \delta^{\dot\alpha}_\alpha\, v^\alpha \, . 
\end{equation}
The nine real vectors in $\rep{3}\otimes\obf{3}$ of $\SL{3,\bbC}$ then decompose as 
\begin{equation}
    \rep{3}\otimes\obf{3} \simeq \rep{3}\otimes\rep{3}
        = \rep{6} \oplus \rep{3}^* \, .  
\end{equation}
We see explicitly that this does not include the adjoint and hence $\SL{3,\bbR}$ cannot be gauged. 

Consider now $\SU{3}$ and $\SU{2,1}$. In these two cases 
%For $\SU{p,3-p}$ 
the conjugate and dual representations are isomorphic since we can write
\begin{equation}
    v^*_\alpha = \eta_{\alpha\dot{\alpha}} \bar{v}^{\dot{\alpha}} \,,
\end{equation}
where $\eta_{\alpha\dot{\alpha}}$ is the invariant Hermitian form, with signature $(3,0)$ and $(2,1)$ for $\SU{3}$ and $\SU{2,1}$ respectively. The nine real vectors in $\rep{3}\otimes\obf{3}$ of $\SL{3,\bbC}$ decompose as
\begin{equation}
    \rep{3}\otimes\obf{3} = \rep{8} \oplus \rep{1} 
        \ni \left( \hat{v}^\alpha{}_\beta, v_0 \right) \, , 
\end{equation}
where $\hat{v}^\alpha{}_\alpha=0$ and 
\begin{equation}
    v^{\alpha\dot\alpha} = \hat{v}^\alpha{}_\beta \eta^{\beta\dot\alpha}
        + \tfrac13 \eta^{\alpha\dot\alpha}v_0 \, .
\end{equation}
The eight vectors $\hat{v}^\alpha{}_\beta$ form the adjoint of 
$\SU{3}$ or $\SU{2,1}$.
%$\SU{p,3-p}$. 
Decomposing the intrinsic torsion \eqref{dec351n8b} under $\SU{3} \times\SU2_R$ ($\SU{2,1} \times\SU2_R$) we get 
\begin{equation}
    W_{\text{int}} = (\mbf{1},\mbf{3} )\oplus (\mbf{8},\mbf{3})  \oplus (\mbf{1},\mbf{1})\oplus 3 \cdot (\mbf{8},\mbf{1})\oplus(\mbf{10},\mbf{1})\oplus(\obf{10},\mbf{1}) \, ,
\end{equation} 
where the four  singlet components are 

\begin{equation}
  \tau_{\alpha\dot{\alpha}}{}^a{}_b = \eta_{\alpha\dot{\alpha}} \tau^a{}_b \,, \qquad
  \tau_{\alpha\dot{\alpha}} = \tau\, \eta_{\alpha\dot{\alpha}} \,. 
\end{equation}
Given two vectors $v=\left( \hat{v}^\alpha{}_\beta, v_0 \right)$ and $w=\left( \hat{w}^\alpha{}_\beta, w_0 \right)$, the Leibniz bracket then reads 
\begin{equation}
 \begin{aligned}
&  \Leib{v}{w}^{\alpha}{}_{\beta} = - \tau \left( \hat{v}^{\alpha}{}_{\gamma} \hat{w}^{\gamma}{}_{\beta} - \hat{w}^{\alpha}{}_{\gamma} \hat{v}^{\gamma}{}_{\beta} \right) \,, \\
&  \Leib{v}{w}_0 = 0 \, . 
 \end{aligned}
\end{equation}
Thus, we see that we can gauge  either $\SU{3}$ or $\SU{1,2}$. The extra vector singlet $v_0$ can gauge the $\U{1}_R$.

\paragraph{$\mathbf{n_{\rm VT} = 14}$:}  This is the maximal case, where the invariant vectors span the whole $\mathcal{V}= \rep{15}^*$ of $\SUs{6}$. It does not correspond to any of the 
generalised structures listed at the beginning of this section and therefore must correspond to a discrete structure group. Indeed, since all the $K_{\tilde{I}}$ are stabilised and from \eqref{scalarmanbis} we have 
\be
\mathcal{M} = \frac{ \Com{G_S}{\Ex{6}} }{ \Com{G_S}{\USp{6}} }   = \frac{ \SUs{6}}{\USp{6}}  \, , 
\ee 
it is easy to identify the generalised structure as
\begin{equation}
	G_S = \mathbb{Z}_2 \subset \Ex{6} \,.
\end{equation}
The $ \mathbb{Z}_2$ acts diagonally as $- \mathbb{1}$ in $\USp{6}$, leading to the global isometry group 
\be
\Giso = C_{\Ex{6}}(\mathbb{Z}_2) = \SU{2}_R \cdot \SUs{6} \, .
\ee
%and $C_{\USp{8}}(\mathbb{Z}_2) = \USp{6} \times \SU{2}_R$,
Decomposing under $\Giso$ we can hence write vectors in $\mathcal{V}$ as 
\begin{equation}
    \mathcal{V} = (\mbf{1},\mbf{15}^* ) \ni v_{ij} \, ,
\end{equation}
where $v_{ij}=v_{[ij]}$ and $i,j =1, \ldots, 6$. 

The singlet intrinsic torsion arranges into representations of $\Giso$ as
\be
\begin{aligned}
	\label{dec351n14}
	W_{\rm int}   & = (\mbf{3},\mbf{15} )  \oplus (\mbf{1},\mbf{21}) \oplus (\mbf{1},\mbf{105}) \\ 
	   &\ni (\tau^{ij a}{}_{b }  ,  \tau^{ij}    ,  \tau^{ijk}{}_l ) \,.
\end{aligned}
\ee
where $\tau^{ij a}{}_{b }=\tau^{[ij] a}{}_{b }$, $\tau^{ij}=\tau^{(ij)}$ and $\tau^{ijk}{}_l=\tau^{[ijk]}{}_l$ with $\tau^{ijl}{}_l=0$. The map 
$T  : \, \mathcal{V} \to \mathfrak{g}_{\rm iso} $  is

\be
\begin{aligned}
	T(v)^a{}_b & =  \tfrac{1}{2} v_{ij}   \tau^{ij a}{}_{b } \,, & 
	T(v)^i{}_j & =    v_{kl}    \tau^{ikl}{}_j  + v_{ik}    \tau^{kj}   \, , 
\end{aligned}
\ee
with  bracket 
\be
\begin{aligned}
\label{nv15lb}
\Leib{v}{w}_{[ij]}  &  =  - T(v)^k{}_{[i} w_{ j]k}  =  t^{[kl] [mn]}{}_{[ij]} v_{[kl]}  w_{[mn]} \\ 
& = -   \tau^{klm}{}_{[i}  w_{j]m} v_{kl} + \tau^{(kl)} v_{k [i } w_{ j] l}  \, . 
\end{aligned}
\ee

As there are 15 vectors, the largest semi-simple groups we can gauge  are different real forms of $\SU{4}\simeq\Spin6$. However $\SU{4}$ and 
$\SU{2,2}\simeq\Spin{4,2}$ do not embed in $\SUs{6}$, and we are left with  $\SUs{4}\simeq\Spin{5,1}$ and  $\SU{3,1}\simeq\Spins{6}$.  These are embedded as
\be \label{eq:TwoSUS4}
\SUs{6}  \supset  \SUs{4}  \times \SU{2} \times \mathbb{R}^+ \,, \qquad \text{and} \qquad   \SUs{6}  \supset    \SU{3,1}/\bbZ_2\simeq\SOs6  \, . 
\ee
From the decomposition of the vectors 
\be
\begin{aligned}
	& \rep{15}^* = (\rep{6}, \rep{1})_{-2}  \oplus (\rep{ 4}^*,\rep{2})_{1}  \oplus (\rep{1}, \rep{1})_{4} \,, \\
	&   \rep{15}^* = \rep{15} \,,
\end{aligned}
\ee
under $\SUs{4} \times \SU{2} \times \mathbb{R}^+$ and $\SOs6$ we see that only $\SOs6$ can be gauged. 

The intrinsic torsion contains an $\SOs6$ singlet from the decomposition of the $\mbf{21}$ of $\SUs{6}$ 
\be 
W_{\rm int} \ni (0, \tau\, \delta^{ij}, 0) \,,
\ee
where  $\delta^{ij}$ is the invariant metric of $\SOs{6}$. Then the bracket \eqref{nv15lb} becomes 
\be
\begin{aligned}
\Leib{v}{w}^i{}_j  
    &= \tfrac12 \tau \bigl( v^i{}_k w^k_{}j - v^i{}_k w^k_{}j \bigr) \, , 
\end{aligned}
\ee
where we have raised indices using the $\SOs6$ metric. We easily recognise  the $\SOs{6}$ Lie algebra. Note that the vectors $v_{ij}$ satisfy a reality condition of the form 
\begin{equation}
 (v^*)_{\bar{i}\bar{j}} = J^k{}_{\bar{i}} \, J^l{}_{\bar{j}}\, v_{kl} \,,
\end{equation}
where $J^i{}_{\bar{j}}$ is the complex structure of $\SUs{6}$. If we take $\delta^{ij}$ to have the standard form, the individual components of $v_{ij}$ are not real. This is why the gauging is $\SOs{6}$ not $\SO{6}$.

For compact gaugings, the largest possible subgroups of $\SUs6$ are $\USp{4} \times \SU{2}$ and $\USp{4}$. However their adjoints are not contained in the $\rep{15}$ representation. The next largest possible gauge group is $\SU{3} \times \U{1}$, which we will now investigate. 

To study the $\SU{3} \times \U{1}$ gauge group, it is useful to consider it as a subgroup of $\SOs{6}$ which preserves the $\U{3}$ Hermitian form. This way, we can also consider the gauge group $\SU{2,1} \times \U{1} \subset \SOs{6} \subset \SUs{6}$ by instead choosing a split-signature Hermitian form. 

Thus, both $\SU{3} \times \U{1}$ and $\SU{2,1} \times \U{1}$ are embedded in $\SOs{6}$ via
\be
\begin{aligned}
\rep{15}^* %=  \rep{14} \oplus \rep{1} 
&= \rep{8}_{0}  \oplus   \rep{3}_{1} \oplus  \obf{ 3}_{-1} \oplus \rep{1}_0 \,, \\
v_{ij} &=  (v^{\alpha}{}_{\beta}, v_{\alpha\beta}=\epsilon_{\alpha\beta\gamma}v^\gamma,\bar{v}^{\alpha\beta}=\epsilon^{\alpha\beta\gamma}\bar{v}_\gamma,v^0) \,.
\end{aligned}
\ee
Here $\alpha=1,2,3$ denotes the fundamental representation of $\SU{3}$ or $\SU{2,1}$, respectively, and the vectors satisfy $v^\alpha{}_\alpha = 0$ as well as the reality conditions
\begin{equation}
 (v^\alpha{}_\beta)^* = v^\beta{}_\alpha \,, \qquad 
 (\bar{v}_{\alpha})^* = v^{\alpha} \,, \qquad 
 (v^0)^* = v^0 \,.
\end{equation}
Thus, we can expect that the vectors  $(v^\alpha{}_\beta,v^0)\in\mbf{8}\oplus\mbf{1}$  gauge $ \SU{3}  \times \U{1}$ or $\SU{2,1}\times\U{1}$, respectively, and the other six are dualised into tensors.

To see whether this can arise, we investigate the intrinsic torsion \eqref{dec351n14}. We find that the intrinsic torsion contains singlets under $\SU{3} \times \U{1}$, $\SU{2,1}\times\U{1}$, respectively, given by
\be
\tau^{\alpha\beta}{}_{\gamma,\rho} 
= \tau_{\gamma\rho}{}^{\alpha,\beta} 
= \tilde{\tau}_1 \delta^{\alpha\beta}_{\gamma\rho} \,,\, 
\quad 
\tau^\alpha{}_\beta = \tilde{\tau}_2 \delta^\alpha_\beta \,,
\quad \text{and} \quad
\tau^{\alpha}{}_\beta{}^a{}_b = \delta^\alpha_\beta \tilde{\tau}^{0a}{}_b. 
\ee
Thus, the map $T$ becomes
\begin{equation}
 \begin{split}
  T(v)^{\alpha}{}_{\beta} &= (\tilde{\tau}_2 - \tilde{\tau}_1) v^\alpha{}_\beta + \tfrac13 (\tilde{\tau}_2 + 2 \tilde{\tau}_1 ) \delta^\alpha{}_\beta v^0 \,, \\
  T(v)^{\alpha\beta} &= (\tilde{\tau}_1 + \tilde{\tau}_2) \epsilon^{\alpha\beta\gamma}\bar{v}_\gamma  \,, \\
  T(v)^a{}_b &= v^0 \tilde{\tau}^{0a}{}_b \,, 
 \end{split}
\end{equation}
with the others following from the above by complex conjugation. This leads to the Leibniz bracket
\begin{equation}
\begin{split}
  \Leib{v}{w}^\alpha{}_\beta &= \tfrac12 ( \tilde{\tau}_2 -  \tilde{\tau}_1 ) ( v^\gamma{}_\beta w^\alpha{}_\gamma - w^\gamma{}_\beta v^\alpha{}_\gamma ) + \tfrac12 (\tilde{\tau}_1 + \tilde{\tau}_2) (v^\alpha \bar{w}_{\beta} - w^\alpha \bar{v}_\beta) \\
  & \quad - \tfrac16 \delta^\alpha{}_\beta (\tilde{\tau}_1 + \tilde{\tau}_2) (v^{\gamma} \bar{w}_{\gamma} - w^{\gamma} \bar{v}_{\gamma} ) \,, \\
  \Leib{v}{w}^\alpha &= \tfrac12\tilde{\tau}_1 (v^\alpha{}_\beta w^{\beta} + w^\alpha{}_{\beta} v^{\beta}) - \tfrac12\tilde{\tau}_2 (v^\alpha{}_{\beta} w^{\beta} - w^\alpha{}_{\beta} v^{\beta}) \\
  & \quad + \tfrac13 \tilde{\tau}_2 (v_0 w^\alpha - w_0 v^{\alpha}) + \tfrac13 \tilde{\tau}_1 (2v_0 w^{\alpha} - w_0 v^{\alpha} ) \,, \\
  \Leib{v}{w}^0 &= (\tilde{\tau}_1 + \tilde{\tau}_2) (v^{\alpha} \bar{w}_{\alpha} - w^{\alpha} \bar{v}_{\alpha} ) \,,
 \end{split}
\end{equation}
with $\Leib{v}{w}_{\alpha} = (\Leib{v}{w}^{\alpha})^*$.

The Leibniz condition now reduces to
\begin{equation}
\tilde{\tau}_1(\tilde{\tau}_1 + \tilde{\tau}_2) = 0 \,.
\end{equation}
The two solutions $\tilde{\tau}_1 = 0$ and $\tilde{\tau}_1 = - \tilde{\tau}_2$ lead to two different gaugings. When $\tilde{\tau}_1 = 0$ we recover the previous case where $\SU{3,1}$ is gauged. For $\tilde{\tau}_1 = - \tilde{\tau}_2$ the Leibniz bracket becomes
 \begin{equation}
 \begin{split}
  \Leib{v}{w}^\alpha{}_\beta &= \tilde{\tau}_2 ( v^\gamma{}_\beta w^\alpha{}_\gamma - w^\gamma{}_\beta v^\alpha{}_\gamma )\,, \\
  \Leib{v}{w}^\alpha &= - \tilde{\tau}_2 v^\alpha{}_\beta w^{\beta} - \tfrac13 \tilde{\tau}_2 v_0 w^\alpha \,, \\
  \Leib{v}{w}^0 &= 0 \,.
\end{split}
\end{equation}
We recognise the Lie algebra of $\SU{3} \times \U{1}$ and $\SU{2,1} \times \U{1}$, respectively, generated by the $v^\alpha{}_\beta$ and $v^0$. The $v^{\alpha}$ and $\bar{v}_{\alpha}$ are in the image of the symmetrised Leibniz bracket and therefore correspond to tensor multiplets which transform in the $\rep{3}_1 \oplus \rep{\bar{3}}_{-1}$ of $\SU{3} \times \U{1}$ or $\SU{2,1} \times \U{1}$. The $\U{1}$ generator $v^0$ can also gauge the $\U{1}_R$ via $\tau^{0a}{}_b$.
 
In Table \ref{summ-gaugingsspe} we summarise the maximal reductive and compact gauge groups
for the special cases of purely vector/tensor multiplet truncations of this section. As in the previous table, whenever we list a product of groups, the individual factors can also be gauged separately even though they are not listed as such. Whenever there are abelian factors in $\Gred$, the $\U{1}_R$ can also be gauged diagonally with some combination of these factors.

\begin{table}[h!]
%\begin{equation} \label{eq:sumtable}
\centering
\begin{tabular}{|c|c|c|c|}
\hline
$n_{\rm VT}$ & $G_{\rm iso}$  & $\Gred$ & $n_{\rm T}$ \\
\hline
\multirow{2}{*}{5} & \multirow{2}{*}{$\SU{2}_R  \times \SL{3,\mathbb{R}}$ }  & $\SL{2,\bbR} \ltimes \bbR^2$   & -- \\
&  &  $\SL{2,\mathbb{R}} \times \U{1}_R $ & 2 \\
\hline
\Tstrut\Bstrut
\multirow{1}{*}{8} & \multirow{1}{*}{$\SU{2}_R  \times \SL{3,\mathbb{C}}$ }  & $\SU{3} \times \U{1}_R $, \; $\SU{2,1} \times \U{1}_R $ & --  \\
%& &  $\SU{2,1} \times \U{1}_R $ & -- \\
\hline
\multirow{2}{*}{14} & \multirow{2}{*}{$\SU{2}_R  \times \SUs{6} $}  & $\SU{3,1}$ & -- \\
 & &  $\SU{3} \times %\U{1}\times  
 \U{1}_R$,\; $\SU{3} \times \U1$& \multirow{1}{*}{6} \\
 %& &  $\SU{3} \times \U{1}$ &  \\
\hline
\end{tabular}
\caption{Maximal reductive and compact gauge groups in the special cases of purely vector/tensor multiplet truncations. The first column gives the total number of vectors and tensor multiplets, the second the global isometry group, the third the allowed gaugings and the last one the number of vectors that are dualised to tensors in each case.} 
\label{summ-gaugingsspe}
\end{table}

\subsection{Truncations with only hypermultiplets}
\label{sec:hyperonly}

Let us now analyse which consistent truncations are possible with only hypermultiplets and no vector multiplets.

Truncations of this kind are associated  to a generalised structures $G_S$  that is defined by a single generalised vector $K$ in the $\mbf{27}^*$ of $\Ex{6}$, 
defining a ${\rm V}$-structure, and a set of adjoint tensors $J_A$, $A = 1, \ldots, {\rm dim}(\Gh)$, satisfying
\begin{equation}
	J_A \cdot K = 0 \,.
\end{equation}
Since the stabiliser of the V-structure  is   $\Fff \subset \Ex{6}$, 

we must  have $G_S \subset \Fff$.
Finally,  by construction, the scalar manifold must be symmetric (see Section \ref{sec:genN2})
\be
\label{symhyper}
\mathcal{M}_{\rm H} = \frac{\Gh}{\SU{2}_R \cdot \Com{G_S}{\USp{6}}} \,,
\ee
where  $\Gh =  \Com{G_S}{\Ex{6}}$ is the group generated by the singlets $J_A$.

The above considerations already restrict the possible scalar manifolds for the hypermultiplets to the following list~\cite{Alek68,AC05}
\begin{equation}
\begin{aligned}
	\label{eq:list-hyperscal}
	\mathcal{M}_{\rm H} &=  \frac{\mathrm{F}_{4(4)}}{\SU{2}\cdot \USp{6}} \,, & \qquad n_{\rm H}  &=   7   \, , \\
	{\cal M}_{\rm H} &= \frac{\SOo{4,p}}{\SO{4} \times \SO{p}} \,, & \qquad n_{\rm H} &= p\,, \quad p \leq 5 \,, \\
	\mathcal{M}_{\rm H} &=  \frac{\mathrm{G}_{2(2)}}{\SO{4}} \,, & \qquad n_{\rm H}  &=   2   \,, \\
	\mathcal{M}_{\rm H} &=  \frac{\SU{2,1}}{\mathrm{S}(\U2 \times \U1)} \,, & \qquad n_{\rm H}  &=   1   \,,
\end{aligned}
\end{equation}
where $\SOo{4,p}$ denotes the connected component of the $\SO{4,p}$. 

However the first two  manifolds  do not arise from  truly $\mathcal{N}=2$ truncations. This is because they correspond to generalised structure groups that lead to extra singlets in the decomposition of the  $\mbf{6}$ of $\USp{6}$.
For ${\cal M}_{\rm H} = \frac{{\rm F}_{4(4)}}{\SU{2} \cdot \USp{6}}$,  the structure group is trivial, $\Gst = \mathbb{1}$, since it is given by the commutant in $\Fff$  of  the isometry group. Thus this truncation  always comes from a sub-truncation of five-dimensional maximal supergravity.
Similarly, for the ${\cal M}_{\rm H} = \frac{\SOo{4,p}}{\SO{4} \times \SO{p}}$, with $p \leq 5$, the structure group has to be
\begin{equation}
	\Gst = \Spin{5-p} \,,
\end{equation}
with $\Spin{0} = \Spin{1} = \mathbb{Z}_2$. The decomposition of the $\mathbf{6}$ of $\USp{6}$ under $G_S$ always contains two extra singlets, so that these cases are sub-truncations of half-maximal gauged supergravity. Indeed,  from  the commutant of $G_S$ in the full $\Ex{6}$ and $\USp{8}$ groups,
\begin{equation}
	\Com{\Spin{5-p}}{\Ex{6}} = \Spin{5,p} \times \mathbb{R}^+ \,, \qquad \Com{\Spin{5-p}}{\USp{8}} = \USp{4} \times \Spin{p} \,,
\end{equation}
one can easily check that  $G_S = \Spin{5-p}$ actually allows for a half-maximal truncation with $p$ vector multiplets and scalar manifold
\begin{equation}
	{\cal M} = \frac{\Spin{5,p}}{\USp{4} \times \Spin{p}} \times \mathbb{R}^+ \,.
\end{equation}

This leaves only the  two last manifolds in \eqref{eq:list-hyperscal} as truly ${\cal N}=2$ truncations.  %
\begin{itemize}

\item  The case with $\mathbf{n_{\rm H}=2}$ hypermultiplets  corresponds to a  $\Gst = \SO3$ that is obtained from (\ref{GstSU2}).
The structure group embeds as
\begin{equation}
\label{branchhyper2}
	\begin{split}
	\Fff & \supset \SU{2} \times  \Gtt \,, \\
	\USp{6} & \supset  \SU{2} \times \SU{2}\,.% \SU{2}_S \, .
	\end{split}
\end{equation}
Decomposing the $\mbf{78}$  of $\Ex{6}$ in representations of   $G_S= \SU{2}$ gives  6 compact and 8 non-compact singlets.
Altogether they correspond to the generators of $\Gtt$, while the compact ones give its $\SO{4}$ maximal compact subgroup. Then
 \eqref{symhyper} gives the expected scalar manifold 
\begin{equation}
	{\cal M} = \mathcal{M}_{\rm H} = \frac{\mathrm{G}_{2(2)}}{\SO{4}} \,.
\end{equation}
It is also easy to check that there are  no vector/tensor multiplets in the truncation, since there are no singlets in the $\rep{26}$ of $\Fff$ under the branching   \eqref{branchhyper2}.

\item The case with  $\mathbf{n_{\rm H}=1}$ %$\mathbf{n_{\rm H}=1}$ 
tensor multiplet corresponds to the generalised structure $\Gst = \SU{3}$ (\ref{GstSU3U1}). This is embedded as
\begin{equation}
\label{branchhyper1}
	\begin{split}
	\Fff & \supset \SU{3} \times \SU{2,1}  \,, \\
	\USp{6} & \supset  \SU{3} \times  \U{1} \, . 
	\end{split}
\end{equation}
In the decomposition of the $\mbf{78}$ of   $\SU{3}$ one finds 4 compact and 4 non-compact singlets, which generate $\SU{2,1}$. The compact ones give the compact subgroup 
$\SU{2} \times \U{1}$ so that we recover the hyperscalar manifold 
\begin{equation}
	{\cal M} = \mathcal{M}_{\rm H} = \frac{\SU{2,1}}{\mathrm{S}(\U2 \times \U1)} \,.
\end{equation}

As, again, there are no singlets in the $\mathbf{26}$ of $\Fff$ under the branching to $\Gst = \SU{3}$,  there are no vector multiplets.
\end{itemize}

The study of the intrinsic torsions and the gauging for the truncations  with only hypermultiplets is very simple.  As the only vector in the theory  is the graviphoton in  the universal multiplet, only  abelian gaugings are possible. Moreover, in all cases,  the intrinsic torsion only contains the adjoint representation of the isometry group 
\be
W_{\rm int} = \adj \Gh \ni \tau_0^A{}_B  \,,
\ee
with $A,B = 1, \ldots  \dim\Gh$ so that the map $T : \mathcal{V} \to \mathfrak{g}_{\rm iso}$ is
\be
T(v^0) = v^0 \tau^A_{0  \, B} \,, \qquad  \quad A= 1, \ldots , \dim \Gh \, . 
\ee
The generalised Lie derivative on the adjoint singlets is
\begin{equation}
	L_{K_0}J_A= [ J_{K_0},J_A] = - T(K_0) \cdot  J_A  = p_{0A}{}^B J_B \,,
\end{equation}
with the component of the embedding tensor
\be
p_{0 A}{}^B = \tau^A_{0  \, B}  \, ,
\ee
and the graviphoton can gauge any one-dimensional subgroup of $\Gh$.

\subsection{Truncations with vector/tensor and hypermultiplets}
\label{sec:mixed}

The last class of truncations that can arise consists of truncations with both vector/tensor and hypermultiplets.  One way to study this class is to start from the truncations with only hypermultiplets discussed in the previous section and  look for a subgroup of the structure group $G_S$ that allows for extra singlet vectors but no extra singlets in the branching of the $\mathbf{6}$ under $\USp{6} \supset \Gst$. This last condition is necessary to have a truly $\mathcal{N}=2$ truncation and leaves only two possible cases: $n_{\rm H} =2$ with  $\Gst = \SU{2}$ (\ref{GstSU2}) or  $n_{\rm H}=1$ and $\Gst = \SU{3}$ (\ref{GstSU3U1}).

The case with $n_{\rm H} =2$ hypermultiplets and hyperscalar manifold 
\begin{equation}
\label{hypermanSU2}
	{\cal M}_{\rm H} = \frac{\Gtt}{\SO{4}} \,,
\end{equation}
is immediately ruled out since any further reduction of the  $\Gst = \SU{2}$   structure group necessarily gives rise to a singlet in the $\mathbf{6}$ of $\USp{6}$. This can be easily see from \eqref{Su2str} by breaking the second $\SU{2}$ factor. Therefore consistent truncations with  hypermultiplets forming the scalar manifold \eqref{hypermanSU2} 
and vector/tensor multiplets necessarily arise from subtruncations of ${\cal N} > 2$ gauged supergravity.

We are left with the case with $n_{\rm H}=1$ hypermultiplet and hyperscalar manifold
\begin{equation}
	{\cal M}_{\rm H} = \frac{\SU{2,1}}{\mathrm{S}(\U2 \times \U1)} \, . 
\end{equation}
The structure group is $\SU{3}$ and we can consider  two non-trivial subgroups  $\Gst = \SU{2} \times \U{1} $ (\ref{GstSU2U1}) and $\Gst = \U{1}$ (\ref{GstU1a}). 
As we will discuss below, they allow for $n_{\rm VT}=1$ and  $n_{\rm VT}=4$ vector multiplets, respectively. Cases with $n_{\rm VT}=2,3$ can only be obtained as sub-truncations of the $n_{\rm VT}=4$ case and therefore we will not discuss them here. 

Recall that the scalar manifold of the vector/tensor multiplets in the truncation can now be computed from the commutant of $\Gst$ within the stabiliser groups $G_{\mathcal{U}}$ and $H_{\mathcal{U}}$, in $\Ex6$ and $\USp8/\bbZ_2$ respectively, of the space $\mathcal{U}$ of $J_A$ that define the hypermultiplet moduli. One finds 
\begin{equation}
	G_{\mathcal{U}} = \SL{3,\mathbb{C}} \subset \Ex{6} \,,
\end{equation}
with compact subgroup
\begin{equation}
	H_{\mathcal{U}} = \SU{3} \subset \USp{8}/\mathbb{Z}_2 \,.
\end{equation}
The scalar manifold of the vector/tensor multiplets is then
\begin{equation}
\label{vecmEm}
	{\cal M}_{\rm VT} = \frac{\Com{\Gst}{G_{\mathcal{U}}}}{\Com{\Gst}{H_{\mathcal{U}}}} = \frac{\Com{\Gst}{\SL{3,\mathbb{C}}}}{\Com{\Gst}{\SU{3}}} \,.
\end{equation}
We thus find the two following possible truncations.

\paragraph{$\mathbf{n_{\rm VT} = 1, n_{\rm H} = 1}$:}  Consider first the structure group $\Gst = \SU{2} \times \U{1}$. 

The $\mbf{27}^*$ of $\Ex{6}$ contains two $G_S$ singlets so that $\mathcal{V}$ is two-dimensional and $n_{\rm VT}=1$. Thus, the scalar manifold is 
\begin{equation}
{\cal M}_{\rm VT} = \mathbb{R}^+ \,, \qquad 
		{\cal M}_{\rm H} = \frac{\SU{2,1}}{\mathrm{S}(\U2 \times \U1)} \,.
\end{equation}
The decomposition of the adjoint of $\Ex{6}$ gives four compact and five non-compact $G_S$ singlets that are the generators of the isometry group
\be
G_{\rm iso} =  \mathbb{R}^+ \times \SU{2,1} \,.
\ee
Under this group the vectors decompose as 
\begin{equation}
    \mathcal{V} = \mbf{1}_{2} \oplus \mbf{1}_{-1} \ni (v^0, v^{1}) \, . 
\end{equation}
where the subscripts denote the $\mathbb{R}^+$ charges. 

To determine the possible gaugings we find that the intrinsic torsion has components 
\be
W_{\rm int} = \mbf{8}_{-2} \oplus \mbf{8}_1 \oplus \mbf{1}_{1}
    \ni (\tau^A_{0  \, B} , \tau^A_{1 \, B} \,, \tau_{1} ) \, , %\tau_0
\ee
which  give the adjoint action
\begin{equation}
\label{torsionvh11}
\begin{split}
	T(v)^A{}_B &=    v^0  \tau^A_{0  \, B } + v^{1}  \tau^A_{1  \, B }  \,,	
\end{split}
\end{equation}
and $T(v)_{(0)} =    v^1  \tau_{1}=0$ by the Leibniz condition. Furthermore \eqref{eq:tp-alg} implies $\tau^A_{0  \, B }$ and $\tau^A_{1  \, B }$ commute. Thus, the two vectors can gauge a one- or two-dimensional abelian subgroup of $\SU{2,1}$, while the $\mathbb{R}^+$ symmetry cannot be gauged. 
\paragraph{$\mathbf{n_{\rm VT} = 4, n_{\rm H} = 1}$:}  Keeping only $\Gst = \U{1} \subset \SU{2}  \times \U{1}$ as structure group the $\mbf{27}^*$ contains five $G_S$ singlets so that $\mathcal{V}$ is five-dimensional and $n_{\rm VT}=4$. The commutators of $\Gst = \U{1}$  in $\SL{3,\mathbb{C}}$ and $\SU{3}$ 
\begin{equation}
	\begin{split}
		\Com{\U{1}}{\SL{3,\mathbb{C}}} &= \SL{2,\mathbb{C}} \times \U{1} \times \mathbb{R}^+ \,, \\
		\Com{\U{1}}{\SU{3}} &= \SU{2} \times \U{1} \,.
	\end{split}
\end{equation}
and hence, from  \eqref{vecmEm}, the scalar manifold is 
\begin{equation}
{\cal M}_{\rm VT} = \frac{\SO{3,1}}{\SO{3}} \times \mathbb{R}^+ \,,\qquad \qquad 
		{\cal M}_{\rm H} = \frac{\SU{2,1}}{\mathrm{S}(\U2 \times \U1)} \,.
\end{equation} 
The adjoint  of $\Ex{6}$  contains seven compact and seven non-compact $G_S$ singlet elements  corresponding to the isometry group
\be
G_{\rm iso} =  \SO{3,1}  \times \mathbb{R}^+ \times \SU{2,1} \, . 
\ee

The intrinsic torsion components arrange themselves in representations of the isometry group  $G_{\rm iso}$

\be
W_{\rm int} = (\mbf{4, 8})_{-1}  \oplus  (\mbf{1,8})_{2}  \oplus (\mbf{6,1})_{-2}  \oplus 2 \cdot (\mbf{4,1})_{1}
\ni  %=
( \tau^A_{i  \, B }  ,   \tau^A_{0  \, B },  \tau^i_{0  \, j },  \tau_i,  \tau^{[ijk]} )\,,
\ee
where  $A,B=1,2,3$ and $i=1,\dots,4$  are $\SU{2,1}$ and $\SO{3,1}$ indices and the subscript denotes the  $\mathbb{R}^+$ charges. The $T$ map is defined as:
\begin{equation}
\begin{split}
	T(v)^A{}_B &=    v^0  \tau^A_{0  \, B } + v^i  \tau^A_{i  \, B }  \,, \\	%
	T(v)^i{}_j  &=    v^0   \tau^i_{0  \, j }  + v^k    \tau_{k}{}^{i}{}_{j} \,, 
\end{split}
\end{equation}
where again $\tau_i=0$ because of the Leibniz condition \eqref{eq:GenLeib}. 

The analysis of the gauging of the vector/tensor multiplet isometries is the same as for the  $n_{\rm VT}=4$ generic case without hypermultiplets, so that the  possible gauge groups are $\SO{2,1}$, $\SO{3}$, $\ISO{2}$,  when there are no tensor multiplets, and  $\SO{2}$ or $\SO{1,1}$ with tensor multiplets.  

The gauging of $\SU{2}_R$ or $\U{1}_R$ subgroups of $\SU{2,1}$
global symmetry group of $\mathcal{M}_{\rm H}$
are also given by the analysis of the case with only $n_{\rm VT}=4$ vector multiplets. 

To see whether other subgroups of the $\SU{2,1}$ 
are possible one has to analyse condition \eqref{eq:tp-alg}, which now implies 
\begin{equation}\label{eq:SU21Leib}
\tau_{0\,i}{}^j \tau_j^A{}_B = 0 \,, \qquad \tau_0^A{}_B \tau_i^B{}_C - \tau_i^A{}_B \tau_0^B{}_C = 0 \,, \qquad \tau_i^A{}_C \tau_j^C{}_B - \tau_j^A{}_C \tau_i^C{}_B = \tau_{ij}{}^k \tau_k^A{}_B \,.
\end{equation}
We again consider two cases, to solve the constraints \eqref{eq:GenLeib}:
\begin{itemize}
\item $\tau_{ijk} = 0$ and $\tau_{0\,i}{}^j \neq 0$. In this case we only have abelian gaugings  and the rank of $\tau_{0\,i}{}^j$ determines the number of tensor multiplets. If it has rank 4, then  $\tau_i{}^A{}_B=0$ and the only possibility is that $v^0$ gauges a one-dimensional subgroup of $\SU{2,1}$ via $\tau_0^A{}_B$. If
$\tau_{0\,i}{}^j$ has rank 0 or 2, any two linearly independent combinations of $v^0$ and $v^i$ can gauge a 1- or 2-dimensional abelian subgroup of $\SU{2,1}$, with the embedding determined by $\tau_0^A{}_B$ and $\tau_i^A{}_B$.

\item $\tau_{ijk} \neq 0$, $\tau_{0\,i}{}^j = 0$. In this case the first equation of \eqref{eq:SU21Leib} is trivially verified. The gauge groups are the same as for the generic case. Indeed in the generic case the gauge groups are given by how the tensor $\tau_{ijk}$ decomposes. Here this tensor gives in the third equation of \eqref{eq:SU21Leib} directly the structure constant of the gauging inside $\SU{2,1}$. So we obtain the same possible gaugings as for the generic case but with two different embedding. We could  either gauge a subgroup of $\SO{3,1}$ or a diagonal subgroup of $\SO{3,1}$ subgroup and $\SU{2,1}$ subgroup.
\end{itemize}

In Table \ref{summ-gaugingsmixed} we give the list of possible gauging for truncations with vector/tensor and hypermultiplets.
For simplicity we give a list of product groups, but the individual factors can also be gauged separately. $\Ggauge$, the $U(1)_R$ can also be gauged diagonally with some combination of these factors.

\begin{table}[h!]
%\begin{equation} \label{eq:sumtable}
\centering
\makebox[\textwidth][c]{\begin{tabular}{|c|c|c|c|c|}
\hline
$n_{\rm VT}$ & $n_{\rm H}$  & $G_{\rm iso}$  & $\Gred$ & $n_{\rm T}$ \\
\hline
\Tstrut\Bstrut
1 & 1  & $\SU{2,1}  \times \SO{1,1}\times\mathbb{R}^+ $  & $\U{1}_R \times \mathbb{R}^+$  &  -- \\
\hline
\multirow{4}{*}{4} & \multirow{4}{*}{1}   & \multirow{4}{*}{$\SU{2,1}  \times \SO{3,1}\times\mathbb{R}^+ $ }  & $\SO{2,1} \times \mathbb{R}^+ \times\U{1}_R$, \; 
$\SO{3} \times \mathbb{R}^+ \times\U{1}_R$, & \multirow{2}{*}{--}
\\
& & & $\ISO{2} \times \mathbb{R}^+ \times\U{1}_R$,\; $\SU{2}_R \times \mathbb{R}^+  $  & 
\\[0.1cm]

 & & &  $\SO{2} \times \U{1}_R \times \bbR^+$, \; $\SO{1,1} \times \U{1}_R\times \bbR^+$  &  2 \\[0.1cm]
  & & &  $\SO{1,1}$  & 4  \\
\hline
\end{tabular}}
\caption{Summary of the gauge groups in the mixed cases.  The first column gives the total number of vectors and tensor multiplets, the second the global isometry group, the third the allowed gaugings and the last one the number of vectors that are dualised to tensors in each case.} 
\label{summ-gaugingsmixed}
\end{table}
%%%%%%%%%%%%%%%%%%%%%%%%%%%%%%%%%%%%%%%%%%%%%%%%%%%%%%%%%%%%%%%%%%%%%%%%%%%%%%%%%%%%%
\section{Conclusions}\label{sec:Conclusions}

In this paper, we used exceptional generalised geometry to classify which  five-dimensional ${\cal N}=2$ gauged supergravities can arise as consistent truncations of 10-/11-dimensional supergravity. 
From the higher-dimensional point of view any  truncation is associated to a generalised $G_S \subset \Ex{6}$ structure on the compactification manifold $M$, with constant intrinsic torsion. The field content of the truncated theory is determined by the nowhere vanishing generalised tensors on $M$ that define the $G_S$ structure, while the embedding tensor is given by the constant singlet intrinsic torsion. 

Requiring that  the $G_S$ structure has constant,  singlet intrinsic torsion 
imposes  differential conditions on the structure on $M$ that we do not analyse in this paper. 
Instead we assume that such conditions are satisfied, and  we show that already the algebraic analysis of the allowed $\Gst$ structures and possible singlet intrinsic torsion severely restricts which five-dimensional ${\cal N}=2$ gauged supergravities can be obtained by a consistent truncation.

In particular, we find that the scalar manifolds must necessarily be symmetric spaces and that there is a very limited number of possible truncations. If there are just vector/tensor multiplets, we can only have
\be
\begin{aligned}
    {\cal M}_{\rm VT} &= \bbR^+ \times \frac{\SO{n_{\rm VT}-1,1}}{\SO{n_{\rm VT}-1}} \,, & \qquad n_{\rm VT} &\leq 6 \,, \\
	\mathcal{M}_{\rm VT} &=  \frac{\SL{3, \bbR}}{\SO{3}} \,,  &  \qquad n_{\rm VT}  &=5   \, ,  \\
	\mathcal{M}_{\rm VT} &=  \frac{\SL{3 ,\bbC}}{\SU{3}}  \,, &  \qquad n_{\rm VT}  &=  8  \, ,  \\
	\mathcal{M}_{\rm VT} &=  \frac{\SUs6}{\USp{6}}  \,,  & \qquad n_{\rm VT}  &=  14   \, ,
\end{aligned}
\ee
while if there are just hypermultiplets, the only possibilities are
\be
\begin{aligned}
	\mathcal{M}_{\rm H} &=  \frac{\mathrm{G}_{2(2)}}{\SO{4}} \,, & \qquad n_{\rm H}  &=   2   \,, \\
	\mathcal{M}_{\rm H} &=  \frac{\SU{2,1}}{\mathrm{S}(\U2 \times \U1)} \,, & \phantom{\times \mathbb{R}^+} \qquad n_{\rm H}  &=   1   \,.
\end{aligned}
\ee
Finally, for vector/tensor and hypermultiplets, the only theories with higher-dimensional origin are of the form
\begin{equation}
 \begin{aligned}
  \mathcal{M}_{\rm H} &=  \frac{\SU{2,1}}{\mathrm{S}(\U2 \times \U1)} \,, & \qquad n_{\rm H}  &=   1 \,, \ \\
  {\cal M}_{\rm VT} &= \bbR^+ \times \frac{\SO{n_{\rm VT}-1,1}}{\SO{n_{\rm VT}-1}} \,, & \qquad n_{\rm VT} &\leq 4 \,,
 \end{aligned}
\end{equation}
Any other five-dimensional ${\cal N}=2$ gauged supergravity cannot be uplifted via a consistent truncation to 10-/11-dimensional supergravity\footnote{The theories with hyperscalar manifolds ${\cal M}_{\rm H} = \frac{{\rm F}_{4(4)}}{\SU{2} \cdot \USp{6}}$, or ${\cal M}_{\rm H} = \frac{\SOo{4,p}}{\SO{4} \times \SO{p}}$, as well as with hyperscalar manifolds $\mathcal{M}_{\rm H} =  \frac{\mathrm{G}_{2(2)}}{\SO{4}}$ and some vector multiplets are necessarily subtruncations of ${\cal N}>2$ supergravities.}.

For each of the above cases, we give the corresponding $G_S$ structure and study what gaugings can arise. Algebraically, these are encoded in the  singlets of the intrinsic torsion subject to the Leibniz condition. The results are summarised in Tables \ref{summ-gengaugings}, \ref{summ-gaugingsspe} and \ref{summ-gaugingsmixed}.  For gauged supergravities with only vector/tensor multiplets, we recover many of the results of  \cite{Gunaydin:1999zx}, where the allowed gaugings are discussed from a purely five-dimensional point of view. However, we can also exclude certain of the five-dimensional gaugings that appear in \cite{Gunaydin:1999zx}. For example in the case of ${\cal M}_{\rm VT} = \mathbb{R}^+ \times \frac{\SO{n_{\rm VT}-1,1}}{\SO{n_{\rm VT}-1}}$, we find that gaugings where the tensors are charged under a vector transforming non-trivially under $\SO{n_{\rm VT}-1,1}$ cannot arise from consistent truncations. For truncations with only hypermultiplets the gaugings are trivial since they reduce to gauging the $\U{1}_R$ symmetry. What is probably more surprising is the very limited number of truncations with both vector- and hypermultiplets.

Our findings are particularly important for the study of gauged supergravities containing AdS vacua. Since no AdS vacuum is believed to admit scale separation \cite{Lust:2019zwm}, those gauged supergravities that cannot be uplifted by a consistent truncation cannot have a higher-dimensional  string theory origin. Therefore, they should belong to the \textit{swampland} of lower-dimensional theories.

An important issue that we do not address here is whether we can actually solve the differential conditions imposed by the intrinsic torsion, that are required for the consistent truncation to exist. 
This would involve constructing explicit examples of background that admit the $G_S$ structure listed in this paper and checking that the 
intrinsic torsion has only singlet constant components. We leave this analysis for future work. It would also be interesting to see whether the approach of \cite{Inverso:2017lrz,Bugden:2021wxg,Bugden:2021nwl}
can be extended to non-maximally supersymmetric truncations and to use the  the five-dimensional embedding tensor to determine what the uplifted geometry should be.
In any case, we expect that imposing the differential conditions from the intrinsic torsion  will further restrict which consistent truncations exist.

It is also worth stressing that we scanned all possible generalised structures where $G_S$ is a Lie group. It is possible that looking at discrete structure groups might increase the number of possible truncations.  We leave this as a problem for the future.

Another direction of future research is to extend our analysis to other dimensions and amounts of supersymmetry. For example, it would be interesting to classify which four-dimensional ${\cal N}=2$ gauged supergravities can be uplifted by consistent truncations to 10-/11-dimensional supergravity.  More ambitious would be to extend our method to three dimensions, where ${\cal N}=1$ and ${\cal N}=2$ gauged supergravities admit deformations corresponding to real/holomorphic superpotentials that are not induced by gaugings \cite{deWit:2003ja}. It would be interesting to explore which of these can arise from consistent truncations. The appropriate framework would be $\Ex{8}$ Exceptional Field Theory \cite{Hohm:2014fxa}, where the generalised Lie derivative does not close without the addition of shift symmetries, leading to technical challenges. Similar questions can be asked in two dimensions, where subgroups of affine global symmetry groups, such as $\Ex{9}$ for maximal supersymmetry, can be gauged and scalar and vector fields transform in infinite-dimensional representations of the affine symmetry. This question can in principle be addressed with $\Ex{9}$ Exceptional Field Theory \cite{Bossard:2017aae,Bossard:2018utw,Bossard:2021jix}, which however requires infinitely-large generalised tangent bundles.

%%%%%%%%%%%%%%%%%%%%%%%%%%%%%%%%%%%%%%%%%%%%%%%%%%%%%%%%%%%%%%%%%%%%%%%%%%%%%%%%%%%%%
\section*{Acknowledgments}

DW is supported in part by the STFC Consolidated Grant ST/T000791/1 and the EPSRC New Horizons Grant EP/V049089/1. EM is supported by the Deutsche Forschungsgemeinschaft (DFG, German Research Foundation) via the Emmy Noether program ``Exploring the landscape of string theory flux vacua using exceptional field theory'' (project number 426510644). We acknowledge the Mainz Institute for Theoretical Physics (MITP) of the Cluster of Excellence PRISMA+ (Project ID 39083149) for hospitality and support during part of this work.

%%%%%%%%%%%%%%%%%%%%%%%%%%%%%%%%%%%%%%%%%%%%%%%%%%%%%%%%%%%%%%%%%%%%%%%%%%%%%%%%%%%%%
%%%%%%%%%%%%%%%%%%%%%%%%%%%%%%%%%%%%%%%%%%%%%%%%%%%%%%%%%%%%%%%%%%%%%%%%%%%%%%%%%%%%%

\appendix

%%%%%%%%%%%%%%%%%%%%%%%%%%%%%%%%%%%%%%%%%%%%%%%
\section{ $\mathbf{E}_{6(6)}$ generalised geometry for M-theory}
\label{PreliminariesE66_Mth}

This section is a brief recall of the main features of the generalised geometry of M-theory compactifications on a six-dimensional manifold $M$. 
For a more detailed discussion we refer to \cite{Coimbra:2011ky} and~\cite[App.~E]{Ashmore:2015joa}. 
\vspace{0.2cm}

For M-theory on a six-dimensional manifold we use $\Ex{6} \times \bbR^+$ generalised geometry.  
 The generalised  tangent bundle $E$ and the dual bundle $E^*$ are
\begin{equation} 
\begin{aligned}
\label{app:gentan}
   E \,\simeq \, TM \oplus \Lambda^2T^*M \oplus \Lambda^5T^*M \, , \\ 
   E^* \,\simeq \, T^* M \oplus \Lambda^2T M \oplus \Lambda^5T M \, ,
 \end{aligned}
\end{equation}
where we decompose the various bundles in representations of 
 $\GL{6}$, the geometric subgroup of  $\Ex{6}$. The sections of $E$ and $E^*$, the  generalised vectors and its dual, transform in the  $\mbf{27}^*$ and the $\rep{27}$ of $\Ex{6}$ and can be written as 
\begin{equation}
\begin{aligned}
\label{app:genvec}
V = v + \omega + \sigma\,, \\
Z =  \hat{v} +  \hat{\omega} +  \hat{\sigma} \,,
\end{aligned}
\end{equation}
where $v$ is an ordinary vector field, $\omega$ is a two-form,  $\sigma$  is a five-form\footnote{The generalised tangent bundle $E$  has a non-trivial  structure that  takes into account the non-trivial gauge potentials of M-theory.  
To be more precise the sections of $E$ are defined as 
\be
\label{twist_Mth}
 V \,=\, \rme^{A + \tilde A}  \cdot \check{V} \, , 
\ee 
where $A + \tilde A$ is an element of the adjoint bundle,  $\check{V}  = v + \omega + \sigma$, with $v \in \Gamma(TM)$ are vectors, $\omega\in \Gamma(\Lambda^2 T^*M)$ and $\sigma\in \Gamma(\Lambda^5T^*M)$, and $\cdot$ defines the adjoint action defined in \eqref{adj27}.   The patching condition on the overlaps $U_{\alpha} \cap U_{\beta}$ is 
\be
V_{(\alpha)} \,=\, \rme^{\dd \Lambda_{(\alpha \beta)} + \dd \tilde \Lambda_ {(\alpha \beta)}} \cdot V_{(\beta)} \, , 
\ee
where $\Lambda_{(\alpha \beta)}$ and $ \tilde\Lambda_ {(\alpha \beta)}$ are a two- and five-form, respectively. This corresponds to the gauge-transformation of  the three- and six-form potentials in~\eqref{twist_Mth} as 
\begin{align}
A_{(\alpha)} \,&=\, A_{(\beta)} + \dd \Lambda_{(\alpha \beta)}\ , \nn \\
\tilde A_{(\alpha)} \,&=\, \tilde A_{(\beta)} + \dd \tilde \Lambda_{(\alpha \beta)}  -\frac{1}{2}   \dd \Lambda_{(\alpha \beta)}  \wedge A_{(\beta)} \ .
\end{align}
The respective gauge-invariant field-strengths reproduce the supergravity ones:
\begin{align}
F \,&=\, \dd A \, ,  \nn \\
\tilde F \,&=\, \dd \tilde A - \frac{1}{2} A \wedge F \, .
\end{align}
},
 $\hat{v}$ is one-form,  $\hat{\omega}$ is a two-vector and  $ \hat{\sigma}$ is a five-vector.  Generalised vectors and dual generalised vectors have a natural pairing given by
\be
\label{pairing_vector_dualvector}
\GM{Z}{V} = \hat v_m v^m + \tfrac{1}{2}\,  \hat\omega^{mn} \omega_{m n} +  \tfrac{1}{5!}\, \hat\sigma^{mnpqr}\sigma_{mnpqr}  \, . 
\ee

We will also need the bundle  $N\simeq \det T^*M \otimes E^*$. In terms of $\\GL{6}$ tensors, $N$ decomposes as 
\be
N \simeq T^*M \oplus \Lambda^4T^*M \oplus  (T^*M \otimes  \Lambda^6 T^*M )  \,,
\ee
and correspondingly its sections $Z_{\flat}$ decompose as
\be
Z_{\flat}  = \lambda  +   \rho  + \tau  \, . 
\ee
The bundle $N$ is  obtained from the symmetric product of two generalised vectors via the  map $\otimes_{N}: E \otimes E \to N$  with
\be
\begin{aligned}
\label{N'prod_Mth}
\lambda \,&= \, v \,\lrcorner\, \omega' + v'\,\lrcorner\,\omega\,,  \\
\rho \,&=\, v \,\lrcorner\, \sigma' + v' \,\lrcorner\, \sigma - \omega \wedge \omega' \,,\\
\tau \,&=\,    j \omega \wedge \sigma' + j \omega' \wedge \sigma   \,.
\end{aligned}
\ee
Wedges and contractions among tensors on $M$ are defined with the following conventions:
\begin{align}
( v \wedge u)^{a_1 \ldots a_{p+p^\prime}} & = \frac{(p+p^\prime)!}{p!\, p^\prime!}\, v^{[a_1 \ldots a_p}  u^{a_{p+1} \ldots a_{p+p^\prime} ]}  , \nn \\
( \lambda \wedge \rho)_{a_1 \ldots a_{q+q^\prime}}  & = \frac{(q+q^\prime)!}{q! \,q^\prime!}\, \lambda_{[a_1 \ldots a_q}  \rho_{a_{q+1} \ldots a_{q+q^\prime} ]} , \nn \\
(v \,\lrcorner\, \lambda)_{a_1 \ldots a_{q-p}}  & = \frac{1}{p!} v^{b_1 \ldots b_p} \lambda_{b_1 \ldots b_p a_1 \ldots a_{q-p}} \,, \quad {\rm if}\  p \leq q , \nn \\
(v \,\lrcorner\, \lambda)^{a_1 \ldots a_{p-q} } & = \frac{1}{q!} v^{a_1 \ldots a_{p-q}b_1\ldots b_q} \lambda_{b_1 \ldots b_q } \,, \quad  {\rm if}\  p \geq q , \nn \\
(j v \,\lrcorner\, j  \lambda)^a{}_b   &  = \frac{1}{(p-1)!} v^{a c_1 \ldots c_{p-1}} \lambda_{b  c_1 \ldots c_{p-1}} ,  \nn \\
\left(j \lambda \wedge \rho\right)_{a,\,a_1\ldots a_d} &= \frac{d!}{(q-1)!(d+1-q)!}\,\lambda_{a[a_1\ldots a_{q-1}}\rho_{a_q\ldots a_d]}\ .
\end{align}

The $\Ex{6}$ cubic invariant is defined  on $E$ and $E^*$as\footnote{This is 6 times the cubic invariant given in \cite{Ashmore:2015joa}. Because of this, we introduced a compensating factor of 6 in the formulae \eqref{eq:cKKK} and \eqref{compJK}.}
\begin{align}
c(V,V,V)  &= - \, 6\,\iota_v\, \omega \wedge \sigma -  \omega\wedge \omega\wedge\omega \, , \nn\\
c^*(Z,Z,Z) &= - \, 6\,\iota_{\hat v}\,\hat \omega \wedge \hat\sigma -  \hat\omega\wedge \hat\omega\wedge\hat\omega\,.
\end{align}

The adjoint bundle is defined as 
\be
{\adj} F \,\simeq \, \bbR \oplus (TM \otimes T^* M) \oplus \Lambda^3 T^* M \oplus \Lambda^6 T^* M  \oplus \Lambda^3 T M \oplus \Lambda^6 T  M  \, , 
\ee
with sections
\be
\label{sec78}
R = l + r + a + \tilde a + \alpha + \tilde \alpha \, ,
\ee
where locally $l \in \bbR$,  $r \in \End(TM)$, $a \in   \Lambda^3 T^* M$, etc.  In order to obtain the $\mathfrak{e}_{d(d)}$ sub-algebra 
we need to fix the factor $l$  in terms of the trace of  $r$ as $ l =  \frac{1}{3}  \tr r$. This choice 
 fixes the weight of the generalised tensors under the $\bbR^+$ factor. In particular it implies that a scalar of weight $k$ is a section of 
$(\det T^* M)^{k/3}$: $ \mathbb{1}_k \in \Gamma ((\det T^* M)^{k/3})$. 

It is also useful to introduce the weighted adjoint bundle 
\be 
(\det T^*M) \otimes {\adj} \,  F    \,\supset\,  \bbR  \oplus  \Lambda^3 T^* M    \oplus (TM \otimes \Lambda^5 T  M)   \, , 
\ee
whose sections are  locally given by the sum 
\be
R_{\flat} =   \tilde{ \phi} + \phi     + \psi    \, , 
\ee
where $ \tilde{ \phi} $, $\phi$   and $\psi$  are obtained from  the adjoint elements $r \in TM \otimes T^*M$,  $\alpha \in  \Lambda^3 T M$,  $\tilde \alpha \in  \Lambda^3 T M$ as 
\be
\tilde  \phi  = \tilde \alpha \lrcorner {\rm vol}_6  \qquad 
 \phi   =  \alpha \lrcorner {\rm vol}_6  \qquad 
  \psi    = r  \cdot {\rm vol}_6   \, , 
\ee
where $\vol_6$ is a reference volume form. We denote by  $\cdot$ the $\mathfrak{gl}(6)$ action on tensors:  given a frame $\{ \hat{e}_a \}$ for $TM$ and a co-frame  $\{ e_a \}$ for $T^*M$, $a=1, \ldots, 6$, the action,  for 
instance, on a vector and a two-form is 
\be
\label{app:gl6ac}
(r \cdot  v )^a = r^a{}_b v^b \qquad  \quad ( r \cdot  \omega)_{ab} = - r^c{}_a \omega_{c b} - r^c{}_b \omega_{a c } \, . 
\ee

The action of an adjoint element $R$ on another adjoint element $R^\prime$ is given by the commutator, $R^{\prime \prime} = [ R , R^\prime]$. In components, $R^{\prime \prime}$ reads 
\be
\label{adjcomm}
\begin{aligned}
l^{\prime \prime} & = \tfrac{1}{3} (\alpha\, \lrcorner \,a^\prime - \alpha^\prime \,\lrcorner\, a )  +\tfrac{2}{3}(\tilde{\alpha}'\,\lrcorner\, \tilde{a} - \tilde{\alpha}\,\lrcorner\,\tilde{a}') \, ,\\ 
r^{\prime \prime} &= [r, r^\prime] + j \alpha \,\lrcorner\, j a^\prime  -  j \alpha^\prime \,\lrcorner\, j a - \tfrac{1}{3} ( \alpha\, \lrcorner\,  a^\prime  -  \alpha^\prime\, \lrcorner\, a )\,\id   \, , \\
& \quad  +  j  \tilde \alpha^\prime \,\lrcorner\, j  \tilde a  -  j \tilde  \alpha \,\lrcorner\, j \tilde  a^\prime  - \tfrac{2}{3} ( \tilde \alpha^\prime \,\lrcorner\,  \tilde a -  \tilde  \alpha \,\lrcorner\,  \tilde  a^\prime  )\, \id  \, ,\\
a^{\prime \prime} & = r \cdot a^\prime - r^\prime \cdot a + \alpha^\prime \,\lrcorner\, \tilde a - \alpha  \,\lrcorner\, \tilde a^\prime     \, ,  \\ 
\tilde a^{\prime \prime} & = r \cdot \tilde a^\prime - r^\prime \cdot  \tilde a - a \wedge a^\prime    \, ,  \\
\alpha^{\prime \prime} & = r \cdot \alpha^\prime - r^\prime \cdot \alpha + \tilde  \alpha^\prime \,\lrcorner\, a -  \tilde \alpha  \,\lrcorner\,  a^\prime    \, , \\
\tilde \alpha^{\prime \prime} & = r \cdot \tilde \alpha^\prime - r^\prime \cdot  \tilde \alpha - \alpha \wedge \alpha^\prime  \, , 
\end{aligned} 
\ee

where   $\cdot$ denotes the $\mathfrak{gl}(6)$ action defined in \eqref{app:gl6ac}. 

The $\mathfrak{e}_{6(6)}$ Killing form on two elements of the adjoint bundle is given by
\be
{\rm tr}(R,R') \,=\, \tfrac{1}{2} \left( \tfrac{1}{3}\,{\rm tr}(r){\rm tr}(r') + {\rm tr}(rr') + \alpha \,\lrcorner\, a' + \alpha' \,\lrcorner\, a -\tilde\alpha \,\lrcorner\, \tilde a' - \tilde \alpha' \,\lrcorner\, \tilde a \right)\,.
\ee

An element $R$ of the adjoint bundle can act on a generalised vector $V \in \Gamma( E)$  and on a dual generalised vector $Z$ as
\be
V^\prime = R \cdot V \,, \qquad \quad  Z^\prime = R \cdot Z \, ,
\ee
where the components of $V^\prime$ are 
\be
\label{adj27}
\begin{aligned}
v^{\prime} & =  l v +  r \cdot v + \alpha \,\lrcorner\, \omega - \tilde \alpha \,\lrcorner\, \sigma   \, ,  \\ 
\omega^{\prime} &=  l \omega +  r \cdot \omega  + v \,\lrcorner\, a  +  \alpha \,\lrcorner\, \sigma   \, ,  \\ 
\sigma^{\prime} & = l \sigma +  r \cdot \sigma  + v \,\lrcorner\, \tilde a  + a \wedge \omega    \, ,  
\end{aligned} 
\ee
and those  of $Z^\prime$ are
\be
\label{adj27b}
\begin{aligned}
\hat v^{\prime} & = -  l  \hat v +  r \cdot  \hat v - \hat \omega  \,\lrcorner\, a  + \hat  \sigma \,\lrcorner\, \tilde  a    \, ,   \\ 
\hat \omega^{\prime} &= -  l \hat\omega +  r \cdot  \hat \omega  - \alpha \,\lrcorner\, \hat{v}   -  \hat  \sigma \,\lrcorner\, a    \, ,  \\ 
\hat \sigma^{\prime} & = - l  \hat \sigma +  r \cdot  \hat \sigma  - \tilde \alpha  \,\lrcorner\, \hat v   - \alpha  \wedge  \hat \omega     \, . 
\end{aligned} 
\ee

In this formalism, diffeomorphisms and gauge transformations by the three-form and six-form potentials combines to define the generalised diffeomorphisms. 
The action of an infinitesimal generalised diffeomorphism is generated by the generalised Lie (or Dorfman) derivative along a generalised vector. The generalised Lie derivative is defined in an analogous way as the Lie derivative between two ordinary vectors $v$ and $v'$ on $TM$, written in components as a $\mathfrak{gl}(6)$ action 
\be
( \mathcal{L}_v v')^m \,=\,  v^n \,\partial_n v'^{\,m} - (\partial \times v)^m{}_n \,v'^{\,n}  \, ,
\ee
the symbol $ \times$ is the projection onto the adjoint bundle of the product of the fundamental and dual representation of $\GL{6}$ . We introduce the operators $\partial_M = \partial_m$ as sections of the dual tangent bundle and we define the generalised Lie derivative as
\be 
\label{eq:Liedefgapp}
(L_V V')^M \,=\,  V^N \partial_N  V'^M - (\partial \times_{\adj} V)^M{}_N V'^N \, , 
\ee
where  $V^M$, $M=1,\ldots,27$, are the components of $V$ in a standard coordinate basis, and $ \times_{\adj}$ is the projection onto the adjoint bundle,
\be
 \times_{\adj} \, : \, E^* \otimes E \rightarrow  {\adj} F\, , 
\ee
whose explicit expression can be found in~\cite[Eq.$\:$(C.13)]{Coimbra:2011ky}.
In terms of  $\\GL{6}$  tensors, \eqref{eq:Liedefgapp} becomes
\begin{equation}
\label{dorf27}
L_{V} V' = \mathcal{L}_{v} v' + \left(\mathcal{L}_{v}  \omega^{\prime} -\iota_{v^\prime}\mathrm{d} \omega\right) + \left(\mathcal{L}_{v} \sigma' -\iota_{v^\prime}\mathrm{d}\sigma -  \omega^{\prime}\wedge \mathrm{d} \omega\right) .
\end{equation}
 The action of the generalised Lie  derivative on a section of the adjoint bundle  \eqref{sec78} is 
\begin{align}
\label{dor78} 
L_{V} R  &=  (\mathcal{L}_v r + j \alpha \,\lrcorner\, j {\rm d} \omega - \tfrac{1}{3}\, \mathbb{1}  \alpha \,\lrcorner\, {\rm d} \omega - 
 j \tilde \alpha \,\lrcorner\, j {\rm d} \sigma + \tfrac{2}{3}\, \mathbb{1} \tilde  \alpha \,\lrcorner\, {\rm d} \sigma) + (\mathcal{L}_v a + r \cdot {\rm d} \omega - \alpha \,\lrcorner\,  {\rm d} \sigma) \nonumber \\
 &\quad +  (\mathcal{L}_v  \tilde a +  r \cdot {\rm d} \sigma +  {\rm d} \omega  \wedge  a ) +    (\mathcal{L}_v  \alpha  - \tilde \alpha \,\lrcorner\,  {\rm d} \omega ) +  \mathcal{L}_v  \tilde  \alpha   \, . 
\end{align}
Given a section $Z_\flat = \lambda + \rho + \tau$ of $N$, its Lie derivative along the 
 generalised vector $V$ is 
\be
\label{dorN} 
L_V Z_\flat = \mathcal{L}_v \lambda + (  \mathcal{L}_v  \rho -\lambda  \wedge \dd \omega ) +  (  \mathcal{L}_v  \tau - j  \rho  \wedge \dd \omega + j \lambda \wedge \dd \sigma ) \,.
\ee
 Since  $Z_\flat = V^\prime \otimes_N V^{\prime \prime}$,   this is obtained by applying the Leibniz rule for $L_V$.
\be
L_V (Z_\flat) =   L_V V^\prime   \otimes_N   V^{\prime \prime} + V^\prime \otimes_N  L_V V^{\prime \prime} \,.
\ee
It is  also  straightforward to verify that 
\be
\label{derid}
\dd Z_\flat = L_V  V^\prime + L_{V^\prime}  V \, ,
\ee
for any element $Z_\flat =V \otimes_{N} V'   \in N$.
%%%%%%%%%%%%%%%%%%%%%%%%%%%%%%%%%%%%%%%%%%%%%%%%%%%%%%%%%%%%%%%%%%%%%%%%%%%%%%%%%%%%%

%%%%%%%%%%%%%%%%%%%%%%%%%%%%%%%%%%%%%%%%%%%%%%%%%%%%%%%%%%%%%%%%%%%%%%%%%%%%%%%%%%%%%
\section{ Intrinsic torsion for $G_S = \SU2\times\Spin{6 - n_{\rm VT}}$ structures}
\label{app:intors}

The intrinsic torsion of a given $G_S$ structure plays an important role in the derivation of the truncated theory as it determines the embedding tensors and the possible gaugings.

As discussed in Section \ref{sec:gentors}, the generalised intrinsic torsion of a $G_S$ structure is given by quotient
 \be
 W_{\text{int}}^{G_S}=W/W_{G_S} \,,
\ee
where $W$ is the bundle of the generalised torsion, which in our case is in   the $\rep{351}$ of $\Ex{6}$,  and $W_{G_S}$ is the image
of the map $\tau  \, : \, K_{G_S}\to W$ from the space  $K_{G_S} =  E^\ast \otimes {\rm ad} G_S$  of $G_S$ compatible connections to $W$.
Moreover, since  in all our cases the $G_S \subset \USp{8}$, one can define a generalised metric $G$  and use the norm defined by $G$ to decompose the bundles $W$ and $K_{\Gst}$ as \cite{Coimbra:2014uxa}  
\be 
\begin{aligned}
\label{ortdect}
W & =  W_{\Gst} \oplus W_{\rm int} \,, \\
K_{\Gst}  &= W_{\Gst} \oplus U_{\Gst} \, . 
\end{aligned}
\ee

\medskip

In this appendix we show how to compute $W_{\text{int}}$ for two  of the examples discussed in Section \ref{sec:GenV}. These two cases allow to illustrate all the subtleties one might encounter in this kind of computation. 

\medskip

We consider first the truncation to $n_{\rm VT} = 6 $ vector multiplets. The structure group is  $G_S = \SU{2} \times \mathbb{Z}_2$ and the isometry group is $\Giso= \SU{2}_R \times  \SO{5,1} \times \mathbb{R}^+$.  We use \eqref{ortdect}  to compute  the intrinsic torsion $W_{int}$ of the $G_S$ structure.  

We first decompose the generalised torsion under $G_S \times \Giso$  and keep only the $G_S$ singlets 
\be
\label{dec351genv6}
\left. W \right|_s  =  (\mbf{3},\mbf{1} )_{-2}  \oplus   (\mbf{3} ,\mbf{6})_{1}   \oplus  (\mbf{1},\mbf{n})_1  \oplus (\mbf{1},\mbf{15})_{-2}     \oplus  (\mbf{1},\mbf{10})_{1}  \oplus  (\mbf{1},\mbf{\overline{10}})_{1}    \, , 
\ee
where  the first entries are  $\SU{2}_R$ representations and the second ones   $ \SO{5,1}$  representations  and
the subscripts are the $ \mathbb{R}^+$ charges.

Then we look for  the $G_S$ singlets in the space of $G_S$ connections,  $K_{G_S}$. The intrinsic torsion will be given by the terms in \eqref{dec351genv6} that are not contained in $K_G$. 
Since the $\rep{27}$ does not contain terms in the adjoint of $G_S$, the product\footnote{In this expression the last entries denote the representations of the structure group.}
\be
 K_{G_S}  =[ (\rep{1}, \rep{1}, \rep{1})_{-2}  \oplus  (\rep{1},\rep{6},  \rep{1})_{1} \oplus  (\rep{1},\rep{4},  \rep{2})_{-1/2} 
 \oplus  (\rep{2},\rep{\bar{4}},  \rep{1})_{-1/2}  \oplus  (\rep{2},\rep{1},  \rep{2})_{-1/2}   ] \otimes [(\rep{1}, \rep{1} , \rep{3})_0 ]    \nn 
\ee
can never contain $G_S$ singlets. This means that the intrinsic torsion of the $G_S$ structure is entirely given by $\left. W \right|_s$
\be
\label{intgenv}
W_{int}   =  (\mbf{3},\mbf{1} )_{-2}  \oplus   (\mbf{3} ,\mbf{6})_{1}   \oplus  (\mbf{1},\mbf{n})_1  \oplus (\mbf{1},\mbf{15})_{-2}     \oplus  (\mbf{1},\mbf{10})_{1}  \oplus  (\mbf{1},\mbf{\overline{10}})_{1}    \,,
\ee
and we do not have to bother about possible kernels of the map $\tau : K_{G_S}  \to W$. 

\medskip

Consider now the case with $n_{\rm VT} = 4$ vector multiplets, which has structure group   $G_S = \SU{2} \times \U{1}$ and  isometry group is $\Giso= \SU{2}_R \times  \SO{3,1} \times \mathbb{R}^+$.  The $G_S$ singlets in the generalised torsion are 
\be
\left. W \right|_s  =  (\mbf{3},\mbf{1} )_{-2}  \oplus   (\mbf{3} ,\mbf{4})_{1}   \oplus 2 \cdot  (\mbf{1},\mbf{4})_1  \oplus (\mbf{1},\mbf{6})_{-2}     \oplus  (\mbf{1},\mbf{4})_{-1}  \oplus  (\mbf{1},\mbf{1})_{-2}    \, , 
\ee
where again  the first entries are  $\SU{2}_R$ representations and the second ones   $ \SO{3,1}$  representations, while 
the subscripts are the $ \mathbb{R}^+$ charges. The $G_S$ singlets in the generalised connection are
\begin{eqnarray}
\label{decKGgenv4}
 K_{G_S} & = &[ (\rep{1}, \rep{2}, \rep{2})_{-1/2, 1}   \oplus (\rep{1}, \rep{\bar 2}, \rep{2})_{-1/2, -1}   \oplus (\rep{1}, \rep{4}, \rep{1})_{1, 0}   \nn \\ 
 & &  \oplus (\rep{1}, \rep{1}, \rep{1})_{1, 2}   \oplus  (\rep{1}, \rep{1}, \rep{1})_{1, - 2}   \oplus   (\rep{1}, \rep{1}, \rep{1})_{-2, 0} \nn \\
 & &    \oplus (\rep{2}, \rep{2}, \rep{1})_{-1/2, 1}    \oplus (\rep{2}, \rep{\bar 2}, \rep{1})_{-1/2, - 1}     \oplus (\rep{2}, \rep{1}, \rep{2})_{1, 0}  ]
  \otimes [(\rep{1}, \rep{1} , \rep{3})_{0,0} \oplus (\rep{1}, \rep{1} , \rep{1})_{0,0}  ]    \nn \\
  & \to &  [  (\rep{1}, \rep{4}, \rep{1})_{1, 0}     \oplus   (\rep{1}, \rep{1}, \rep{1})_{-2, 0}     ]\,.
\end{eqnarray}
Again from \eqref{ortdect} the intrinsic torsion is given by the elements of $\left. W \right|_s$ that are not contained in \eqref{decKGgenv4} 
\be
W_{\rm int}   \supseteq  (\mbf{3},\mbf{1} )_{-2}  \oplus   (\mbf{3} ,\mbf{4})_{1}   \oplus   (\mbf{1},\mbf{4})_1  \oplus (\mbf{1},\mbf{6})_{-2}     \oplus  (\mbf{1},\mbf{4})_{-1}    \, .
\ee
In this case, one should make sure that the map $\tau$ has no kernel so that the relation above is an equality.
The explicit definition of  the map $\tau : K_G \to W$ is via the generalised Lie derivative. 
Given a $G_S$ compatible connection
\be
\label{appcompc}
\tilde{D}_M W^N = \partial_M W^N + \Omega_M{}^N{}_P W^P \,,
\ee 
the intrinsic torsion can be defined as 
\be
 \label{eq:gen-torsion-app1}
 \begin{aligned}
 T(V)^M{}_N  W^N  & =  \,   \big(\Lgen^{\tilde{D}}_V  W \big)^M - \big( \Lgen_V W \big)^M  \\
 &   =V^P \big( \Omega_P{}^M{}_N  - \Omega_N{}^M{}_P  +  \alpha\, c^{MSQ} c_{RNQ}\, \Omega_S{}^R{}_P  \big)  W^N  =: V^P  T_P{}^M{}_N  W^N\,,
  \end{aligned} 
\ee
where $V$ and $W$ are generalised vectors and, in the second line, we plugged  \eqref{appcompc}  and 
we used the explicit expression for the $\Ex{6}$ adjoint action in the  generalised Lie derivative \eqref{eq:Liedefgapp}
\be
\big( \Lgen_V W \big)^M  = V^N \partial_N W^M -   W^N \partial_N V^M + \alpha\, c^{M P Q} c_{R N Q} \partial_P V^R W^N  \, . 
\ee 
The  second line in  \eqref{eq:gen-torsion-app1}  defines the map $\tau$ as 
\be
\label{exp-tau}
\tau(\Omega)_P{}^M{}_N  = T_P{}^M{}_N  \, . 
\ee

By computing \eqref{exp-tau} one can check  the  that there is indeed no kernel, as can also be seen in terms of representations

\be
\begin{aligned}
 & T(v^0)^a{}_b &\quad  &\longleftrightarrow  \quad  (\mbf{3},\mbf{1} )_{0}  \in \rep{1}_2 \otimes   (\mbf{3},\mbf{1} )_{-2}   \,,   \\
 &T(v^0)^i{}_j &\quad &\longleftrightarrow \quad  (\mbf{1},\mbf{ad})_{0}  \in \rep{1}_{2} \otimes   (\mbf{3},\mbf{1} )_{1}  \,,   \\
 &T(v^i)^a{}_b &\quad &\longleftrightarrow  \quad    (\mbf{3},\mbf{1} )_{0} \in \rep{n}_{-1} \otimes   (\mbf{3},\mbf{1} )_{1}   \,,    \\
 &T(v^i)_{(0)} &\quad &\longleftrightarrow  \quad    (\mbf{1},\mbf{ad})_{0} \in \rep{n}_{-1} \otimes   (\mbf{1},\mbf{n} )_{1}   \,,   \\
 &T(v^i)_{jk}  &\quad &\longleftrightarrow  \quad    (\mbf{1},\mbf{ad})_{0} \in  \rep{n}_{-1} \otimes   (\mbf{1},\mbf{X} )_{1} \,.      
\end{aligned}
\ee 

We have not directly checked that there are no singlet intrinsic torsion kernels for the other $G_S$ structures that appear in paper, although our expectation is that there are not. 

%%%%%%%%%%%%%%%%%%%%%%%%%%%%%%%%%%%%%%%%%%%%%%%%%%%%%%%%%%%%%%%%%%%%%%%%%%

\section{The truncation ansatz}
\label{app:fielddec}

In this section we discuss the truncation ansatz for ruductions of eleven-dimensional supergravity to five dimensions. The ansatz gives the explicit relation between the eleven-dimensional fields and those of the reduced theory.
The discussion for type IIB reduction follows the same lines.

We consider eleven-dimensional supergravity  on a background 
$X \times M$, 
where $X$ is a non-compact five-dimensional space-time and $M$ is a six-dimensional compact space. 
We focus on the bosonic sector of eleven-dimensional supergravity, which consists of the metric $\hat{g}$, a three-form potentia $\hat{A}$  and a six-form potential $\hat{\tilde{A}}$.
We  use the conventions  of~\cite{Coimbra:2011ky}. 

\vspace{0.2cm}

The first step of the truncation consists in decomposing the eleven-dimensional fields according to $\GL(6,\bbR)\times \Ex{6}$, where $\GL(6,\bbR)$, the structure group of $X$,  determines  the tensorial structure of the fields in the five-dimensional theory
\begin{eqnarray}
\label{11d_metric_general}
\hat{g}  &=&  \rme^{2\Delta}\, g_{\mu\nu} \,\dd{x}^\mu\dd{x}^\nu + g_{mn} Dz^m Dz^n\ , \nn \\[1mm]
\hat{A} &= & \tfrac{1}{3!} A_{mnp} D{z}^{mnp} + \tfrac{1}{2}{A}_{\mu mn}
\dd{x}^\mu\wedge D{z}^{mn} + \tfrac{1}{2}\,{\bar{A}}_{\mu\nu
  m}\dd{x}^{\mu\nu} \wedge Dz^m +
\tfrac{1}{3!}\,{\bar{A}}_{\mu\nu\rho}\,\dd{x}^{\mu\nu\rho}  \, , \nn  \\[1mm]
\hat{\tilde{A}} &= & \tfrac{1}{6!} \tilde{A}_{m_1\ldots m_6}
Dz^{m_1\ldots m_6} + \tfrac{1}{5!} {\tilde A}_{\mu m_1\ldots m_5}
\dd{x}^\mu \!\wedge\! Dz^{m_1\ldots m_5} \nn \\[1mm] 
& & \quad  + \, \tfrac{1}{2\cdot 4!} \bar{{\tilde A}}_{\mu\nu m_1\ldots m_4} \dd{x}^{\mu\nu} \!\wedge\! Dz^{m_1\ldots m_4} +\, \ldots  \,,
\end{eqnarray}
with   $x^\mu$, $\mu=0,\ldots,4$, and $y^m$, $m=1,\ldots,6$, 
the coordinates on $X$ and $M$, respectively, and   $Dy^m \,=\, \dd{y}^m - h_\mu{}^m \dd{x}^\mu$. 
All the components in \eqref{11d_metric_general}  may depend both on $x^\mu$ and $y^m$, the only exception being the  external metric, which only depends on the external coordinates only, $g_{\mu\nu}= g_{\mu\nu}(x)$.

Then we arrange the fields in \eqref{11d_metric_general} according to $\Ex{6}$ representations.\footnote{Note that, in order to reproduce the gauge transformation of the reduced theory, the barred components of  three- and six-form potentials must be redefined, Appendix C of \cite{Cassani:2020cod}.  The expressions for the  redefined fields, which we denote by unbarred $A$ and $\tilde A$  are not relevant for this work.}

The field with all components on the internal manifold $M$
arrange into the inverse generalised metric
\be
\label{gmscalars}
G^{MN} \ \longleftrightarrow\ \{\Delta,\,g_{mn}, \, A_{mnp}, \, \tilde A_{m_1\ldots m_6}  \}\ \, . 
\ee
The explicit embedding is given by 
\begin{eqnarray}
\label{gen_metr_general}
 ( G^{-1})^{mn}  & = &  \rme^{2 \Delta} g^{mn}  \,, \nn \\[1mm]
(G^{-1})^m{}_{n_1 n_2}  &  = &  \rme^{2 \Delta} g^{mp} A_{p n_1 n_2}  \,, \nn  \\[1mm]
(G^{-1})^m{}_{n_1 \ldots  n_5}  & = &   \rme^{2 \Delta} g^{mp}  ( A_{p [n_1 n_2}  A_{n_3 n_4 n_5]} + \tilde{A}_{p n_1 \ldots n_5} )   \,,\nn  \\[1mm]
(G^{-1})_{m_1 m_2 \, n_1  n_2}  & =&  \rme^{2 \Delta} (  g_{m_1 m_2, n_1 n_2} +  g^{pq}   A_{p m_1 m_2}  A_{q n_1 n_2]} )   \,, \nn  \\[1mm]
(G^{-1})_{m_1 m_2 \,  n_1  \ldots n_5}  & = &  \rme^{2 \Delta} [ g_{m_1
  m_2 , [n_1 n_2} A_{n_3 n_4 n_5]}
+ g^{pq}  ( A_{p m_1 m_2}  ( A_{q [n_1 n_2}  A_{n_3 n_4 n_5]} + \tilde{A}_{q n_1 \ldots n_5} ) ]  \,, \nn  \\[1mm]
(G^{-1})_{m_1 \ldots m_5 \,  n_1  \ldots n_5}  & = &
\rme^{2 \Delta}  g^{pq}  ( A_{p [m_1 m_2} A_{m_3 m_4 m_5]} + \tilde{A}_{p m_1 \ldots m_5} ) 
( A_{q [n_1 n_2}  A_{n_3 n_4 n_5]} + \tilde{A}_{q n_1 \ldots n_5} ) \nn   \\ 
& &  + \, \rme^{2 \Delta}  g_{m_1 \ldots m_5 ,\,  n_1  \ldots n_5} \, , 
\end{eqnarray}
where $g_{m_1m_2,\,n_1n_2}= g_{m_1[n_1} g_{|m_2|n_2]}$, and similarly for $g_{m_1 \ldots m_5 ,\,  n_1  \ldots n_5}$.
 
The tensors with one external leg arrange into a generalised vector $\mathcal{A}_\mu$ on $M$, with components
\be
\label{Avec}
\mathcal{A}_\mu{}^M(x,y)  = \{ h_\mu{}^m ,\, A_{\mu mn} , \,\tilde{A}_{\mu m_1\ldots m_5}\,\}\, \in \Gamma( T^* M \otimes E) \,  ,
\ee
while those with two external anti-symmetric indices  define a weighted dual vector in the bundle $N$ 
\be
\label{Btens}
\mathcal{B}_{\mu\nu\,M} =\{  A_{\mu\nu m}, \,   \tilde{A}_{\mu\nu m_1\ldots m_4}  ,\, \tilde g_{\mu\nu m_1\ldots m_6,n} \} \in \Gamma(\Lambda^2 T^*M \otimes N)  \,,
\ee
The last term in \eqref{Btens} is  related to the dual graviton and is not necessary in the truncation. 
Finally, the  tensors with three antisymmetrised external indices arrange into the generalised tensor
\be
\mathcal{C}_{\mu\nu\rho}{}^{\hat\alpha} = \{ A_{\mu\nu \rho}, \,\tilde{A}_{\mu\nu \rho m_1m_2 m_3},\, \tilde g_{\mu\nu\rho m_1\ldots m_5,n}\} \in\Gamma(C^\prime)\,,
\ee
where $C^\prime$ is a sub-bundle of the weighted adjoint bundle $\det T^*M \otimes \adj F$, whose components are labeled by $\hat{\alpha}=1,\ldots,78$. See e.g.~\cite{Riccioni:2007ni,deWit:2008ta} for more details on this tensor hierarchy.

The truncation ansatz for the bosonic sector of eleven-dimensional supergravity is obtained by expanding the generalised tensors define above into singlets of the $G_S$ structure. 

The scalars of the truncated theory are determined by the generalised metric. To obtain the ansatz for the scalars one first needs to construct a family of HV structures in terms of the $G_S$ singlets as described in Section \ref{mod-space}
 \begin{equation}
\begin{aligned}
    K(x,y) &= h^{\tilde I}(x) K_{\tilde I}(y) \,, \\
    J_\alpha(z,y) &= L(x) j_\alpha(y)  L(x)^{-1} \,,
  \end{aligned} 
\end{equation}
where $L$ is the representative of the 
coset $\mathcal{M}_{\rm H}$ and $h$ parameterise $\mathcal{M}_{\rm VT}$.
Then  plugging $K$ and $J_\alpha$  in the  expression \eqref{USp6_gm} gives the generalised metric, which now depends 
 on the H and V structure moduli. These are 
identified with the hyperscalar and vector multiplet scalar fields of the truncated theory.
Comparing the generalised metric obtained this way with its general form \eqref{gen_metr_general}, we obtain the truncation ansatz for $\Delta$, $g_{mn}$, $A_{mnp}$,  $\tilde A_{m_1\ldots m_6}$ (if needed). 

The gauge potential of the five-dimensional theory are given by expanding the generalised vector \eqref{Avec} on the $G_S$ invariant vectors $K_{\tilde I}$
\be\label{ansatz_Avec}
\mathcal{A}_\mu(x,y)  \,=\, \mathcal{A}_\mu{}^{\tilde I}(x)\, K_{\tilde I}(y) 
\,. \ee

As for the metric, identifying the components on the two sides of the equation above gives the truncation ansatz for $h_\mu^m$,  $A_{\mu mn}$ and ${\tilde A}_{\mu m_1 \ldots m_5}$. 

Similarly the two-form fields and the ansatz for the field with two antisymmetrised external indices are obtained from 
\be
\label{ansatz-Bform}
\mathcal{B}_{\mu\nu}(x,y) \,=\, \mathcal{B}_{\mu\nu\,{\tilde I}}(x)\, K^{\tilde I}_\flat(y) \,,
\ee
where $K_\flat^{\tilde I}$  are the $G_S$ singlet  weighted dual basis vectors, which are defined by 
 $K_\flat^{\tilde I}(K_{\tilde J})=3\kk^2\,\delta^{\tilde I}{}_{\tilde J}$.We can also give the ansatx for the three-forms of the reduced theory 
\be
\label{ansatz-Cform}
\mathcal{C}_{\mu\nu\rho} \,=\, \mathcal{C}_{\mu\nu\rho}{}^A(x)\, J^\flat_A  \,,
\ee
where $J^\flat_A=\kk^2 J_A$ are the  $G_S$ singlets in the weighted adjoint
bundle.

%%%%%%%%%%%%%%%%%%%%%%%%%%%%%%%%%%%%%%%%%%%%%%%%%%%%%%%%%%%%%%%%%%%%%%%%%%

\bibliographystyle{JHEP}
\bibliography{Bibliography}

\providecommand{\href}[2]{#2}\begingroup\raggedright\begin{thebibliography}{10}

\bibitem{Duff:1984hn}
M.~J. Duff, B.~E.~W. Nilsson, C.~N. Pope and N.~P. Warner, \emph{{On the
  Consistency of the {Kaluza-Klein} Ansatz}},
  \href{http://dx.doi.org/10.1016/0370-2693(84)91558-2}{\emph{Phys. Lett.} {\bf
  149B} (1984) 90--94}.

\bibitem{Lust:2019zwm}
D.~L\"ust, E.~Palti and C.~Vafa, \emph{{AdS and the Swampland}},
  \href{http://dx.doi.org/10.1016/j.physletb.2019.134867}{\emph{Phys. Lett. B}
  {\bf 797} (2019) 134867}, [\href{http://arxiv.org/abs/1906.05225}{{\tt
  1906.05225}}].

\bibitem{Cvetic:2000dm}
M.~Cvetic, H.~Lu and C.~N. Pope, \emph{{Consistent Kaluza-Klein sphere
  reductions}}, \href{http://dx.doi.org/10.1103/PhysRevD.62.064028}{\emph{Phys.
  Rev.} {\bf D62} (2000) 064028},
  [\href{http://arxiv.org/abs/hep-th/0003286}{{\tt hep-th/0003286}}].

\bibitem{Malek:2020mlk}
E.~Malek, H.~Nicolai and H.~Samtleben, \emph{{Tachyonic Kaluza-Klein modes and
  the AdS swampland conjecture}},
  \href{http://dx.doi.org/10.1007/JHEP08(2020)159}{\emph{JHEP} {\bf 08} (2020)
  159}, [\href{http://arxiv.org/abs/2005.07713}{{\tt 2005.07713}}].

\bibitem{Giambrone:2021zvp}
A.~Giambrone, E.~Malek, H.~Samtleben and M.~Trigiante, \emph{{Global Properties
  of the Conformal Manifold for S-Fold Backgrounds}},
  \href{http://dx.doi.org/10.1007/JHEP06(2021)111}{\emph{JHEP} {\bf 06} (2021)
  111}, [\href{http://arxiv.org/abs/2103.10797}{{\tt 2103.10797}}].

\bibitem{Scherk:1979zr}
J.~Scherk and J.~H. Schwarz, \emph{{How to Get Masses from Extra Dimensions}},
  \href{http://dx.doi.org/10.1016/0550-3213(79)90592-3}{\emph{Nucl. Phys.} {\bf
  B153} (1979) 61--88}.

\bibitem{Gauntlett:2009zw}
J.~P. Gauntlett, S.~Kim, O.~Varela and D.~Waldram, \emph{{Consistent
  Supersymmetric Kaluza-Klein Truncations with Massive Modes}},
  \href{http://dx.doi.org/10.1088/1126-6708/2009/04/102}{\emph{JHEP} {\bf 04}
  (2009) 102}, [\href{http://arxiv.org/abs/0901.0676}{{\tt 0901.0676}}].

\bibitem{Cassani:2010uw}
D.~Cassani, G.~Dall'Agata and A.~F. Faedo, \emph{{Type IIB Supergravity on
  Squashed Sasaki-Einstein Manifolds}},
  \href{http://dx.doi.org/10.1007/JHEP05(2010)094}{\emph{JHEP} {\bf 05} (2010)
  094}, [\href{http://arxiv.org/abs/1003.4283}{{\tt 1003.4283}}].

\bibitem{Gauntlett:2010vu}
J.~P. Gauntlett and O.~Varela, \emph{{Universal Kaluza-Klein Reductions of Type
  IIB to ${\mathcal{N}}\!=4$ Supergravity in Five Dimensions}},
  \href{http://dx.doi.org/10.1007/JHEP06(2010)081}{\emph{JHEP} {\bf 06} (2010)
  081}, [\href{http://arxiv.org/abs/1003.5642}{{\tt 1003.5642}}].

\bibitem{Liu:2010sa}
J.~T. Liu, P.~Szepietowski and Z.~Zhao, \emph{{Consistent massive truncations
  of IIB supergravity on Sasaki-Einstein manifolds}},
  \href{http://dx.doi.org/10.1103/PhysRevD.81.124028}{\emph{Phys. Rev.} {\bf
  D81} (2010) 124028}, [\href{http://arxiv.org/abs/1003.5374}{{\tt
  1003.5374}}].

\bibitem{Cassani:2011fu}
D.~Cassani and P.~Koerber, \emph{{Tri-Sasakian Consistent Reduction}},
  \href{http://dx.doi.org/10.1007/JHEP01(2012)086}{\emph{JHEP} {\bf 01} (2012)
  086}, [\href{http://arxiv.org/abs/1110.5327}{{\tt 1110.5327}}].

\bibitem{deWit:1986oxb}
B.~de~Wit and H.~Nicolai, \emph{{The Consistency of the S**7 Truncation in D=11
  Supergravity}},
  \href{http://dx.doi.org/10.1016/0550-3213(87)90253-7}{\emph{Nucl. Phys.} {\bf
  B281} (1987) 211--240}.

\bibitem{Nastase:1999kf}
H.~Nastase, D.~Vaman and P.~van Nieuwenhuizen, \emph{{Consistency of the
  Ad$S^7$ $\times$ $S^4$ Reduction and the Origin of Selfduality in Odd
  Dimensions}},
  \href{http://dx.doi.org/10.1016/S0550-3213(00)00193-0}{\emph{Nucl. Phys.}
  {\bf B581} (2000) 179--239}, [\href{http://arxiv.org/abs/hep-th/9911238}{{\tt
  hep-th/9911238}}].

\bibitem{Nastase:1999cb}
H.~Nastase, D.~Vaman and P.~van Nieuwenhuizen, \emph{{Consistent nonlinear K K
  reduction of 11-d supergravity on AdS(7) x S(4) and selfduality in odd
  dimensions}},
  \href{http://dx.doi.org/10.1016/S0370-2693(99)01266-6}{\emph{Phys. Lett. B}
  {\bf 469} (1999) 96--102}, [\href{http://arxiv.org/abs/hep-th/9905075}{{\tt
  hep-th/9905075}}].

\bibitem{Coimbra:2014uxa}
A.~Coimbra, C.~Strickland-Constable and D.~Waldram, \emph{{Supersymmetric
  Backgrounds and Generalised Special Holonomy}},
  \href{http://dx.doi.org/10.1088/0264-9381/33/12/125026}{\emph{Class. Quant.
  Grav.} {\bf 33} (2016) 125026}, [\href{http://arxiv.org/abs/1411.5721}{{\tt
  1411.5721}}].

\bibitem{Cassani:2019vcl}
D.~Cassani, G.~Josse, M.~Petrini and D.~Waldram, \emph{{Systematics of
  consistent truncations from generalised geometry}},
  \href{http://dx.doi.org/10.1007/JHEP11(2019)017}{\emph{JHEP} {\bf 11} (2019)
  017}, [\href{http://arxiv.org/abs/1907.06730}{{\tt 1907.06730}}].

\bibitem{Lee:2014mla}
K.~Lee, C.~Strickland-Constable and D.~Waldram, \emph{{Spheres, generalised
  parallelisability and consistent truncations}},
  \href{http://dx.doi.org/10.1002/prop.201700048}{\emph{Fortsch. Phys.} {\bf
  65} (2017) 1700048}, [\href{http://arxiv.org/abs/1401.3360}{{\tt
  1401.3360}}].

\bibitem{Hohm:2014qga}
O.~Hohm and H.~Samtleben, \emph{{Consistent Kaluza-Klein Truncations via
  Exceptional Field Theory}},
  \href{http://dx.doi.org/10.1007/JHEP01(2015)131}{\emph{JHEP} {\bf 01} (2015)
  131}, [\href{http://arxiv.org/abs/1410.8145}{{\tt 1410.8145}}].

\bibitem{Baguet:2015sma}
A.~Baguet, O.~Hohm and H.~Samtleben, \emph{{Consistent Type IIB Reductions to
  Maximal 5D Supergravity}},
  \href{http://dx.doi.org/10.1103/PhysRevD.92.065004}{\emph{Phys. Rev.} {\bf
  D92} (2015) 065004}, [\href{http://arxiv.org/abs/1506.01385}{{\tt
  1506.01385}}].

\bibitem{Ciceri:2016dmd}
F.~Ciceri, A.~Guarino and G.~Inverso, \emph{{The exceptional story of massive
  IIA supergravity}},
  \href{http://dx.doi.org/10.1007/JHEP08(2016)154}{\emph{JHEP} {\bf 08} (2016)
  154}, [\href{http://arxiv.org/abs/1604.08602}{{\tt 1604.08602}}].

\bibitem{Cassani:2016ncu}
D.~Cassani, O.~de~Felice, M.~Petrini, C.~Strickland-Constable and D.~Waldram,
  \emph{{Exceptional Generalised Geometry for Massive IIA and Consistent
  Reductions}}, \href{http://dx.doi.org/10.1007/JHEP08(2016)074}{\emph{JHEP}
  {\bf 08} (2016) 074}, [\href{http://arxiv.org/abs/1605.00563}{{\tt
  1605.00563}}].

\bibitem{Inverso:2017lrz}
G.~Inverso, \emph{{Generalised Scherk-Schwarz reductions from gauged
  supergravity}}, \href{http://dx.doi.org/10.1007/JHEP12(2017)124}{\emph{JHEP}
  {\bf 12} (2017) 124}, [\href{http://arxiv.org/abs/1708.02589}{{\tt
  1708.02589}}].

\bibitem{Bugden:2021wxg}
M.~Bugden, O.~Hulik, F.~Valach and D.~Waldram, \emph{{G-Algebroids: A Unified
  Framework for Exceptional and Generalised Geometry, and
  Poisson\textendash{}Lie Duality}},
  \href{http://dx.doi.org/10.1002/prop.202100028}{\emph{Fortsch. Phys.} {\bf
  69} (2021) 2100028}, [\href{http://arxiv.org/abs/2103.01139}{{\tt
  2103.01139}}].

\bibitem{Bugden:2021nwl}
M.~Bugden, O.~Hulik, F.~Valach and D.~Waldram, \emph{{Exceptional algebroids
  and type IIB superstrings}},  \href{http://arxiv.org/abs/2107.00091}{{\tt
  2107.00091}}.

\bibitem{Malek:2017njj}
E.~Malek, \emph{{Half-Maximal Supersymmetry from Exceptional Field Theory}},
  \href{http://dx.doi.org/10.1002/prop.201700061}{\emph{Fortsch. Phys.} {\bf
  65} (2017) 1700061}, [\href{http://arxiv.org/abs/1707.00714}{{\tt
  1707.00714}}].

\bibitem{Malek:2019ucd}
E.~Malek, H.~Samtleben and V.~Vall~Camell, \emph{{Supersymmetric AdS$_7$ and
  AdS$_6$ vacua and their consistent truncations with vector multiplets}},
  \href{http://dx.doi.org/10.1007/JHEP04(2019)088}{\emph{JHEP} {\bf 04} (2019)
  088}, [\href{http://arxiv.org/abs/1901.11039}{{\tt 1901.11039}}].

\bibitem{Cassani:2020cod}
D.~Cassani, G.~Josse, M.~Petrini and D.~Waldram, \emph{{$\mathcal{N} $ = 2
  consistent truncations from wrapped M5-branes}},
  \href{http://dx.doi.org/10.1007/JHEP02(2021)232}{\emph{JHEP} {\bf 02} (2021)
  232}, [\href{http://arxiv.org/abs/2011.04775}{{\tt 2011.04775}}].

\bibitem{Faedo:2019cvr}
A.~F. Faedo, C.~Nunez and C.~Rosen, \emph{{Consistent truncations of
  supergravity and $\frac{1}{2}$-BPS RG flows in $4d$ SCFTs}},
  \href{http://dx.doi.org/10.1007/JHEP03(2020)080}{\emph{JHEP} {\bf 03} (2020)
  080}, [\href{http://arxiv.org/abs/1912.13516}{{\tt 1912.13516}}].

\bibitem{Maldacena:2000mw}
J.~M. Maldacena and C.~Nunez, \emph{{Supergravity description of field theories
  on curved manifolds and a no go theorem}},
  \href{http://dx.doi.org/10.1142/S0217751X01003935,
  10.1142/S0217751X01003937}{\emph{Int. J. Mod. Phys.} {\bf A16} (2001)
  822--855}, [\href{http://arxiv.org/abs/hep-th/0007018}{{\tt
  hep-th/0007018}}].

\bibitem{Bah:2012dg}
I.~Bah, C.~Beem, N.~Bobev and B.~Wecht, \emph{{Four-Dimensional SCFTs from
  M5-Branes}}, \href{http://dx.doi.org/10.1007/JHEP06(2012)005}{\emph{JHEP}
  {\bf 06} (2012) 005}, [\href{http://arxiv.org/abs/1203.0303}{{\tt
  1203.0303}}].

\bibitem{Malek:2015hma}
E.~Malek and H.~Samtleben, \emph{{Dualising consistent IIA/IIB truncations}},
  \href{http://dx.doi.org/10.1007/JHEP12(2015)029}{\emph{JHEP} {\bf 12} (2015)
  029}, [\href{http://arxiv.org/abs/1510.03433}{{\tt 1510.03433}}].

\bibitem{Gunaydin:1999zx}
M.~Gunaydin and M.~Zagermann, \emph{{The Gauging of five-dimensional, N=2
  Maxwell-Einstein supergravity theories coupled to tensor multiplets}},
  \href{http://dx.doi.org/10.1016/S0550-3213(99)00801-9}{\emph{Nucl. Phys. B}
  {\bf 572} (2000) 131--150}, [\href{http://arxiv.org/abs/hep-th/9912027}{{\tt
  hep-th/9912027}}].

\bibitem{Ceresole:2000jd}
A.~Ceresole and G.~Dall'Agata, \emph{{General matter coupled N=2, D = 5 gauged
  supergravity}},
  \href{http://dx.doi.org/10.1016/S0550-3213(00)00339-4}{\emph{Nucl. Phys. B}
  {\bf 585} (2000) 143--170}, [\href{http://arxiv.org/abs/hep-th/0004111}{{\tt
  hep-th/0004111}}].

\bibitem{Bergshoeff:2004kh}
E.~Bergshoeff, S.~Cucu, T.~de~Wit, J.~Gheerardyn, S.~Vandoren and
  A.~Van~Proeyen, \emph{{N = 2 supergravity in five-dimensions revisited}},
  \href{http://dx.doi.org/10.1088/0264-9381/23/23/C01}{\emph{Class. Quant.
  Grav.} {\bf 21} (2004) 3015--3042},
  [\href{http://arxiv.org/abs/hep-th/0403045}{{\tt hep-th/0403045}}].

\bibitem{deWit:1991nm}
B.~de~Wit and A.~Van~Proeyen, \emph{{Special geometry, cubic polynomials and
  homogeneous quaternionic spaces}},
  \href{http://dx.doi.org/10.1007/BF02097627}{\emph{Commun. Math. Phys.} {\bf
  149} (1992) 307--334}, [\href{http://arxiv.org/abs/hep-th/9112027}{{\tt
  hep-th/9112027}}].

\bibitem{Gunaydin:1983bi}
M.~Gunaydin, G.~Sierra and P.~K. Townsend, \emph{{The Geometry of N=2
  Maxwell-Einstein Supergravity and Jordan Algebras}},
  \href{http://dx.doi.org/10.1016/0550-3213(84)90142-1}{\emph{Nucl. Phys. B}
  {\bf 242} (1984) 244--268}.

\bibitem{Gunaydin:1984ak}
M.~Gunaydin, G.~Sierra and P.~K. Townsend, \emph{{Gauging the d = 5
  Maxwell-Einstein Supergravity Theories: More on Jordan Algebras}},
  \href{http://dx.doi.org/10.1016/0550-3213(85)90547-4}{\emph{Nucl. Phys. B}
  {\bf 253} (1985) 573}.

\bibitem{Gunaydin:1986fg}
M.~Gunaydin, G.~Sierra and P.~K. Townsend, \emph{{More on $d=5$
  Maxwell-einstein Supergravity: Symmetric Spaces and Kinks}},
  \href{http://dx.doi.org/10.1088/0264-9381/3/5/007}{\emph{Class. Quant. Grav.}
  {\bf 3} (1986) 763}.

\bibitem{Wolf65}
J.~Wolf, \emph{Complex homogeneous contact manifolds and quaternionic symmetric
  spaces}, \href{http://dx.doi.org/10.1512/iumj.1965.14.14065}{\emph{J. Appl.
  Math. Mech.} {\bf 14} (1965) }.

\bibitem{Alek68}
D.~V. Alekseevski\u{\i}, \emph{Compact quaternion spaces}, {\emph{Funkcional.
  Anal. i Prilo\v{z}en} {\bf 2} (1968) 11--20}.

\bibitem{AC05}
D.~V. Alekseevsky and V.~Cort\'{e}s, \emph{Classification of
  pseudo-{R}iemannian symmetric spaces of quaternionic {K}\"{a}hler type},  in
  \emph{Lie groups and invariant theory}, vol.~213 of \emph{Amer. Math. Soc.
  Transl. Ser. 2}, pp.~33--62.
\newblock Amer. Math. Soc., Providence, RI, 2005.
\newblock \href{http://dx.doi.org/10.1090/trans2/213/03}{DOI}.

\bibitem{Bergshoeff:2002qk}
E.~Bergshoeff, S.~Cucu, T.~De~Wit, J.~Gheerardyn, R.~Halbersma, S.~Vandoren
  et~al., \emph{{Superconformal N=2, D = 5 matter with and without actions}},
  \href{http://dx.doi.org/10.1088/1126-6708/2002/10/045}{\emph{JHEP} {\bf 10}
  (2002) 045}, [\href{http://arxiv.org/abs/hep-th/0205230}{{\tt
  hep-th/0205230}}].

\bibitem{Louis:2016qca}
J.~Louis and C.~Muranaka, \emph{{Moduli spaces of AdS$_{5}$ vacua in $
  \mathcal{N} $ = 2 supergravity}},
  \href{http://dx.doi.org/10.1007/JHEP04(2016)178}{\emph{JHEP} {\bf 04} (2016)
  178}, [\href{http://arxiv.org/abs/1601.00482}{{\tt 1601.00482}}].

\bibitem{Samtleben:2008pe}
H.~Samtleben, \emph{{Lectures on Gauged Supergravity and Flux
  Compactifications}},
  \href{http://dx.doi.org/10.1088/0264-9381/25/21/214002}{\emph{Class. Quant.
  Grav.} {\bf 25} (2008) 214002}, [\href{http://arxiv.org/abs/0808.4076}{{\tt
  0808.4076}}].

\bibitem{Trigiante:2016mnt}
M.~Trigiante, \emph{{Gauged Supergravities}},
  \href{http://dx.doi.org/10.1016/j.physrep.2017.03.001}{\emph{Phys. Rept.}
  {\bf 680} (2017) 1--175}, [\href{http://arxiv.org/abs/1609.09745}{{\tt
  1609.09745}}].

\bibitem{deWit:2011gk}
B.~de~Wit and M.~van Zalk, \emph{{Electric and magnetic charges in N=2
  conformal supergravity theories}},
  \href{http://dx.doi.org/10.1007/JHEP10(2011)050}{\emph{JHEP} {\bf 10} (2011)
  050}, [\href{http://arxiv.org/abs/1107.3305}{{\tt 1107.3305}}].

\bibitem{Louis:2012ux}
J.~Louis, P.~Smyth and H.~Triendl, \emph{{Supersymmetric Vacua in N=2
  Supergravity}}, \href{http://dx.doi.org/10.1007/JHEP08(2012)039}{\emph{JHEP}
  {\bf 08} (2012) 039}, [\href{http://arxiv.org/abs/1204.3893}{{\tt
  1204.3893}}].

\bibitem{Grana:2009im}
M.~Grana, J.~Louis, A.~Sim and D.~Waldram, \emph{{E7(7) formulation of N=2
  backgrounds}},
  \href{http://dx.doi.org/10.1088/1126-6708/2009/07/104}{\emph{JHEP} {\bf 07}
  (2009) 104}, [\href{http://arxiv.org/abs/0904.2333}{{\tt 0904.2333}}].

\bibitem{Ashmore:2015joa}
A.~Ashmore and D.~Waldram, \emph{{Exceptional Calabi-Yau spaces: the geometry
  of $\mathcal{N}=2$ backgrounds with flux}},
  \href{http://dx.doi.org/10.1002/prop.201600109}{\emph{Fortsch. Phys.} {\bf
  65} (2017) 1600109}, [\href{http://arxiv.org/abs/1510.00022}{{\tt
  1510.00022}}].

\bibitem{Ashmore:2016qvs}
A.~Ashmore, M.~Petrini and D.~Waldram, \emph{{The exceptional generalised
  geometry of supersymmetric AdS flux backgrounds}},
  \href{http://dx.doi.org/10.1007/JHEP12(2016)146}{\emph{JHEP} {\bf 12} (2016)
  146}, [\href{http://arxiv.org/abs/1602.02158}{{\tt 1602.02158}}].

\bibitem{LeDiffon:2008sh}
A.~Le~Diffon and H.~Samtleben, \emph{{Supergravities without an Action: Gauging
  the Trombone}},
  \href{http://dx.doi.org/10.1016/j.nuclphysb.2008.11.010}{\emph{Nucl. Phys. B}
  {\bf 811} (2009) 1--35}, [\href{http://arxiv.org/abs/0809.5180}{{\tt
  0809.5180}}].

\bibitem{deWit:2003ja}
B.~de~Wit, I.~Herger and H.~Samtleben, \emph{{Gauged locally supersymmetric D =
  3 nonlinear sigma models}},
  \href{http://dx.doi.org/10.1016/j.nuclphysb.2003.08.022}{\emph{Nucl. Phys. B}
  {\bf 671} (2003) 175--216}, [\href{http://arxiv.org/abs/hep-th/0307006}{{\tt
  hep-th/0307006}}].

\bibitem{Hohm:2014fxa}
O.~Hohm and H.~Samtleben, \emph{{Exceptional field theory. III. E$_{8(8)}$}},
  \href{http://dx.doi.org/10.1103/PhysRevD.90.066002}{\emph{Phys. Rev.} {\bf
  D90} (2014) 066002}, [\href{http://arxiv.org/abs/1406.3348}{{\tt
  1406.3348}}].

\bibitem{Bossard:2017aae}
G.~Bossard, M.~Cederwall, A.~Kleinschmidt, J.~Palmkvist and H.~Samtleben,
  \emph{{Generalized diffeomorphisms for $E_9$}},
  \href{http://dx.doi.org/10.1103/PhysRevD.96.106022}{\emph{Phys. Rev. D} {\bf
  96} (2017) 106022}, [\href{http://arxiv.org/abs/1708.08936}{{\tt
  1708.08936}}].

\bibitem{Bossard:2018utw}
G.~Bossard, F.~Ciceri, G.~Inverso, A.~Kleinschmidt and H.~Samtleben,
  \emph{{E$_{9}$ exceptional field theory. Part I. The potential}},
  \href{http://dx.doi.org/10.1007/JHEP03(2019)089}{\emph{JHEP} {\bf 03} (2019)
  089}, [\href{http://arxiv.org/abs/1811.04088}{{\tt 1811.04088}}].

\bibitem{Bossard:2021jix}
G.~Bossard, F.~Ciceri, G.~Inverso, A.~Kleinschmidt and H.~Samtleben,
  \emph{{E$_{9}$ exceptional field theory. Part II. The complete dynamics}},
  \href{http://dx.doi.org/10.1007/JHEP05(2021)107}{\emph{JHEP} {\bf 05} (2021)
  107}, [\href{http://arxiv.org/abs/2103.12118}{{\tt 2103.12118}}].

\bibitem{Coimbra:2011ky}
A.~Coimbra, C.~Strickland-Constable and D.~Waldram, \emph{{$E_{d(d)} \times
  \mathbb{R}^+$ generalised geometry, connections and M theory}},
  \href{http://dx.doi.org/10.1007/JHEP02(2014)054}{\emph{JHEP} {\bf 02} (2014)
  054}, [\href{http://arxiv.org/abs/1112.3989}{{\tt 1112.3989}}].

\bibitem{Riccioni:2007ni}
F.~Riccioni and P.~C. West, \emph{{E(11)-extended spacetime and gauged
  supergravities}},
  \href{http://dx.doi.org/10.1088/1126-6708/2008/02/039}{\emph{JHEP} {\bf 02}
  (2008) 039}, [\href{http://arxiv.org/abs/0712.1795}{{\tt 0712.1795}}].

\bibitem{deWit:2008ta}
B.~de~Wit, H.~Nicolai and H.~Samtleben, \emph{{Gauged Supergravities, Tensor
  Hierarchies, and M-theory}},
  \href{http://dx.doi.org/10.1088/1126-6708/2008/02/044}{\emph{JHEP} {\bf 02}
  (2008) 044}, [\href{http://arxiv.org/abs/0801.1294}{{\tt 0801.1294}}].

\end{thebibliography}\endgroup

\end{document}